\documentclass[
reprint, prd, aps, twocolumn, nofootinbib, preprintnumbers, superscriptaddress, balancelastpage, longbibliography,
 amsmath,amssymb,
 aps,
]{revtex4-2}

\newcommand{\Fermi}{\textit{Fermi}}

\newcommand{\FermiLAT}{\textit{Fermi}-LAT}

\newcommand{\beq}{\begin{equation}}
\newcommand{\eeq}{\end{equation}}

\newcommand\orcid[1]{\href{https://orcid.org/#1}{$\!$\includegraphics[scale=0.0045]{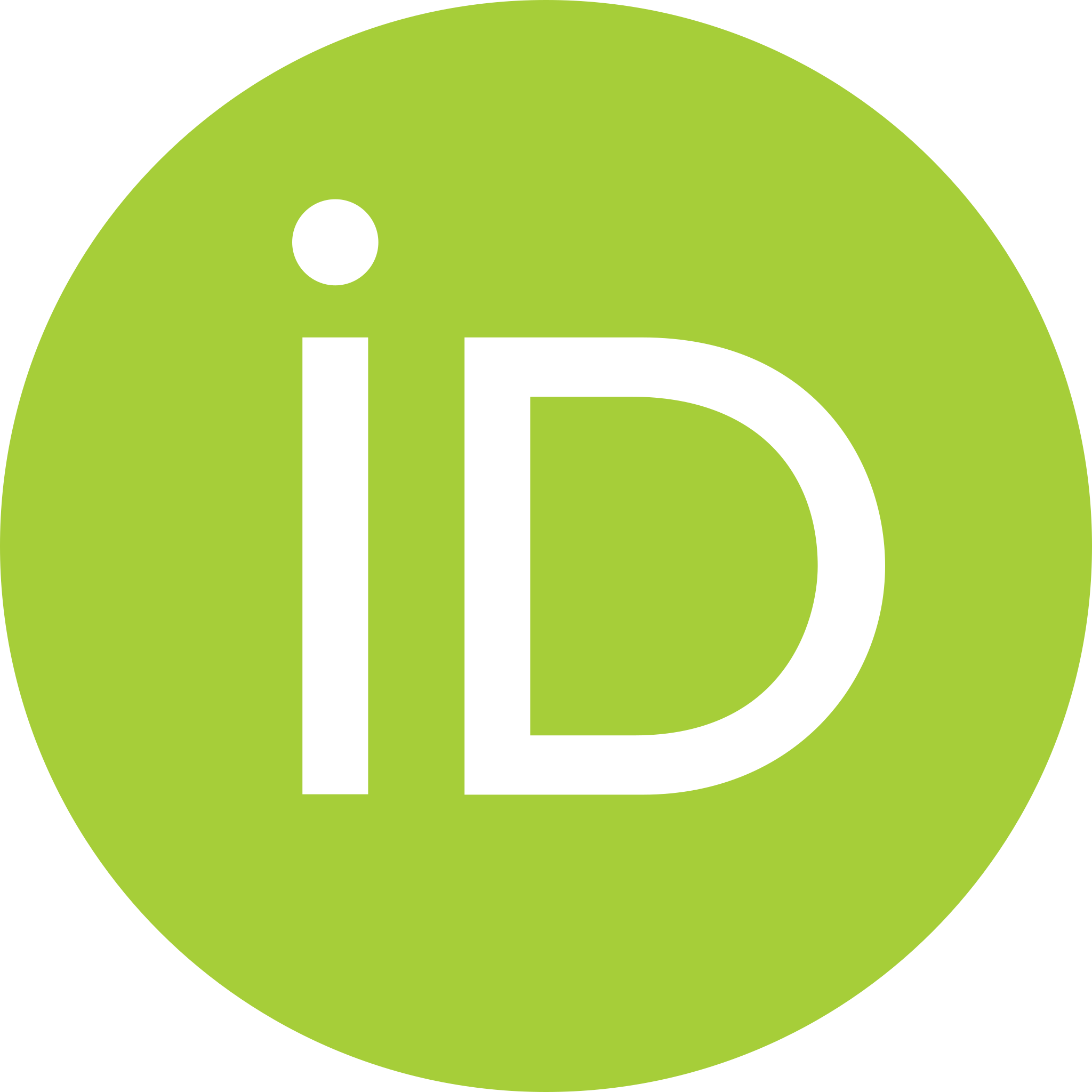} $\!\!$}}

\usepackage{graphicx}
\usepackage{dcolumn}
\usepackage{bm}
\usepackage{hyperref}
\usepackage{color}
\usepackage[dvipsnames]{xcolor}
\usepackage{todonotes}
\usepackage{pifont}
\usepackage{longtable,booktabs}
\usepackage{lineno}

\definecolor{lilac}{HTML}{b875eb}
\definecolor{green-blue}{HTML}{2DBC94}
\definecolor{slate-color}{HTML}{1b3644}
\definecolor{salvia-blue}{HTML}{96bfe6}
\hypersetup{
    colorlinks   = true,
    citecolor    = lilac,
    urlcolor     = lilac,
    linkcolor    = lilac
}

\begin{document}




\title{On the $\gamma$-ray Efficiency of Superluminous Supernovae: \\ Potential Detections and Population-Level Constraints}

\author{Milena Crnogor\v{c}evi\'{c} \orcid{0000-0002-7604-1779}}
\email{milena.crnogorcevic@fysik.su.se}
\affiliation{The Oskar Klein Centre, Department of Physics, Stockholm University, Stockholm 106 91, Sweden}

\author{Tim Linden \orcid{0000-0001-9888-0971}}
\email{linden@fysik.su.se}
\affiliation{The Oskar Klein Centre, Department of Physics, Stockholm University, Stockholm 106 91, Sweden}
\affiliation{Erlangen Centre for Astroparticle Physics (ECAP), Friedrich-Alexander-Universität \\ Erlangen-Nürnberg, Nikolaus-Fiebiger-Str. 2, 91058 Erlangen, Germany}

\author{Ariel Goobar \orcid{0000-0002-4163-4996}}
\email{ariel@fysik.su.se}
\affiliation{The Oskar Klein Centre, Department of Physics, Stockholm University, Stockholm 106 91, Sweden}

\author{Brian D.~Metzger \orcid{0000-0002-4670-7509}}
\email{bdm2129@columbia.edu}
\affiliation{Department of Physics and Columbia Astrophysics Laboratory, Columbia University, New York, NY 10027, USA}
\affiliation{Center for Computational Astrophysics, Flatiron Institute, 162 5th Ave, New York, NY 10010, USA}

\date{\today}

\begin{abstract}
\noindent \noindent Superluminous supernovae (SLSNe) are among the most energetic stellar explosions, yet their central power source remains uncertain. Models invoking magnetar spin-down or circumstellar interaction predict GeV $\gamma$-ray emission once the ejecta becomes transparent to high-energy photons. We search for such emission from 223 hydrogen-poor SLSNe using 17 years of \textit{Fermi}-LAT data, defining source-specific search windows based on the Bethe--Heitler transparency time. We find no significant ($\geq5\sigma$) GeV emission. A joint-likelihood analysis constrains the GeV-to-optical efficiency to $\eta < 1.3\times10^{-3}$, two orders of magnitude below the predictions for weakly magnetized magnetar nebulae. A hierarchical population analysis shows that fewer than $0.7\%$ of SLSNe-I can have $\eta > 10^{-2}$. SN~2017egm, however, shows a suggestive excess ($\sim$4~$\sigma$). In the 0.1--500\,GeV band, the observed $L_\gamma/L_{\rm opt} \sim 0.68$ for SN\,2017egm exceeds hadronic expectations by over an order of magnitude, favoring a magnetar origin. The non-detection of the similarly nearby SN~2018bsz disfavors simple uniform-efficiency scenarios, or potentially points to diversity in the underlying powering mechanisms. We also note a possible excess from SN~2024jlc, though continued \textit{Fermi}-LAT monitoring is needed because the source may still be within its transparency window.
\end{abstract}      

\maketitle
\section{Introduction}
\label{sec:intro}
\vspace{-0.2cm}

\noindent What powers the most luminous supernovae in the Universe? Superluminous supernovae (SLSNe) are an exceptionally bright class of transient objects, with intrinsic luminosities up to a hundred times greater than those of canonical core-collapse supernovae \citep{Quimby+07,GalYam18}. Over the past decade, wide-field transient surveys---including the Palomar Transient Factory (PTF), Sloan Digital Sky Survey (SDSS), Optical Gravitational Lensing Experiment (OGLE), Pan-STARSS1 (PS1), Catalina Sky Survey (CSS), SuperNova Legacy Survey (SNLS), Gaia, Dark Energy Survey (DES) and, most recently, the Zwicky Transient Facility (ZTF)---have discovered several hundred SLSNe extending to redshifts of $\sim$4 \cite{Palomar,SDSSII-1, SDSSII-2,OGLE,Panstarss.7733E..0EK, Catalina, SNLS:2013qua, Gaia2021A&A...649A...1G, DES_2016MNRAS.460.1270D, ZTF_2019PASP..131a8002B}. Despite the remarkable observational progress, the physical mechanisms powering these extreme systems remain uncertain.
\begin{figure}[t!]
\centering
\includegraphics[width=\columnwidth]{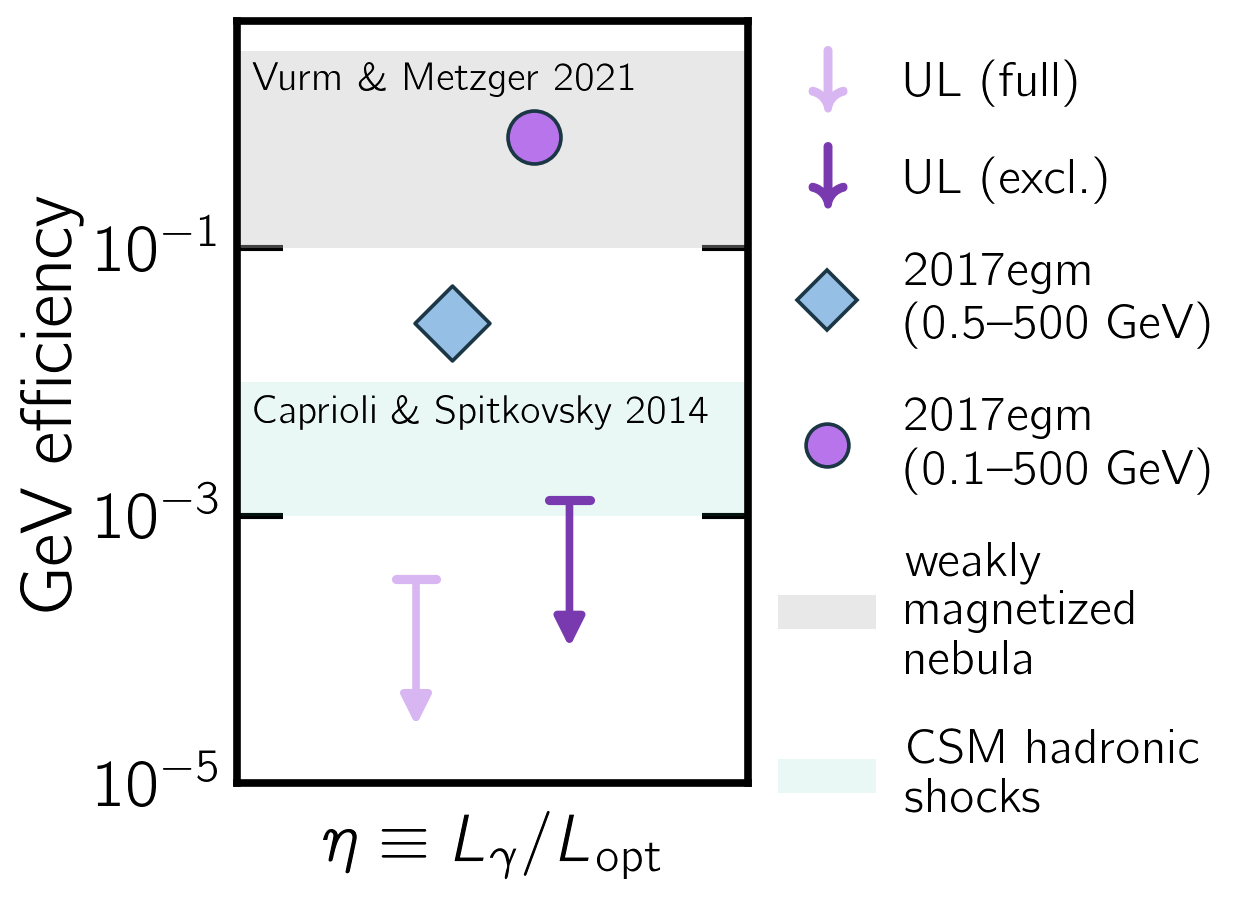}
\caption{GeV-to-optical efficiency for SLSNe-I, $\eta \equiv L_\gamma/L_{\rm opt}(t_{\rm BH})$. The gray band marks the range expected for weakly magnetized magnetar nebulae \citep{Vurm&Metzger21}, while the green band shows the approximate range $\eta \sim 10^{-3}$--$10^{-2}$ expected for hadronic shock scenarios \citep{Caprioli&Spitkovsky14}. Downward arrows show 95\% C.L.\ upper limits from the joint likelihood in the 0.5--500\,GeV energy band: light purple for the full 223-source sample, and dark purple after removing the 13 highest-weight sources in the optical-luminosity weighting model. The blue diamond and pink circle mark the SN~2017egm best-fit values in the 0.5--500\,GeV and 0.1--500\,GeV bands, respectively. The 0.1--500\,GeV value for SN~2017egm lies well above the range expected for CSM-powered shocks, favoring a magnetar interpretation if the excess is real.}
\label{fig:gev_efficiency}
\end{figure}

Two leading models have emerged to explain the extraordinary optical SLSN luminosities. In the \textbf{magnetar spin-down} scenario, a rapidly rotating, highly magnetized neutron star injects rotational energy into the expanding ejecta, sustaining the optical lightcurve over time \cite{Kasen&Bildsten10,Woosley10, Metzger+15}. In contrast, the \textbf{circumstellar medium (CSM) interaction} model attributes the emission to the conversion of kinetic energy into radiation as the ejecta collides with dense material surrounding the progenitor \cite{Smith+07,Chevalier&Irwin11}. Both mechanisms can produce GeV $\gamma$-rays observable with the \textit{Fermi} Large Area Telescope (LAT): magnetars via synchrotron and inverse Compton (IC) emission from the magnetar wind nebula, and CSM interaction through hadronic processes as shock-accelerated cosmic rays collide with the surrounding gas. While both mechanisms predict comparable flux in the 0.1--1\,GeV band, they are potentially distinguishable through the temporal profile of the $\gamma$-ray emission and the spectral shape \cite{Fang+20,Vurm&Metzger21}. 

In both scenarios, the emergent GeV emission depends not only on the energy injection mechanism but also on the evolving transparency of the ejecta. A well-motivated $\gamma$-ray search must therefore account for time-dependent effects. In the case of SLSNe, the relevant timescales---months to years post-explosion---are accessible to continuous $\gamma$-ray monitoring with the \textit{Fermi}-LAT, unfolding on human timescales rather than cosmological ones, and making SLSNe among the few classes of astrophysical objects where the onset of high-energy transparency can be observed in real time.

The emergent $\gamma$-ray flux from SLSNe is governed by the pair-production opacity within the ejecta. During the early phases of expansion, GeV photons undergo Bethe--Heitler (BH) pair production with the dense field of nuclei and thermal electrons, resulting in strong attenuation and effectively suppressing any escaping $\gamma$-ray emission. For photons of energy $E_\gamma \gg m_ec^2 \simeq 0.511$~MeV, this attenuation can be characterized by the BH optical depth through the oxygen-rich supernova ejecta, 
\begin{equation}
\label{eq:BH_tau}
\tau_{\rm BH} \simeq \frac{21}{2\pi}\alpha_{\rm fs} \left[\ln\left(\frac{2E_\gamma}{m_e c^2}\right) - \frac{109}{42} \right]\tau_{\rm T} \mathrel{\underset{E_\gamma = 1~{\rm GeV}}{\simeq}} 0.14\tau_{\rm T},
\end{equation}
where 
\begin{equation}
\label{eq:Th_xsec}
\tau_{\rm T} \simeq \frac{M_{\rm ej} \kappa_{\rm T}}{4\pi R_{\rm ej}^2}
\end{equation}
is the Thomson optical depth through the ejecta of mass $M_{\rm ej}$ and radius $R_{\rm ej}$, $\alpha_{\rm fs} \simeq 1/137$ is the fine-structure constant, and $\kappa_{\rm T}\simeq 0.2$~cm$^2$~g$^{-1}$ is the Thomson opacity for hydrogen-poor matter with an electron fraction $Y_e=0.5$. As the ejecta expands at velocity $v_{\rm ej} = R_{\rm ej}/t$, $\tau_{\rm BH}$ decreases, allowing $\gamma$-rays to escape once $\tau_{\rm BH}\approx 1$. The corresponding transparency time is 
\begin{equation}
\label{eq:transparency_time}
t_{\rm BH} \approx 91~{\rm d}
\left(\frac{M_{\rm ej}}{5~M_\odot}\right)^{1/2}
\left(\frac{v_{\rm ej}}{6000~{\rm km~s^{-1}}}\right)^{-1},
\end{equation}
typically corresponding to a few months post-explosion. The resulting timescale of $\sim(1-3)t_{\rm BH}$ defines the optimal search window for the GeV emission.

One of the key diagnostics distinguishing the two powering mechanisms of $\gamma$-ray emission is $\eta$,
\begin{equation}
\label{eq:Lratio}
\eta \equiv \frac{L_\gamma}{L_{\rm opt}(t \sim t_{\rm BH})},
\end{equation}
where $L_\gamma$ is the $\gamma$-ray luminosity and $L_{\rm opt}(t \sim t_{\rm BH})$ is the optical luminosity during the transparency window. 

If the magnetar nebula is relatively weakly magnetized, then its non-thermal emission is dominated by the IC process, peaking in the GeV band. At $t\sim t_{\rm BH}$, the ejecta are partially transparent ($\tau_{\rm BH}\approx1$), so the spin-down power is split between escaping GeV emission and reprocessed optical radiation in roughly comparable fractions, giving $\eta\sim1$ \citep{Vurm&Metzger21}.
By contrast, for more highly magnetized nebulae, a greater fraction of the nebula emission occurs via synchrotron radiation peaked at lower energies that fall within the X-ray band; the latter are efficiently absorbed and reprocessed into optical radiation, giving $\eta\ll1$. In CSM interaction models, only a small fraction ($p \lesssim 0.1$) of the shock power is channeled into relativistic particles, giving $\eta \lesssim p$ \cite{Caprioli&Spitkovsky14, Marti-Devesa:2024hic, Fang+20}. Measuring $\eta$ therefore provides a direct way to discriminate between these scenarios. 

Previous \textit{Fermi}-LAT searches for GeV counterparts to SLSNe have generally adopted uniform time windows rather than ejecta-dependent transparency periods. ~\citet{RT_2018A&A...611A..45R} conducted the first systematic analysis of 45 SLSNe using \textit{Fermi}-LAT data, finding no significant detections but setting upper limits that disfavor rapidly rotating magnetars with $B \sim 10^{12}$–$10^{13}$~G as the dominant power sources. Although the sample was largely composed of hydrogen-poor (Type~I) SLSNe, the limits were driven by a hydrogen-rich (Type~II) event, CSSS140222 which, as the authors note, was more plausibly powered by ejecta-CSM interaction than by a magnetar. A follow-up study combining \textit{Fermi}-LAT and VERITAS observations of SN~2015bn and SN~2017egm found no signal but demonstrated that SLSNe become transparent to GeV--TeV photons only a few months post-explosion \cite{VERITAS:2023msf}. 

Most recently, \citet{Li:2024ics} reported a tentative GeV excess from SN~2017egm approximately two months after explosion, consistent with magnetar model predictions. Our independent analysis, using a different time-window tied to the ejecta-specific BH transparency time, corroborates the presence of this excess at the position of SN~2017egm (Sec.~\ref{subsec:2017egm}). The agreement between two analyses is reassuring and strengthens the case that this excess is not an artifact of a particular analysis choice. Nevertheless, the tentative SN~2017egm excess remains below the conventional $5\sigma$ detection threshold. If real, SN~2017egm would constitute \textit{the first detection of GeV $\gamma$-rays from a SLSN}, providing direct evidence that a central engine remains active months after the explosion and opening a new electromagnetic window into the physics of SLSN powering mechanisms. 

In a concurrent study, the \textit{Fermi}-LAT Collaboration presents a dedicated analysis of six nearby SLSNe ($d < 200$~Mpc), including both Type~I and Type~II events \citep{Acero:2026slsn}. Using a summed PSF-type likelihood in the 0.1--100~GeV band and a Bayesian blocks algorithm to identify the optimal time window, they report a $\geq 5\sigma$ detection of SN~2017egm ($\mathrm{TS}=26$--33 depending on the time window), with emission concentrated between $\sim$50 and 160~days post-explosion. The temporal and spectral properties are consistent with predictions from a magnetar wind nebula model with low nebular magnetization ($\epsilon_B \lesssim 10^{-6}$), while a hadronic CSM interaction scenario is disfavored on the basis of both the timing of the $\gamma$-ray signal relative to the CSM shell crossings and the observed ratio $L_\gamma/L_{\rm opt} \sim 1$, which exceeds CSM expectations by one to two orders of magnitude. No other source in their sample is detected. 

In this paper, we build on these results and perform a systematic search for GeV emission from 223 \textbf{Type~I} SLSNe (hereafter SLSNe-I) using 17 years of \textit{Fermi}-LAT data. Unlike previous studies that included hydrogen-rich events, our analysis focuses exclusively on hydrogen-poor supernovae, for which GeV photons are expected to escape within months of explosion. 
For each source, we compute the ejecta-specific transparency time based on optical modeling of $M_{\rm ej}$ and $v_{\rm ej}$, defining physically motivated search windows. We perform both individual and joint likelihood analyses. We find no statistically significant ($\geq5\sigma$) GeV emission from either the individual or joint likelihood analysis. The tightest common-parameter constraint, $\eta < 1.3\times10^{-3}$ (95\% C.L.), lies two orders of magnitude below the predictions of weakly magnetized magnetar nebula models \citep{Vurm&Metzger21}, implying that either the magnetar-driven GeV production is strongly suppressed in SLSNe or only a fraction of events host GeV-bright central engines (Fig.~\ref{fig:gev_efficiency}). The nearest and most constraining source, SN~2017egm, shows a suggestive excess at $\sim$4~$\sigma$. A simple scaling argument using the non-detection of another nearby SN~2018bsz disfavors the possibility that SN~2017egm is representative of the population, although this comparison may be complicated if SN~2018bsz is not a fully canonical SLSN-I. This conclusion is nevertheless reinforced by the null results of the joint-likelihood analysis. We additionally note a suggestive excess from SN~2024jlc ($\sim2.7~\sigma$) in a supplementary sample of recent ZTF discoveries, though the source may still be within its transparency window and need continued \textit{Fermi}-LAT monitoring for confirmation.

The paper is organized as follows. Section~\ref{sec:data-selection} describes 
the SLSN sample, data selection, and \textit{Fermi}-LAT observations and Sec.~\ref{sec:analysis} outlines the analysis procedures. 
Section~\ref{sec:results} presents the resulting flux limits, joint-likelihood 
constraints, and inferred efficiency values. Section~\ref{sec:ztf} presents 
a supplementary analysis of recent ZTF-discovered SLSNe-I not included in the 
main sample. Section~\ref{sec:disc} discusses the implications of our findings 
for the dominant power source of SLSNe and prospects for future $\gamma$-ray 
and optical observations. Finally, Sec.~\ref{sec:concl} summarizes our results.

\section{Data Selection}
\label{sec:data-selection}
\subsection{SLSNe}

Our SLSN-I sample is drawn from the comprehensive catalog of \citet{Gomez:2024xce}, which compiles 265 spectroscopically confirmed hydrogen-poor events discovered through December~31,~2022. The catalog integrates data from the \textit{Open Supernova Catalog}, the \textit{Transient Name Server} (TNS)\footnote{\url{https://www.wis-tns.org/}, accessed on October 30, 2025.}, WISeREP\footnote{\url{https://www.wiserep.org/}, accessed on October 30, 2025.}, published literature, and the \textit{FLEET} follow-up program, providing over 30,000 individual photometric detections spanning the ultraviolet through near-infrared with uniform calibration and quality control \citep{Guillochon:2016rhj, Gomez:2020jyq, Gomez:2022xsq}.

Events in the catalog are assigned to three quality tiers: \textit{gold} (171), \textit{silver} (70), and \textit{bronze} (24). Gold SLSNe have spectra consistent with the SLSN-I classification and multi-band photometry covering both the rise and decline of the optical emission. Silver objects satisfy at least one relaxed criterion---but may lack pre- or post-peak coverage, have noisy or single-band photometry, uncertain redshift estimates, or moderate extinction---while bronze events lack public spectra or have ambiguous classifications. For our \FermiLAT\ analysis, we retain only the \textbf{gold} and \textbf{silver} subsamples (\textbf{241 sources}), which provide reliable explosion times and well-constrained positions. We then analyze \FermiLAT\ $\gamma$-ray data after applying the following selection cuts:


\begin{figure*}
\centering
\includegraphics[width=\textwidth]{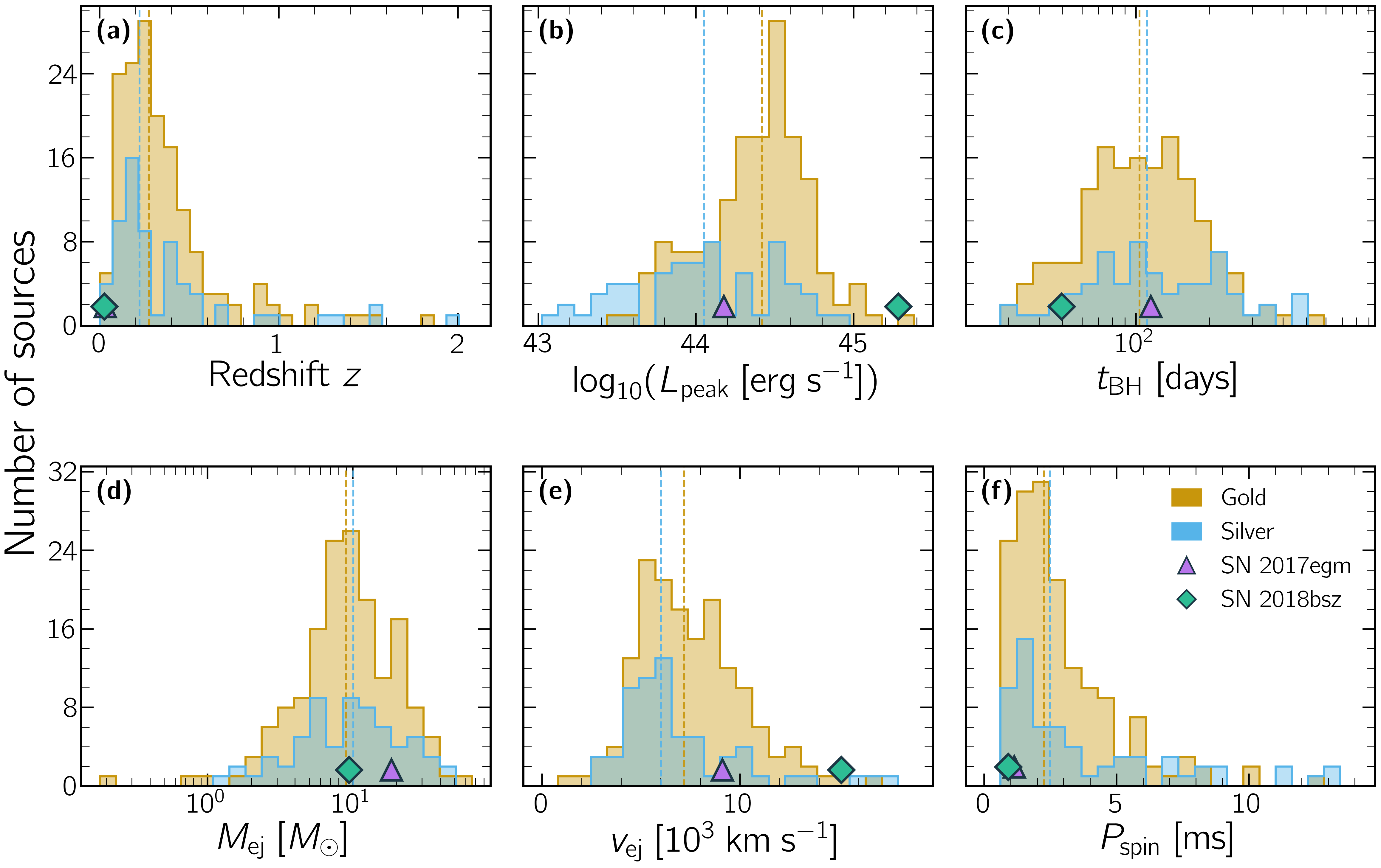}
\caption{Properties of the final 223 SLSN-I sample. Gold (orange) and silver (blue) sources are drawn from the \citet{Gomez:2024xce} catalog. Dashed vertical lines mark the median of each subsample: \textbf{(a)}~redshift distribution, peaking at $z\approx0.26$; \textbf{(b)}~peak bolometric luminosity from \texttt{MOSFiT} fits; \textbf{(c)}~BH transparency time $t_{\rm BH}$ (Eq.~\ref{eq:transparency_time}), which defines the onset of the GeV search window; \textbf{(d)}~ejecta mass $M_{\rm ej}$, with a median of $\sim$10~$M_\odot$; \textbf{(e)}~ejecta velocity $v_{\rm ej}$, peaking near 6000~km~s$^{-1}$; and \textbf{(f)}~magnetar spin period $P_{\rm spin}$, strongly peaked at $\sim$2--3~ms. The properties of SN~2017egm (\textcolor{lilac}{\ding{115}}) and SN~2018bsz (\textcolor{green-blue}{\ding{117}}) are indicated for reference in each panel to show how these two nearby, high-weight events compare to the full sample.}
\label{fig:sample_properties}
\end{figure*}

\begin{enumerate}
    \item \textbf{Temporal overlap with \FermiLAT.} We require that the
    observer-frame $\gamma$-ray search window $[t_0 + 0.5\,t_{\rm BH}(1+z),\, t_0 + 3.0\,t_{\rm BH}(1+z)]$ is fully contained within the \FermiLAT\ science data span (MJD~54682.65--60860.0), where the $(1+z)$ factor converts rest-frame to observer-frame timescales (see Sec.~\ref{subsec:time_window}). Sources whose window either predates the \Fermi\ launch or extends beyond the analysis cutoff are excluded. Fifteen sources are removed by this criterion, leaving 226 events.
    \item \textbf{Galactic latitude.} We exclude sources within $|b| < 10^{\circ}$ of the Galactic plane, where the bright and structured diffuse emission substantially degrades the \FermiLAT\ point-source sensitivity and increases systematic errors in the diffuse background model. This removes two additional sources (SN~2010hy and SN~2020wnt).
    \item \textbf{Proximity to bright \FermiLAT\ sources.} We exclude SLSNe falling within the 95\% positional containment radius of any 4FGL-DR4 source with TS $>$ 100 \cite{Fermi-LAT:2022byn}. 
    No sources are removed by this criterion.
    \item \textbf{Valid background control window.} For each source we assign a background control window of the same observer-frame duration as the signal window at the event's location, also required to lie fully within the \FermiLAT\ data span (see Sec.~\ref{subsec:time_window}, Eqs.~\ref{eq:control_pre} or \ref{eq:control_post}). This criterion removes one additional event (PS110ahf).   
\end{enumerate}

After these cuts, the final analysis sample comprises \textbf{223 SLSNe-I}. A complete list of sources is provided in the Appendix \ref{app:A}. The properties of the final sample are summarized in Fig.~\ref{fig:sample_properties}. The redshift distribution peaks at $z\approx0.26$, driven by the magnitude limits of the primary discovery surveys (ZTF and ATLAS at $m_r\sim20.5$; DES and PS1 at $m_r\sim23.5$). Population-averaged \texttt{MOSFiT} posteriors from \citet{Gomez:2024xce} yield ejecta masses and velocities of $M_{\rm ej} = 9.3^{+12.9}_{-4.8}~M_{\odot}$ and $v_{\rm ej} = 6800^{+3400}_{-2000}$~kms$^{-1}$, with a median magnetar transparency timescale $t_{\rm BH}\approx110$~days.

We caution that, for individual sources, \textbf{the \texttt{MOSFiT} posteriors can be sensitive to the temporal selection of the input lightcurves}. In particular, for one of the closest candidates---SN~2017egm---\citet{Gomez:2024xce} catalog finds $M_{\rm ej} = 18.5^{+7.6}_{-4.5}~M_\odot$, whereas the earlier fit by \citet{Nicholl+17}, based on data around optical peak, yields $M_{\rm ej} \approx 3~M_\odot$. The difference likely reflects the choice of late-time data exhibiting bumps and plateaus \citep{Zhu:2023ntt, Hosseinzadeh:2021uep} that are not captured by the magnetar-plus-decay model. We adopt the \citet{Gomez:2024xce} posteriors throughout for sample-wide consistency and specifically consider their impact on SN~2017egm in Sec.~\ref{subsec:2017egm}.

\subsection{\FermiLAT}
\FermiLAT\ is a pair-conversion $\gamma$-ray telescope sensitive to photons from $\sim$100~MeV to more than 500~GeV, with a field of view of $\sim$2.4~sr and full-sky coverage every $\sim$3~hours \citep{FermiMission2009}. We consider 17 years of \FermiLAT\ Pass~8 data spanning August~4,~2008 (MJD~54682) through August~4,~2025 (MJD~60860.0), obtained from the publicly accessible \Fermi\ \textit{Science Support Center} (FSSC)\footnote{\url{https://fermi.gsfc.nasa.gov/ssc/data/}, accessed on October 23, 2025.}. We employ the \texttt{fermipy} package (v.~1.2.2), with the underlying \texttt{FermiTools} (v.~2.2.0)\footnote{\url{https://fermi.gsfc.nasa.gov/ssc/data/analysis/software/}, accessed on October 23, 2025.}. We select events in the energy range 500~MeV--500~GeV, divided into eight logarithmic bins per decade. The lower bound of 500~MeV is chosen to limit the broad LAT point-spread function (PSF) effects at lower energies, which would increase source confusion and dilute the signal from our target positions. We use the \texttt{P8R3\_SOURCE\_V3} instrument response functions (IRFs) with \texttt{FRONT+BACK} event type (\texttt{evtype}$=3$). We use the standard LAT Collaboration models for the Galactic diffuse emission (\texttt{gll\_iem\_v07.fits}) and the isotropic background (\texttt{iso\_P8R3\_SOURCE\_V3\_v1.txt})\footnote{\url{https://fermi.gsfc.nasa.gov/ssc/data/access/lat/BackgroundModels.html}, accessed on October 23, 2025.}. For each SLSN-I, we define a $10^{\circ} \times 10^{\circ}$ region of interest (RoI) centered on its optical position, with a pixel size of $0.1^{\circ}$. To account for $\gamma$-ray leakage from sources just outside the RoI, we model all point and extended sources from the 4FGL-DR4 catalog within a $15^{\circ} \times 15^{\circ}$ region \cite{Fermi-LAT:2022byn}. We apply a zenith-angle cut of $100^{\circ}$ to suppress any contamination from the Earth limb. We enable the energy dispersion for the target source and all catalog sources, but disable it for the diffuse components. For the likelihood fit, we free the normalizations of the Galactic and isotropic diffuse models, the normalizations of catalog sources with test statistic, $\mathrm{TS}$, greater than 25 within $5^{\circ}$ of the target, and all spectral parameters of sources with $\mathrm{TS} > 500$ within $7^{\circ}$. The target SLSN-I is modeled as a point source at the optical position with a power-law spectrum,
\begin{equation}
\label{eq:powerlaw}
\frac{dN}{dE} = N_0 \left(\frac{E}{E_0}\right)^{-\Gamma},
\end{equation}
with pivot energy $E_0 = 1$~GeV, initial power-law index $\Gamma = 2.0$, and prefactor $N_0 = 10^{-14}$~cm$^{-2}$~s$^{-1}$~MeV$^{-1}$. The full fitting procedure---including initial RoI optimization, iterative parameter freeing, and spectral energy distribution extraction---is described in Section~\ref{subsec:individual}.

\section{Data analysis}
\label{sec:analysis}

\subsection{Time-Window Selection}
\label{subsec:time_window}

A physically motivated search for GeV emission from SLSNe-I must account for the time-dependent transparency of the ejecta. The GeV photons are initially trapped by the BH pair production and the ejecta become optically thin at the characteristic time $t_{\rm BH}$ (Eq.~\ref{eq:transparency_time}), which depends on the ejecta mass $M_{\rm ej}$ and velocity $v_{\rm ej}$ of each event. Rather than adopting a single, population-averaged time window, we compute $t_{\rm BH}$ individually for each SLSN-I using the posterior distributions from the \texttt{MOSFiT} magnetar-plus-radioactive-decay model fits in~\citet{Gomez:2024xce}, propagating the reported posterior uncertainties on $M_{\rm ej}$ and $v_{\rm ej}$. Across the sample, $t_{\rm BH}$ spans roughly 30--840~days (median $\approx 110$~days), with typical fractional uncertainties $\Delta t_{\rm BH}/t_{\rm BH} \simeq0.2$, reflecting the wide range of ejecta properties (see Fig.~\ref{fig:sample_properties}c).

For each SLSN-I, we define a \FermiLAT\ observation window
\begin{equation}
\big[t_0 + 0.5\,t_{\rm BH}(1+z), t_0 + 3.0\,t_{\rm BH}(1+z)\big],
\label{eq:tbest_window}
\end{equation}
where the $(1+z)$ factor converts the rest-frame transparency timescale to the observer-frame elapsed time recorded by \FermiLAT. The lower bound captures the onset of transparency while allowing for uncertainties in the estimate of $t_{\rm BH}$, and the upper bound encompasses the period over which the central engine or shock power is expected to remain significant. In the observer frame, these windows have durations ranging from $\sim80$ to $\sim$1740~days (median $\sim$350~days).

To assess the rate of spurious TS fluctuations and verify that any candidate signal is temporally coincident with the supernova, we construct a background-only control window for each source. The primary control interval is
\begin{equation}
\label{eq:control_pre}
\big[t_0 - 3.0\,t_{\rm BH}(1+z), t_0 - 0.5\,t_{\rm BH}(1+z)\big],
\end{equation}
which has the same duration as the signal window but precedes the explosion, probing an identical background environment (diffuse emission and catalog point sources) at the same sky position. When this pre-explosion window extends before the start of the \Fermi\ mission, we instead adopt a post-explosion control interval,
\begin{equation}
\label{eq:control_post}
\big[t_0 + 3.5t_{\rm BH}(1+z), t_0 + 6.0t_{\rm BH}(1+z)\big],
\end{equation}
placed well after the expected $\gamma$-ray signal has faded. Of the 223 SLSN-I in our sample, 214 use the pre-explosion window and 9 fall back to the post-explosion interval. The resulting empirical background-only TS distribution provides a direct, data-driven calibration of our detection threshold.

\subsection{Individual-Source Likelihood Analysis}
\label{subsec:individual}

For each SLSN-I in the sample, we perform a binned likelihood analysis within the source-specific time window defined in Section~\ref{subsec:time_window}. We begin with an initial optimization of the RoI using \texttt{fermipy}'s \texttt{gta.optimize()} routine, which iteratively adjusts the spectral parameters of bright sources. We then free the normalizations of the Galactic and isotropic diffuse components, the normalizations of 4FGL-DR4 sources with $\mathrm{TS}>25$ within $5^{\circ}$ of the target, and all spectral parameters of sources with $\mathrm{TS}>500$ within $7^{\circ}$. The target SLSN-I, modeled as a point source with the power-law spectrum of Eq.~\ref{eq:powerlaw}, is included with free normalization and spectral index of 2. A joint likelihood maximization (\texttt{gta.fit()}) then yields the best-fit spectral parameters and global log-likelihood.

For each source, we extract a spectral energy distribution (SED) using \texttt{fermipy}'s \texttt{gta.sed()} method, which performs independent likelihood scans in each energy bin. This produces the profile log-likelihood per energy bin $j$, $\Delta\ln\mathcal{L}_j(F_j)$, as a function of the differential flux $F_j$. These per-bin likelihood profiles serve as input to the power-law grid scan: from the per-bin SED likelihood profiles, we construct a two-dimensional TS surface over the space of power-law parameters. For a given normalization $N_0$ and spectral index $\Gamma$, the predicted differential flux in energy bin $j$ is $F_j(N_0,\Gamma) = N_0(E_j/E_0)^{-\Gamma}\Delta E_j$, and the total log-likelihood is then given by
\begin{equation}
\label{eq:loglike_grid}
\log\mathcal{L}(N_0,\Gamma) = \sum_j \Delta\log\mathcal{L}_j\!\big(F_j(N_0,\Gamma)\big).
\end{equation}
We evaluate this on a grid of 100 logarithmically spaced normalization values from $10^{-18}$ to $10^{-12}$~MeV$^{-1}$~cm$^{-2}$~s$^{-1}$ and 50 linearly spaced index values from $\Gamma=1.5$ to $4.0$. As usual, the TS is defined as
\begin{equation}
\label{eq:TS_def}
\mathrm{TS}(N_0,\Gamma) = 2\big[\log\mathcal{L}(N_0,\Gamma) - \log\mathcal{L}_0\big],
\end{equation}
where $\log\mathcal{L}_0 \equiv \log\mathcal{L}(N_0\to0)$ is the null-hypothesis (background-only) log-likelihood. We then use the matched control-window analyses described in Sec.~\ref{subsec:time_window} to empirically calibrate the null distribution of the maximum $\mathrm{TS}$ for this search. 


\subsection{Joint-Likelihood Analysis}
\label{subsec:jla}

Individual SLSN-I analyses are limited by the low photon statistics available for any single source. To harness the full statistical power of the sample, we combine all sources into a joint-likelihood analysis (JLA) framework that ties their $\gamma$-ray flux normalizations to a shared physical model. This approach mirrors the strategy used in population-level \textit{Fermi}-LAT analyses of low-luminosity active galactic nuclei and dark-matter targets (see, e.g., \citep{Fermi-LAT:2016uux, McDaniel:2023bju, Crnogorcevic:2025gxd}), adapted here to models for SLSN-I central engines and shock interaction.

For a given physical weighting model, the predicted flux normalization of source $i$ is parameterized as $c_i = A\,w_i$, where $A$ is a single shared amplitude and $w_i$ is a physically motivated weight encoding the relative expected $\gamma$-ray brightness of event $i$. The joint log-likelihood is then
\begin{equation}
\label{eq:joint_loglike}
\log\mathcal{L}_{\rm joint}(A,\Gamma) = \sum_{i=1}^{N} \Delta\log\mathcal{L}_i\!\big(c_i = A\,w_i;\,\Gamma\big),
\end{equation}
where $\Delta\log\mathcal{L}_i$ is the individual-source log-likelihood profile from the 2D grid scan described in Section~\ref{subsec:individual}, evaluated at normalization $c_i$ and spectral index $\Gamma$. We maximize Eq.~\ref{eq:joint_loglike} over $(A,\Gamma)$ to obtain joint best-fit parameters and a joint TS. We test \textbf{six} physically motivated weighting prescriptions, spanning the major theoretical families of SLSN-I power sources:

\paragraph{Uniform.} All sources contribute equally: $w_i = 1$. This model assumes identical observed fluxes regardless of distance or physical properties and serves as a control to verify that the joint-likelihood does not introduce spurious correlations.

\paragraph{Standard candle ($d_L^{-2}$).} The simplest physical hypothesis is that all SLSNe-I share a common intrinsic $\gamma$-ray luminosity, so the expected flux scales as $w_i \propto d_{L,i}^{-2}$. This tests a luminosity-independent, purely geometric scenario.

\paragraph{Constant optical efficiency at transparency ($\eta$).} If the GeV emission tracks the optical luminosity at the time when the ejecta becomes transparent, the expected flux scales as
\begin{equation}
w_i \propto \frac{L_{{\rm opt},i}(t_{{\rm BH},i})}{d_{L,i}^{2}}.
\end{equation}
We compute $L_{{\rm opt},i}(t_{\rm BH})$ using the magnetar thermalization model from \citet{Nicholl+17d}, where 
\begin{equation}
    L_{{\rm opt}}(t) = L_{\rm sd}(t)\left[1-e^{-\tau_{\gamma}(t)} \right], \qquad  \tau_\gamma(t) = \frac{3\,\kappa_\gamma\,M_{\rm ej}}{4\pi\,(v_{\rm ej}\,t)^2}
\label{eq:kappa_gamma}
\end{equation}
where $\kappa_{\gamma}$ is the effective $\gamma$-ray opacity, and $L_{\rm sd}(t)$ is the spin-down luminosity given by 
\begin{equation}
\label{eq:Lsd}
L_{\rm sd}(t) = \frac{L_0}{\left(1 + t/T_{\rm sd}\right)^2},
\end{equation}
with $L_0$ representing the initial spin-down power and $T_{\rm sd}$ the characteristic spin-down timescale. All five quantities---$v_{\rm ej}, M_{\rm ej}, \kappa_{\gamma}, L_0$, and $T_{\rm sd}$---are extracted from the \texttt{MOSFiT} fits in \citet{Gomez:2024xce}. In Eq.~\ref{eq:kappa_gamma}, the factor $e^{-\tau_\gamma}$ captures the declining thermalization efficiency as the ejecta expand: at early times ($\tau_\gamma \gg 1$), nearly all of the spin-down power is reprocessed into optical radiation, while at $t \gtrsim t_{\rm BH}$ the ejecta become partially transparent and $L_{\rm opt}$ drops below $L_{\rm sd}$. The corresponding common parameter is $\eta$ which directly probes the fraction of the optical luminosity at transparency that emerges as GeV $\gamma$-rays. 

\paragraph{Magnetar spin-down power at transparency ($\varepsilon_{\rm sd}$).} In magnetar-powered SLSNe, the escaping GeV luminosity should trace the residual spin-down power at the moment the ejecta become transparent. Assuming a constant conversion efficiency, $\varepsilon_{\rm sd}$, the predicted luminosity scales as $L_{\gamma,i}\propto L_{\rm sd}(t_{{\rm BH},i})$, introduced in Eq.~\ref{eq:Lsd}. We compute $L_{\rm sd}(t_{\rm BH})$ from the \texttt{MOSFiT} posteriors for each source and set $w_i \propto L_{\rm sd}(t_{{\rm BH},i})\,d_{L,i}^{-2}$. This model directly encodes the prediction that the GeV signal traces the residual engine power at the moment the ejecta becomes transparent. 

The two magnetar weighting models in \textit{(c--d)} are physically related but not identical. Because $L_{\rm opt}$ is derived from $L_{\rm sd}$ via Eq.~\ref{eq:kappa_gamma}, the two efficiencies are connected by $\varepsilon_{\rm sd} = \eta\,(1-e^{-\tau_{\gamma}})^{-1}$: at the transparency time $\tau_\gamma \approx 1$, so $\varepsilon_{\rm sd} \approx 1.6\,\eta$ and the two models yield similar weights. They diverge, however, for sources where $\tau_\gamma(t_{\rm BH})$ departs significantly from unity---either because the ejecta are already largely transparent ($\tau_\gamma \ll 1$, so $L_{\rm opt} \ll L_{\rm sd}$ and $\varepsilon_{\rm sd} \gg \eta$) or still heavily obscured ($\tau_\gamma \gg 1$, so $L_{\rm opt} \approx L_{\rm sd}$ and $\varepsilon_{\rm sd} \approx \eta$). Including both models therefore tests whether the results are sensitive to this distinction.



\begin{figure*}[t!]
    \centering
    \includegraphics[width=\textwidth]{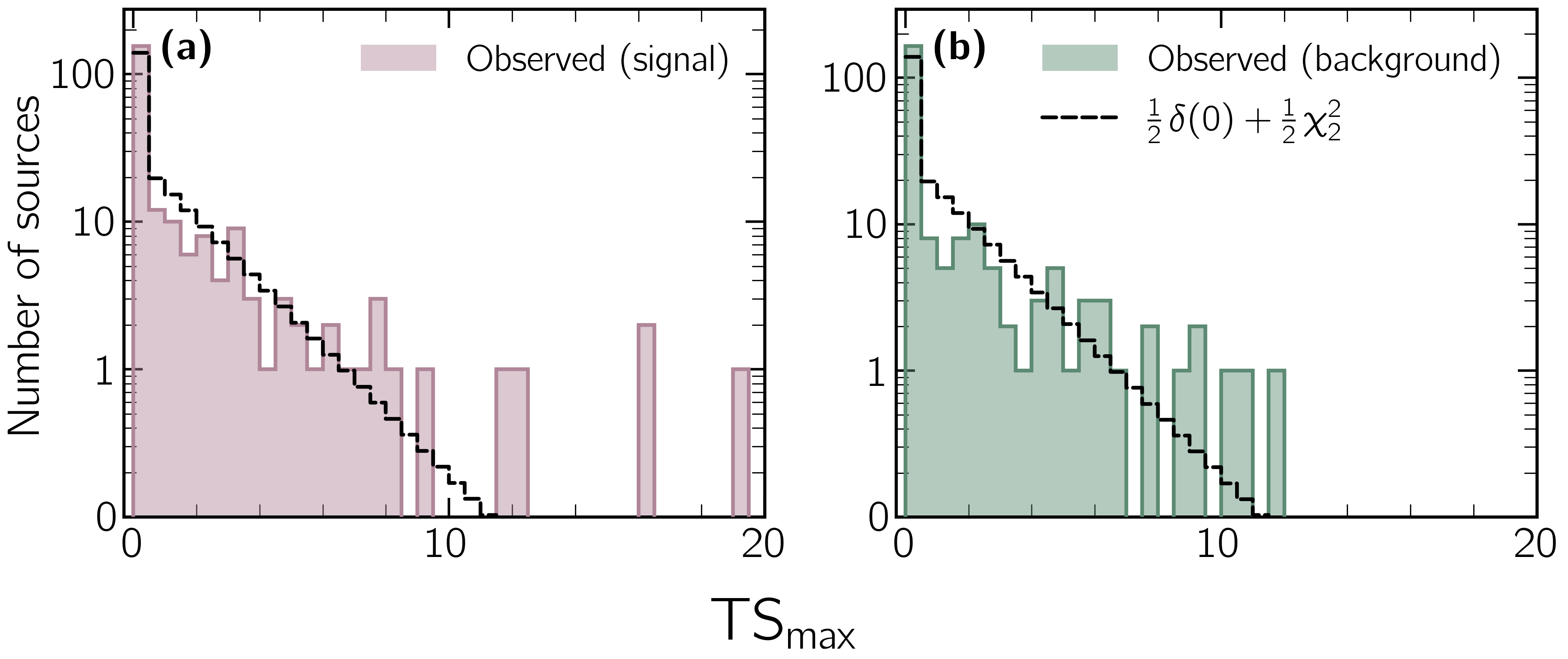}
    \caption{%
        Distribution of $\mathrm{TS}_{\max}$ values for 223 SLSNe-I, evaluated in the signal window (pink; Eq.~\ref{eq:tbest_window}) and in the matched pre/post-explosion control window (green; Eqs.~\ref{eq:control_pre} and \ref{eq:control_post}). Both windows span $2.5\,t_{\rm BH}(1+z)$ observer-frame days by construction. The dashed curve shows the asymptotic bounded-Wilks expectation $\tfrac{1}{2}\delta(0)+\tfrac{1}{2}\chi^{2}_{2}$. We observe 6 (signal) and 5 (control) sources with $\mathrm{TS}_{\max}>9$, indicating a heavier empirical high-TS tail than the asymptotic expectation, which is a known limitation of the bounded-Wilks approximation in LAT analyses with limited photon statistics over short time windows. The most significant candidate is SN~2017egm, with $\mathrm{TS}_{\max}=19.5$.}
    \label{fig:ts_dist}
\end{figure*}

\paragraph{Kinetic-energy.} In CSM interaction-powered scenarios, $\gamma$-rays are produced when shock-accelerated particles collide with the CSM, and the total available energy budget is set by the ejecta kinetic energy $E_{\rm k} = \tfrac{3}{10}M_{\rm ej}v_{\rm ej}^2$. We therefore test $w_i \propto E_{{\rm k},i}\,d_{L,i}^{-2}$.

\paragraph{Shock-power.} If GeV emission traces the instantaneous shock power rather than the integrated kinetic energy, the luminosity should scale with the rate of energy dissipation. In the absence of resolved shock profiles, we adopt $E_{\rm k}/t_{\rm rise}$ as a proxy for the peak shock luminosity, where $t_{\rm rise}$ is the optical rise time from explosion to $r$-band peak as measured and reported in \citet{Gomez:2024xce}. The resulting weight is $w_i \propto (E_{{\rm k},i}/t_{{\rm rise},i})\,d_{L,i}^{-2}$. We define the shock GeV efficiency as the ratio of the $\gamma$-ray luminosity to this shock-power proxy,
\begin{equation}
\label{eq:epssh}
 \varepsilon_{\rm sh} \equiv \frac{L_\gamma}{E_k/t_{\rm rise}}.
\end{equation}
In the calorimetric limit of Refs.~\citep{Murase+19,Fang+20}, this quantity directly probes the hadronic acceleration efficiency at radiative shocks, with $\varepsilon_{\rm sh} \sim 10^{-3}-10^{-1}$ inferred from nova observations. We note, however, that the calorimetric framework relates the $\gamma$-ray production to shock power at optical peak, when the shocks are radiative and hadronic interactions are efficient, but the ejecta are still opaque to GeV photons. Our measurement, on the other hand, probes the escaping GeV flux at $\sim t_{\rm BH}$, typically months after peak, when the shock power has declined and the hadronic interactions may be less efficient. The relationship between $\varepsilon_{\rm sh}$ as defined here and the intrinsic acceleration efficiency of \citet{Fang+20} therefore depends on the time evolution of both the shock power and the ejecta opacity. Direct comparison therefore requires a time-dependent model that is beyond the scope of this work. 

Throughout this paper we thus distinguish between \textbf{three primary efficiency metrics}. First, the GeV-to-optical efficiency $\boldsymbol \eta$ (Eq.~\ref{eq:Lratio}), defined as the ratio of $\gamma$-ray to optical luminosity at the time when the ejecta become transparent to GeV photons. This is a direct observable, independent of the assumed power source. In the magnetar scenario, $\eta$ also constrains the spin-down efficiency $\boldsymbol{\varepsilon_{\rm sd}} \equiv L_{\gamma}/L_{\rm sd} = \eta\,(1-e^{-\tau_{\gamma}})$. Finally, for shock-powered models we define $\boldsymbol{\varepsilon_{\rm sh}}$ (Eq.~\ref{eq:epssh}), the ratio of the $\gamma$-ray luminosity to the peak shock power proxy $E_k/t_{\rm rise}$, which probes the hadronic acceleration efficiency at radiative shocks (subject to the caveats discussed above.)

Together, these six weighing prescriptions span the primary theoretical scenarios for GeV emission from SLSNe-I: purely geometric flux scaling in~\textit{(b)}, magnetar spin-down physics in~\textit{(c--d)}, and shock-interaction models in~\textit{(e--f)}. A joint likelihood maximized independently for each model then yields population-level constraints on the corresponding physical efficiency parameter.

\section{Results}
\label{sec:results}

\subsection{Individual-source TS distribution}

For each of the 223 SLSN-I, we compute the maximum test statistic $\mathrm{TS}_{\max}=\max\mathrm{TS}(N_0,\Gamma)$ over the two-dimensional grid scan described in Sec.~\ref{subsec:individual}, both in the signal window (Eq.~\ref{eq:tbest_window}) and in a matched control window (Eqs.~\ref{eq:control_pre} and \ref{eq:control_post}). The resulting $\mathrm{TS}_{\max}$ distributions are shown in Fig.~\ref{fig:ts_dist}.

\begin{figure}
    \centering
    \includegraphics[width=\columnwidth]{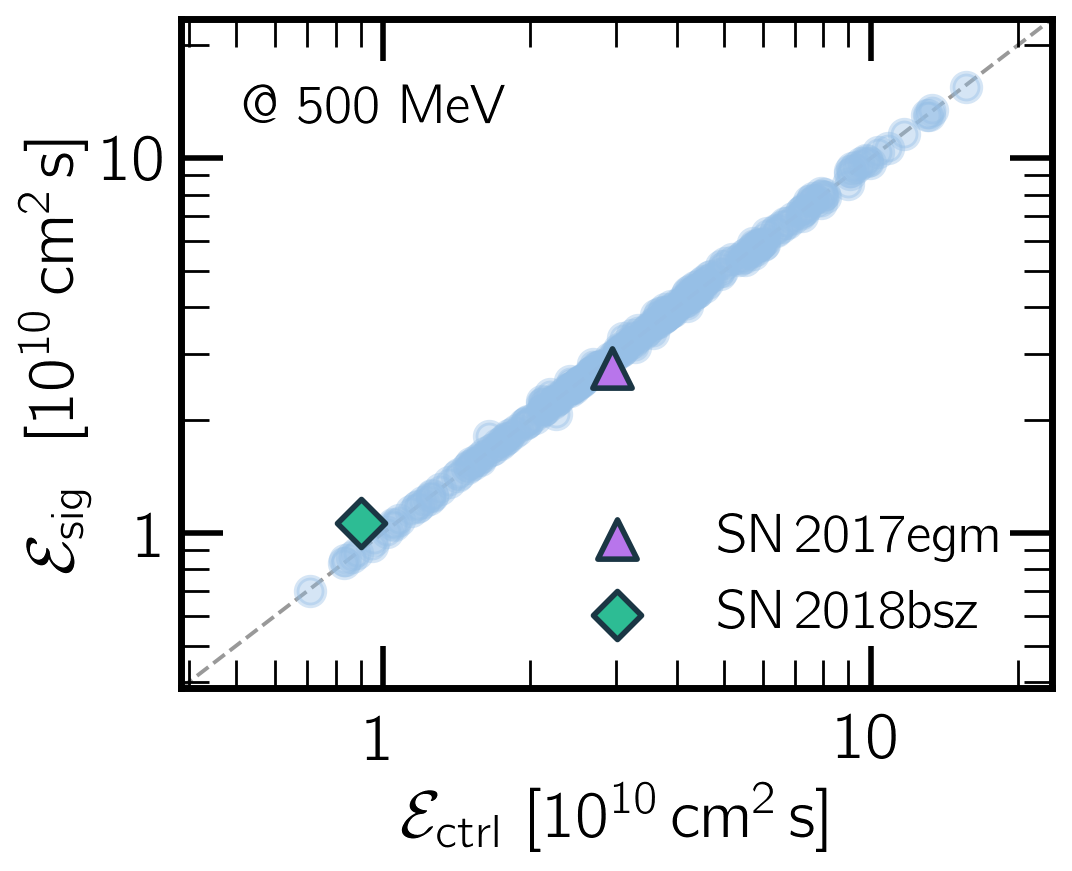}
    \caption{\FermiLAT\ exposure in the signal vs.~control regions at 500~MeV for the 223 SLSNe-I. Each blue circle (\textcolor{salvia-blue}{\ding{108}}) corresponds to one source; the dashed line marks equal signal and control exposure. SN~2017egm (\textcolor{lilac}{\ding{115}}) and SN~2018bsz (\textcolor{green-blue}{\ding{117}}) are highlighted. For the majority of sources the exposures agree to within a few per cent, confirming that GTI filtering and spacecraft-pointing variations introduce no systematic bias between the signal and control (background) windows.}
    \label{fig:exposure_scatter}
\end{figure}

In the signal window, six sources have $\mathrm{TS}_{\max}>9$, compared to an expectation of $\lambda \approx 1.3$ from the asymptotic bounded-Wilks distribution  \citep{wilks1938}. However, the control window provides a more relevant null calibration: it too shows five sources with $\mathrm{TS}_{\max}>9$, indicating a larger high-TS tail in the empirical sample than in the asymptotic expectation. This broadening is not unexpected in LAT analyses over short time windows, driven primarily by imperfections in the Galactic diffuse emission model and, to a lesser extent, by mismodeled or unresolved point sources whose time-averaged fluxes do not match the emission during the analysis time interval.
We therefore assess candidate significance primarily relative to the control distribution. A direct comparison of signal and control counts tests whether the high-TS excesses are temporally associated with the SLSNe-I. By construction, both windows span $2.5t_{\rm BH}(1{+}z)$ at the same sky position. Small exposure differences arise from good-time-interval (GTI) filtering and spacecraft pointing, but these are both modest and unbiased---the signal window has larger exposure for some sources and smaller for others (Fig.~\ref{fig:exposure_scatter})---and therefore do not systematically favor either window. The similar number of high-TS events in the signal (6) and control (5) windows show no evidence for a population-level excess. 

The most significant candidate is SN~2017egm, with $\mathrm{TS}_{\max}=19.5$ and best-fit spectral index $\Gamma=2.04$. As the second-nearest source in the sample ($z=0.031$, $d_L=139~\mathrm{Mpc}$), SN~2017egm is among the most favorable cases for detecting GeV emission. The effective trials factor is thus smaller than the full sample size of 223, since only a handful of nearby events dominate the sensitivity, and the \textit{a priori} probability of finding a marginal excess among them is correspondingly higher than a simple 223-trial correction would suggest. We examine this source in detail in Secs.~\ref{subsec:2017egm} and~\ref{subsec:2017egm_rep}.

\begin{figure*}
\centering
\includegraphics[width=\textwidth]{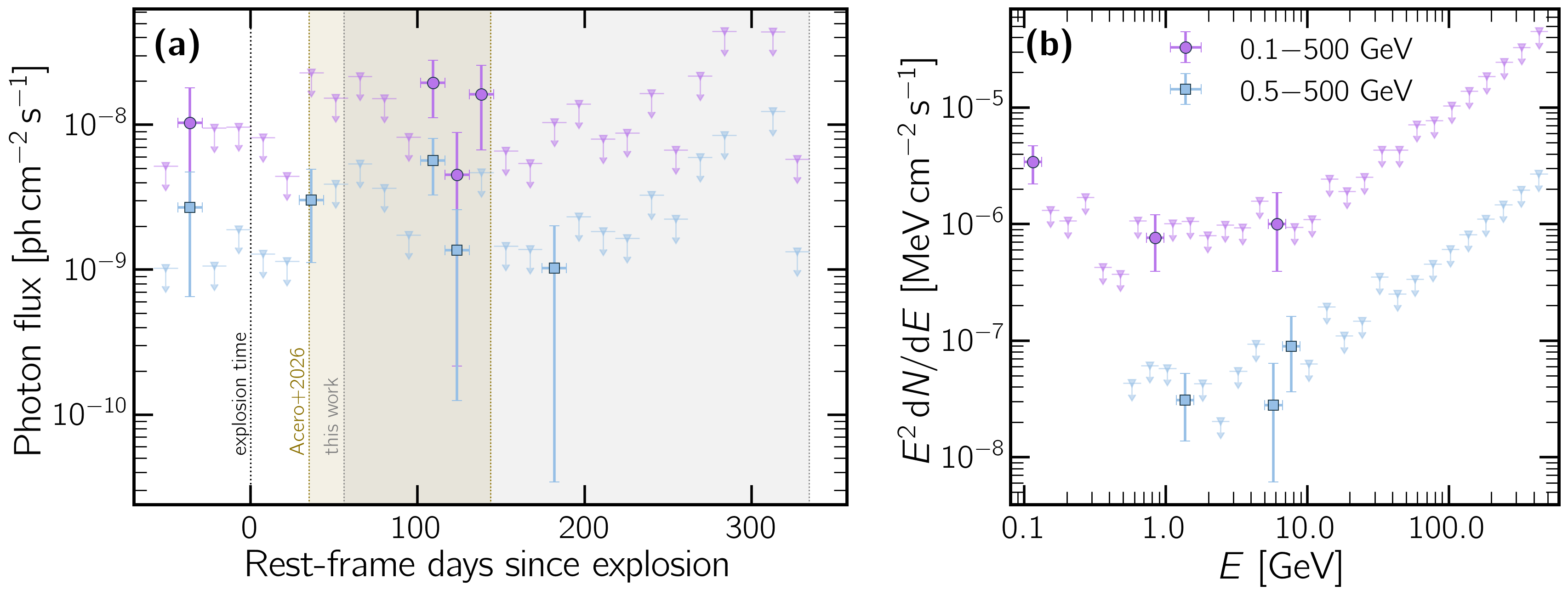}
\caption{\textbf{(a)}~\FermiLAT\ lightcurve of SN~2017egm in 15-day rest-frame bins for two energy ranges: 0.1--500\,GeV (purple circles) and 0.5--500\,GeV (blue squares). Flux points are shown for bins with $\mathrm{TS}\geq4$; downward arrows are 95\% C.L.\ upper limits. The shaded bands mark the Bayesian blocks window identified by Ref.~\citep{Acero:2026slsn} (yellow) and our default signal window from Eq.~\ref{eq:tbest_window} (gray). The excess is concentrated around rest-frame days 109--138 in both energy ranges, with the dominant bin reaching $\mathrm{TS}=19.2$ at 0.5--500\,GeV and $\mathrm{TS}=16.6$ at 0.1--500\,GeV. \textbf{(b)}~Spectral energy distribution of SN~2017egm over the full signal window in both energy ranges. The 0.1--500\,GeV analysis recovers additional flux below 500\,MeV. Above 1\,GeV the two analyses are consistent, with the signal concentrated in a few bins around 2--10\,GeV. The fine spectral binning (eight bins per decade) distributes the signal across many bins, causing most individual bins to fall below the $\mathrm{TS}\geq4$ threshold for displaying flux points even when they contribute cumulatively to the overall excess.}
\label{fig:lightcurve_2017}
\end{figure*}

\subsection{Cross-checks for SN~2017egm}
\label{subsec:2017egm}

The $\mathrm{TS}=19.5$ excess associated with SN~2017egm is the highest in the sample. As one of the nearest sources ($z=0.031$, $d_L=139~\mathrm{Mpc}$), it provides the most likely case for a potential detection. We examine its properties below.

\noindent \textbf{(i) Temporal properties.} We compare the signal window to a background-only control window of equal duration at the same sky position (Sec.~\ref{subsec:time_window}). Because both the signal and the control windows share the same Galactic coordinates and diffuse-model components, any persistent background mismodeling---whether from the Galactic diffuse template, the isotropic component, or an unresolved steady source---would produce a comparable TS in both intervals. The LAT exposures in the two windows differ by $\sim$7\% at 500 MeV ($\mathcal{E}_{\rm sig} = 2.75 \times 10^{10}, \mathcal{E}_{\rm ctrl} = 2.94 \times 10^{10}$~cm$^{2}$~s), with the control window receiving the \textit{larger} exposure (see Fig.~\ref{fig:exposure_scatter}). In the background-dominated regime, TS scales linearly with exposure at a fixed sky position, so a steady emission would yield $\mathrm{TS}_{\rm sig}\sim\mathrm{TS}_{\rm ctrl}$. Instead, we find $\mathrm{TS}_{\rm ctrl}=0.09$,  consistent with zero, compared to $\mathrm{TS}_{\rm sig}=19.5$, ruling out a time-invariant background origin.

To localize the signal in time, we produce a lightcurve in 15-day rest-frame bins across the full signal window using \texttt{fermipy}'s \texttt{lightcurve()} method with $\Gamma=2$ fixed and only the normalization free. As shown in Fig.~\ref{fig:lightcurve_2017}a, the excess is not spread over the full window but is concentrated in a cluster of $\sim$3 bins around rest-frame days 109--138, with the dominant bin at day $\sim$109 reaching $\mathrm{TS}=19.2$. Before day $\sim$100 and after day $\sim$140, the lightcurve is consistent with zero. This temporal concentration is qualitatively consistent with the expected rise and decline of magnetar-powered emission around $t_{\rm BH}$, though the low photon statistics ($n_{\rm pred} \approx 3$) preclude any meaningful constraint on the temporal profile.

Our default analysis adopts the \citet{Gomez:2024xce} posteriors, yielding $t_{\rm BH} \approx 115$\,d for SN~2017egm and a signal window spanning rest-frame days 56--334. Had we instead adopted the \citet{Nicholl+17} parameters ($M_{\rm ej} \approx 3~M_\odot$, $v_{\rm ej} \approx 7000$~km~s$^{-1}$), the transparency time would be $t_{\rm BH} \approx 45$\,d and the signal window would begin at rest-frame day $\sim$23. The observed excess at day $\sim$109 falls well within both windows, but the shorter \citet{Nicholl+17} window would exclude the late, signal-free bins that dilute the time-averaged TS. We note that the Bayesian blocks window identified in the \textit{Fermi}-LAT Collaboration analysis (days $\sim$43--155) is more consistent with the lower ejecta mass ~\citep{Acero:2026slsn}, suggesting that the posteriors in Ref.~\citep{Gomez:2024xce} may place our window onset too late for this source.

\noindent \textbf{(ii) Spectral properties.}~Magnetar or CSM-driven emission is expected to be spectrally coherent across the \textit{Fermi}-LAT energy band, with the flux following a smooth power law or mildly curved spectrum. The SED over the full signal window is shown in Fig.~\ref{fig:lightcurve_2017}b. A single bin centered at $E_{\rm ref} = 7.7$~GeV (6.7--8.9~GeV) contributes $\mathrm{TS}_{\rm bin} = 11.8$, or $\sim$40\% of the summed per-bin test statistic $\Sigma \mathrm{TS}_{\rm bin} = 28.9$. The expected number of source-model counts in this bin is $n_{\rm pred} \approx 3$, placing the measurement squarely in the Poisson regime where a single excess photon above the background prediction can produce $\mathrm{TS} \gtrsim 10$. The global RoI fit, which constrains the flux to follow a single power law across all bins simultaneously, yields $\mathrm{TS}_{\rm RoI} = 19.5$. This is consistent with expectations for a coherent power-law source given the additional degrees of freedom in the bin-by-bin fit.

This spectral morphology, however, is not by itself diagnostic. At the flux level implied by $\mathrm{TS}_{\rm RoI} \approx 20$, any source---real or spurious---would produce only $\mathcal{O}(\text{few})$ detected photons concentrated in a narrow energy range where the LAT signal-to-background ratio peaks. A genuine faint power-law source near the detection threshold can exhibit a similarly ``lumpy'' SED, and the ratio $\Sigma \mathrm{TS}_{\rm bin} / \mathrm{TS}_{\rm ROI} \approx 1.5$ is typical of marginal 4FGL detections \citep{Fermi-LAT:2022byn}.

\noindent \textbf{(iii) Extended energy range.}~As noted in Sec.~\ref{sec:data-selection}, our default analysis uses 500~MeV--500~GeV. For SN~2017egm we additionally perform the lightcurve and SED analysis in the 100~MeV--500~GeV band to probe the energy range where both magnetar IC and hadronic $\pi^0$-decay spectra are predicted to contribute \citep{Vurm&Metzger21,Fang+20}. Both lightcurve and SED in the 100~MeV--500~GeV energy range are shown in Fig.~\ref{fig:lightcurve_2017}.

The excess appears at the same time ($\sim$day 109) in both energy ranges (starting at 100~MeV and 500~MeV), confirming that the signal is unlikely an artifact of a particular energy threshold. The dominant bin actually has \textit{higher} TS at 500~MeV ($\mathrm{TS}=19.2$) than at 100~MeV ($\mathrm{TS}=16.6$), despite the lower photon flux, because the sharper PSF at higher energies provides better signal-to-noise. The only bin where the two analyses disagree is rest-frame day $\sim$138, where the 100~MeV analysis yields $\mathrm{TS}=6.3$ compared to $\mathrm{TS}=1.9$ at 500~MeV. Given the number of time bins ($\sim$19), this level of disagreement is consistent with statistical fluctuations.

In a concurrent study, the \textit{Fermi}-LAT Collaboration reports a $\geq 5\sigma$ detection of SN~2017egm in the 0.1--100~GeV band using a Bayesian blocks time window of 112~days (rest-frame days $\sim$43--155) and a summed PSF-type likelihood \citep{Acero:2026slsn}. When we restrict our 100~MeV--500~GeV lightcurve to the same time interval, the average normalization is $N_0 \approx 8.1 \times 10^{-13}$~ph~cm$^{-2}$~s$^{-1}$~MeV$^{-1}$ at $E_0 = 1$~GeV, consistent with their reported value of $N_0 = 8.2 \times 10^{-13}$ to within $\sim$1\%. The apparent factor of $\sim$45 difference between our default 500~MeV full-window result ($N_0 = 1.8 \times 10^{-14}$) and the value in Ref.~\citep{Acero:2026slsn} is fully accounted for by two effects: the wider energy band extending to 100~MeV inflates $N_0$ by a factor of $\sim$14 (due to soft photons below 500~MeV), and the shorter Bayesian blocks window concentrates the signal rather than diluting it over the full 278-day window informed by \citep{Gomez:2024xce} (a factor of $\sim$3.2). There is no additional methodological discrepancy.

\noindent \textbf{(iv) Nearby blazar contamination.}~Roughly 90\% of high-latitude sky positions lie within $3^\circ$ of a cataloged 4FGL blazar, so the presence of a nearby variable source is not unusual. For SN~2017egm, the relevant neighbor is 4FGLJ1015.0+4926 (1ES1011+496), a BLLac $3.06^\circ$ away with a variability index of 418 \citep{Fermi-LAT:2022byn}. We verify that it does not bias our result.
At 7.7~GeV (the most statistically significant bin in SN~2017egm spectrum), the 4FGLJ1015.0+4926's flux is $\sim$26 times higher than the SN2017egm best-fit flux, raising the possibility of contamination through either direct photon leakage or indirect bias of the background model.

At 7.7~GeV, however, the \textit{Fermi}-LAT 68\% PSF containment radius is $\approx 0.15^{\circ}$, far smaller than the separation \citep{FermiMission2009}. Thus, the two sources are fully resolved and the likelihood fit models them independently. Below $\sim$1 GeV, where the PSF broadens to 1--2$^{\circ}$, the blazar could in principle contribute photons to the SN2017egm ROI; however, in this regime the blazar is explicitly included in the likelihood model with free normalization (Sec.~\ref{subsec:individual}), absorbing any such leakage into the fitted model.

A perhaps more relevant concern is whether blazar variability during the signal window could bias the background model. Our binned likelihood analysis integrates over the full 296-day signal window, so the fitted normalization of 4FGL J1015.0+4926 reflects its time-averaged flux. If the blazar underwent a flare within this interval, the single fitted normalization would underestimate the flux during the flare and overestimate it during quiescence, potentially leaving structured residuals in the ROI. Because BL Lac flares are often accompanied by spectral hardening, a mismodeled flare could preferentially deposit residual counts at multi-GeV energies in the surrounding region. We thus examine the \textit{Fermi}-LAT light curve of 4FGL J1015.0+4926 in weekly bins covering the full mission duration\footnote{\url{https://fermi.gsfc.nasa.gov/ssc/data/access/lat/LightCurveRepository/source.html?source_name=4FGL_J1015.0+4926}, accessed on March 11, 2026.}. Over the 296-day signal window, the blazar flux is consistent with or below its long-term average of $\sim$4--5$\times 10^{-8}$~GeV~ph~cm$^{-2}$~s$^{-1}$ (0.1--100~GeV), with no evidence of flaring activity. The signal window falls between two elevated-activity periods (2014 and 2018--2019), coinciding with a local minimum in blazar output. We therefore conclude that background model bias induced by blazar variability is unlikely to account for the observed excess.

Taken together, these checks do not provide a compelling systematic explanation for the SN~2017egm excess. The agreement between three independent analyses---this work and Refs.~\citep{Li:2024ics, Acero:2026slsn}---using different time windows, energy ranges, and analysis configurations strengthens the case that the excess at the position of SN~2017egm is not an artifact of a particular analysis choice. Nevertheless, $\mathrm{TS}=19.5$ in our primary 0.5--500~GeV analysis falls below the conventional detection threshold ($\mathrm{TS}\gtrsim 25$). We note, however, that this threshold is calibrated for blind all-sky searches, where the source position is unknown and the background model must be extrapolated from the diffuse template alone. In our case, the source position is fixed by the optical transient and the background is directly calibrated using data from the same sky position before the explosion, substantially reducing both the trials factor and the systematic uncertainty in the background model. The low photon count ($n_{\rm pred}\approx 3$) nonetheless means a statistical fluctuation cannot be excluded. We therefore treat this excess as \textbf{suggestive but inconclusive}, and assess its implications for the SLSN-I population in the following section.

\subsection{Can SN2017egm be representative?}
\label{subsec:2017egm_rep}
Suppose the SN2017egm excess is genuine GeV emission. In the background-dominated regime, the \textit{Fermi}-LAT TS scales as $\mathrm{TS}\propto F_\gamma^2\,\mathcal{E}$, where $F_\gamma$ is the photon flux and $\mathcal{E}$ is the exposure, approximately proportional to the signal-window duration $\Delta t$.  The expected TS for source~$i$ relative to SN~2017egm is therefore

\begin{equation}
\label{eq:ts_scaling_general}
\mathrm{TS}_i \approx \mathrm{TS}_{\rm egm}\;
\left(\frac{L_{\gamma,i}}{L_{\gamma,\rm egm}}\right)^{\!2}
\left(\frac{d_{L,\rm egm}}{d_{L,i}}\right)^{\!4}
\frac{\Delta t_i}{\Delta t_{\rm egm}}\,.
\end{equation}
This allows us to approximate how the TS of every other SLSN should scale relative to SN2017egm under various physically distinct scenarios. We consider two such scenarios: (i) a standard candle and (ii) a constant magnetar efficiency.
 
\noindent\textbf{Standard candle.}  Let us assume that all SLSNe-I share a common intrinsic GeV luminosity, $L_{\gamma,i} = L_\gamma$ (while retaining source-specific ejecta properties and hence different transparency timescales). Then, the luminosity ratio in Eq.~(\ref{eq:ts_scaling_general}) reduces to

\begin{equation}
\label{eq:ts_scaling_sc}
\mathrm{TS}_i^{\rm sc} = \mathrm{TS}_{\rm egm}\;
\left(\frac{d_{L,\rm egm}}{d_{L,i}}\right)^{\!4}
\frac{\Delta t_i}{\Delta t_{\rm egm}}\,, 
\end{equation}
where the superscript \textsc{sc} indicates a standard candle. This is the most conservative hypothesis: it requires no knowledge of the explosion physics and encodes only the geometric $d_L^{-2}$ flux scaling.  The only source of uncertainty is $\Delta t_i \propto t_{{\rm BH},i}$, which enters linearly and is typically constrained to $\sim$15\% by the \texttt{MOSFiT} posteriors.

\noindent\textbf{Constant magnetar efficiency.}  In the magnetar-powered scenario, GeV photons escape when the ejecta become transparent at $t_{\rm BH}$, and the available power is the residual spin-down luminosity $L_{\rm sd}(t_{\rm BH})$ (Eq.~\ref{eq:Lsd}).  If a fixed fraction $\varepsilon_{\rm sd}$ of this power is converted to GeV emission---the same for every source---then \mbox{$L_{\gamma,i} = \varepsilon_{\rm sd}\,L_{{\rm sd},i}(t_{{\rm BH},i})$} and
\begin{equation}
\label{eq:ts_scaling_mag}
\mathrm{TS}_i^{\rm mag} \;=\; \mathrm{TS}_{\rm egm}\;
\left(\frac{L_{{\rm sd},i}(t_{{\rm BH},i})}{L_{{\rm sd,egm}}(t_{{\rm BH,egm}})}\right)^{\!2}
\left(\frac{d_{L,\rm egm}}{d_{L,i}}\right)^{\!4}
\frac{\Delta t_i}{\Delta t_{\rm egm}}\,,
\end{equation}
with the superscript indicating the magnetar-powered scenario. Unlike the standard candle, this model rewards sources with a more powerful engine at the moment of transparency. However, the prediction inherits the full uncertainty of the \texttt{MOSFiT} posteriors on $L_0$ and $T_{\rm sd}$.  For each source we propagate errors through Eq.~(\ref{eq:Lsd}). The dominant uncertainty is $\delta\!\log L_0$ ($\sim$0.3--0.8\,dex from the \texttt{MOSFiT} posteriors), which propagates into TS and yields error bars spanning up to two orders of magnitude.
\begin{figure}
\centering
\includegraphics[width=0.9\columnwidth]{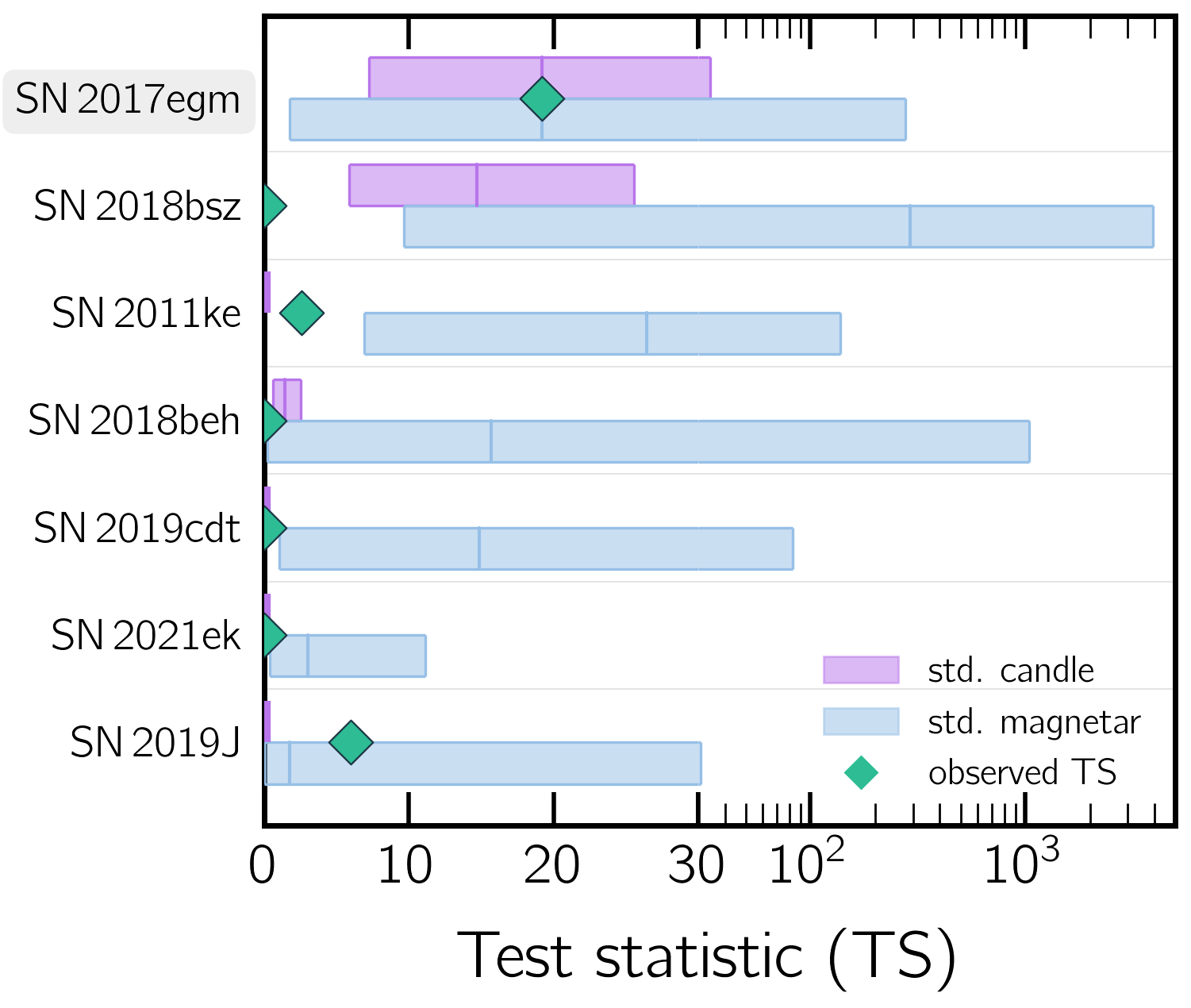}
\caption{Predicted vs. observed \textit{Fermi}-LAT TS for the six most constraining SLSNe-I under two models normalized to the SN2017egm $\gamma$-ray excess: a standard candle model (\textcolor{lilac}{lilac}) and a constant magnetar efficiency model (\textcolor{salvia-blue}{blue}). Bar widths span $\pm1\sigma$ uncertainties propagated from MOSFiT posteriors and the statistical uncertainty on $\mathrm{TS}$. Green diamonds (\textcolor{green-blue}{\ding{117}}) mark the observed TS from the individual LAT analyses above 500~MeV. SN~2018bsz---closer than SN~2017egm and with a more powerful inferred engine---is predicted to yield $\mathrm{TS} \gtrsim 9$ under either model, yet is undetected ($\mathrm{TS}=0$), disfavoring a scenario in which GeV emission is uniform across the SLSNe-I population.}
\label{fig:ts_pred}
\end{figure}

Figure~\ref{fig:ts_pred} compares the predictions to the observed TS values. The decisive case is SN~2018bsz ($z=0.027$, $d_L=121$~Mpc), which is \emph{closer} than SN~2017egm and has an inferred spin-down luminosity ${\sim}5\times$ larger at $t_{\rm BH}$, yielding a median magnetar-model prediction of $\mathrm{TS}^{\rm mag} \approx 290$, yet we observe $\mathrm{TS}=0$. The $\pm0.6$~dex uncertainty on $\log L_0$ propagates into a predicted TS range spanning more than two decades (${\sim}9$--4000, at $\pm1\sigma$), so the non-detection is formally only ${\sim}1\sigma$ tension with the assumed SN~2017egm detection. The SN~2017egm--SN~2018bsz pair therefore \emph{disfavors}, but does not conclusively rule out, a uniform $\varepsilon_{\rm sd}$ across the population. 

At $\Gamma=2$, the SN~2017egm excess implies $\eta \approx 2.8\times10^{-2}$ if interpreted as real emission, whereas the individual 95\% upper limit for SN~2018bsz is $9.5\times10^{-4}$. The implied ratio $\varepsilon_{\rm sd}(2017{\rm egm})/\varepsilon_{\rm sd,UL}(2018{\rm bsz})\approx30$ disfavors a uniform-efficiency interpretation. Three interpretations remain open in order to accommodate both events within a single population: (i) the SN~2017egm excess is a statistical fluctuation, consistent with the control-window tail; (ii) the GeV efficiency varies between sources; or (iii) the two events are powered by fundamentally different central engines, with SN~2017egm hosting a weakly magnetized magnetar nebula while SN~2018bsz is driven by a more highly magnetized engine or CSM interaction, both of which predict $\eta \ll 1$.

We therefore conclude that, \textbf{if the SN~2017egm excess is real}, it is unlikely to be representative of the broader SLSN-I population under simple uniform-efficiency models. The joint-likelihood analysis in Sec.~\ref{subsec:jla_results} quantifies the allowed efficiency at the population level.


\begin{table*}
  \centering
  \caption{Top six sources by fractional weight for each JLA weighting model.}
  \label{tab:top_sources}
  \resizebox{\textwidth}{!}{

  \begin{tabular}{lcccccc}
    \hline\hline
    Model & \#1 & \#2 & \#3 & \#4 & \#5 & \#6 \\
    \hline
    Uniform
      & \multicolumn{6}{c}{all sources equal ($0.45\%$ each)} \\
    Std.\ candle
      & 2018bsz (14.7\%)
      & {2017egm} (11.1\%)
      & 2019ieh (10.2\%)
      & iPTF15eov (3.5\%)
      & 2018beh (2.8\%)
      & 2018hti (2.7\%) \\
    Opt.\ lum.\ ($t_{\rm BH}$)
      & 2018bsz (21.7\%)
      & iPTF15eov (10.1\%)
      & 2021bnw (7.4\%)
      & 2019neq (5.9\%)
      & {2017egm} (5.3\%)
      & PTF12dam (5.0\%) \\
    Magnetar s.d.
      & 2018bsz (19.2\%)
      & iPTF15eov (9.1\%)
      & 2021bnw (8.1\%)
      & 2019neq (6.9\%)
      & PTF12dam (5.8\%)
      & {2017egm} (5.2\%) \\
    Kinetic energy
      & 2018bsz (39.9\%)
      & {2017egm} (23.4\%)
      & 2019ieh (2.2\%)
      & 2018hti (1.9\%)
      & 2019cdt (1.5\%)
      & 2020qlb (1.5\%) \\
    Shock power
      & 2018bsz (63.6\%)
      & {2017egm} (10.6\%)
      & 2019ieh (3.0\%)
      & 2018bgv (1.6\%)
      & iPTF15eov (1.3\%)
      & 2018beh (1.2\%) \\
    \hline
  \end{tabular}
  }
\end{table*}

\begin{figure}
  \centering
  \includegraphics[width=\columnwidth]{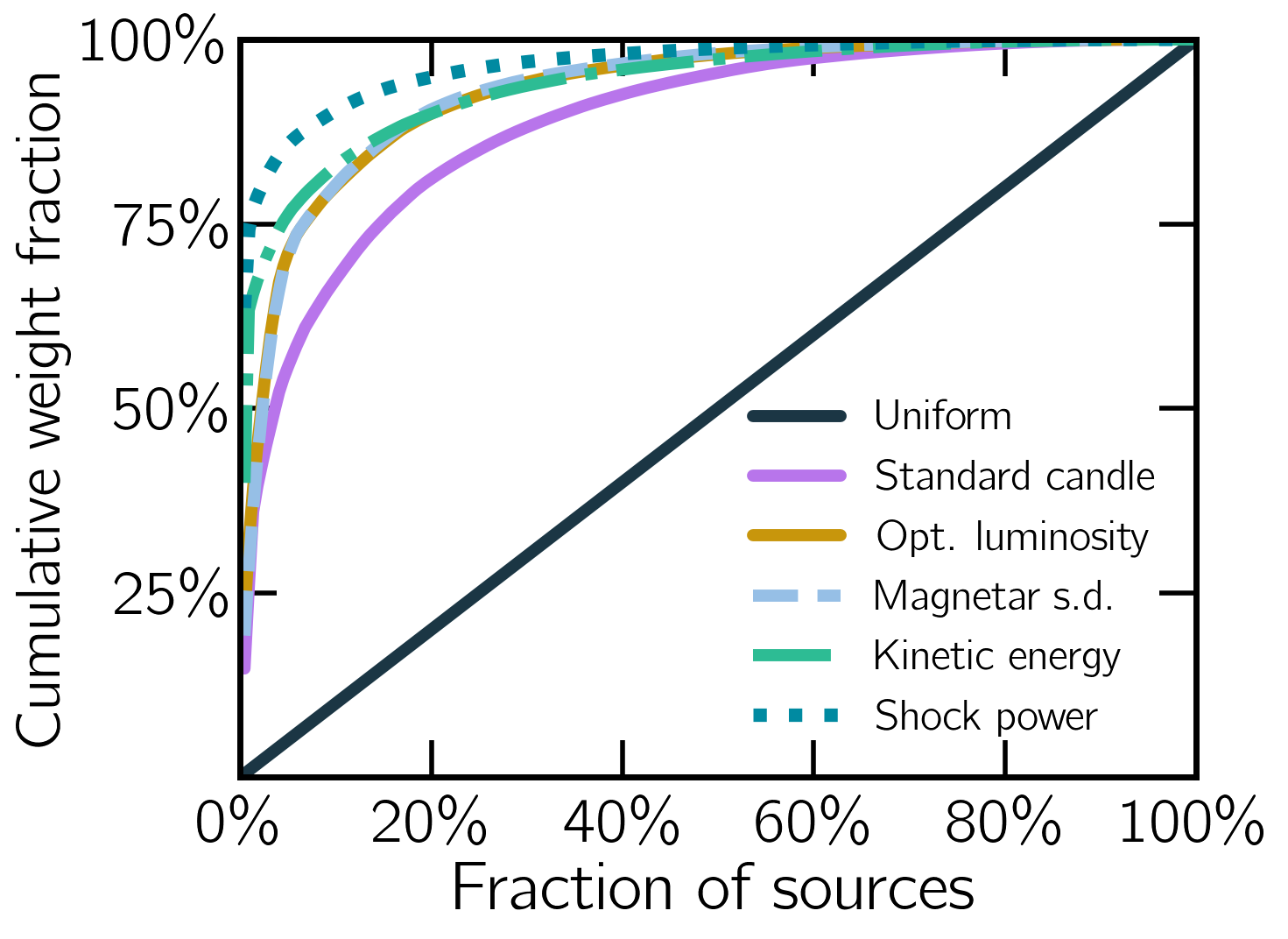}
\caption{The cumulative fraction of total joint-likelihood weight as a function of source fraction (ranked high $\to$ low) for the six weighting models. The uniform model (diagonal) distributes weight equally across all 223 sources. The standard-candle ($N_{\rm eff}=19$), optical-luminosity ($N_{\rm eff}=13$), and magnetar ($N_{\rm eff}=14$) models distribute weight moderately, while the kinetic-energy and shock-power models rise steeply, with only a handful of sources dominating.}
  \label{fig:jla_lorenz}
\end{figure}
\subsection{Side-by-side comparison: SN2017egm vs SN2018bsz}
Both SN2017egm and SN2018bsz are among the nearest 
SLSNe-I discovered to date, with comparable distances ($z \approx 0.03$) and peak luminosities typical of the class. Their proximity has enabled extensive multiwavelength follow-up and makes them particularly valuable for testing models of high-energy emission from SLSNe. Despite these similarities, the two events differ in several important respects in their optical lightcurves and spectra. SN~2017egm exhibits a relatively fast rise to peak brightness and otherwise canonical SLSN-I spectral evolution, including prominent early-time O~II absorption features that later evolve toward spectra resembling those of Type~Ic supernovae \cite{Nicholl2017,Bose2018}. This feature is thought to arise from non-thermal excitation of CNO layers by the magnetar wind nebula \cite{Mazzali:2016pgx}. In contrast, SN~2018bsz displays an unusually slow photometric evolution, characterized by a prolonged pre-maximum plateau lasting several weeks prior to the main rise. Spectroscopically, it also shows unusually strong C~II absorption features indicative of carbon-rich ejecta \cite{Pursiainen2022, Anderson:2018yst}. The absence of the O~II absorption features in SN~2018bsz may indicate that its ejecta are less transparent to the non-thermal radiation from a central engine, or that the engine properties differ from those of SN~2017egm.
The host environments also differ markedly: SN~2017egm occurred in the massive, near-solar metallicity spiral galaxy NGC~3191 \cite{Nicholl2017,Izzo2018}, whereas SN~2018bsz exploded in a lower-mass, sub-solar metallicity star-forming galaxy more representative of the typical SLSN host population \cite{Schulze2018}. These contrasting environments imply differences in progenitor metallicity, stellar mass, and possibly circumstellar structure, factors that may influence engine properties and particle acceleration conditions. Because the two events lie at nearly identical distances and exhibit broadly similar peak luminosities, their differing early-time behavior and host environments may provide important clues to the physical conditions which could alter the efficiency or timing of particle acceleration and associated $\gamma$-ray emission.

Moreover, SN~2018bsz exhibits the late-time appearance of hydrogen emission lines \citep{Pursiainen2022}, which may indicate some degree of interaction between the ejecta and hydrogen-rich CSM. This suggests that SN~2018bsz may not be an ideal comparison object for a purely engine-powered, canonical SLSN-I. Supporting this picture, X-ray and radio detections of SN~2018bsz have also been obtained, both of which are characteristic of CSM interaction \cite{Margutti_priv}. If SN~2018bsz is indeed partly or predominantly CSM-powered, then the non-detection of GeV emission from this source would not necessarily conflict with a magnetar origin for SN~2017egm, since the two events may represent physically distinct explosion channels despite their similarities in distance and peak luminosity. This possibility would lessen the direct tension between SN~2017egm and SN~2018bsz discussed in Sec.~\ref{subsec:2017egm_rep}, though it would not alter the broader conclusion that simple population-wide uniform-efficiency models are disfavored.

\subsection{Joint-likelihood constraints}
\label{subsec:jla_results}
We find no evidence for GeV emission in any of the six weighting models applied to the SLSN-I sample. The joint likelihood yields $\mathrm{TS}_{\rm max} = 0$ across all spectral indices $\Gamma \in [1.5, 4.0]$, fully consistent with the background hypothesis. 

An important diagnostic of each model is the effective number of contributing sources to the joint likelihood,
\begin{equation}
  \label{eq:N_eff}
  N_{\rm eff} = \frac{1}{\sum_i q_i^2}, \qquad q_i = \frac{w_i}{\sum_j w_j},
\end{equation}
which measures how evenly the statistical weight is distributed. The models span a wide range in $N_{\rm eff}$, from 223 (uniform weighting) down to ${\sim}2$ for shock-power model. For the shock-power model, the joint likelihood is dominated by SN~2018bsz ($z=0.027$), which alone contributes $\sim$64\% of the total weight ($N_{\rm eff}=2.4$). The optical-luminosity and magnetar spin-down models are more broadly distributed ($N_{\rm eff}\approx13$--14), with SN~2018bsz contributing $\sim$22\% and $\sim$19\% respectively, followed by several other nearby sources. The standard-candle model distributes weight more broadly ($N_{\rm eff}\approx19$), though it remains strongly biased toward the nearest events. The cumulative fractions of total joint-likelihood weight for each model are shown in Fig.~\ref{fig:jla_lorenz} and the dominant sources in our weighing scheme are listed in Table~\ref{tab:top_sources}.  This is an inevitable consequence of the $d_L^{-2}$ sensitivity weighting: sources at $z \lesssim 0.05$ contribute weights $10^2$--$10^3$ times larger than the sample median. 


\begin{table*}
  \centering
  \caption{Joint-likelihood 95\% C.L.\ upper limits at $\Gamma=2$ for each weighting model. $N_{\rm eff}$ (Eq.~\ref{eq:N_eff}) measures the concentration of statistical weight. UL$_{\rm full}$ uses the full 223-source sample. UL$_{\rm excl}$ is the weakest limit obtained after removing the sources dominating the weighing. All models give $\mathrm{TS}_{\rm max}=0$ over $\Gamma\in[1.5,4.0]$. For comparison, the last two columns show SN~2017egm individual best-fit values in 0.1--500~GeV and 0.5--500~GeV energy ranges.}
  \label{tab:jla_results}
  \begin{tabular}{lrcll|ll}
    \hline\hline
    Model & $N_{\rm eff}$ & Quantity & UL$_{\rm full}$ & UL$_{\rm excl}$ & \multicolumn{2}{c}{SN\,2017egm}   \\
    \hline
    Uniform           & 223  & ---                            & ---                                    & ---                                    & [0.5--500~GeV]&[0.1--500~GeV] \\
    Std.\ candle      & 19.3 & $L_\gamma$                     & $<1.5\times10^{40}$\,erg\,s$^{-1}$   & $<2.4\times10^{40}$\,erg\,s$^{-1}$   & $3.5\times10^{41}$\,erg\,s$^{-1}$ & $8.2\times10^{42}$\,erg\,s$^{-1}$
    \\
    Opt.\ lum.\ ($t_{\rm BH}$)  & 12.8 & $\eta$            & $<3.4\times10^{-4}$                   & $<1.3\times10^{-3}$                   & $2.8\times10^{-2}$ &$6.8\times10^{-1}$
    \\
    Magnetar SD       & 14.3 & $\varepsilon_{\rm sd}$         & $<3.1\times10^{-4}$                   & $<1.3\times10^{-3}$                   & $2.5\times10^{-2}$ & $6.1\times10^{-1}$
    \\
    Kinetic energy    &  4.6 & $L_\gamma/E_k$                 & $<4.4\times10^{-12}$\,s$^{-1}$       & $<3.5\times10^{-11}$\,s$^{-1}$       & $3.6\times10^{-11}$\,s$^{-1}$ & $8.5\times10^{-10}$~s$^{-1}$
    \\
    Shock power       &  2.4 & $\varepsilon_{\rm sh}$         & $<2.2\times10^{-6}$                   & $<9.7\times10^{-6}$                   & $9.5\times10^{-5}$ & $2.3\times10^{-3}$
    \\
    \hline
  \end{tabular}
\end{table*}

\subsection{Constraints on common GeV luminosity and efficiency models}
\label{subsec:jla_constraints}

\begin{figure}[t!]
\centering
\includegraphics[width=\columnwidth]{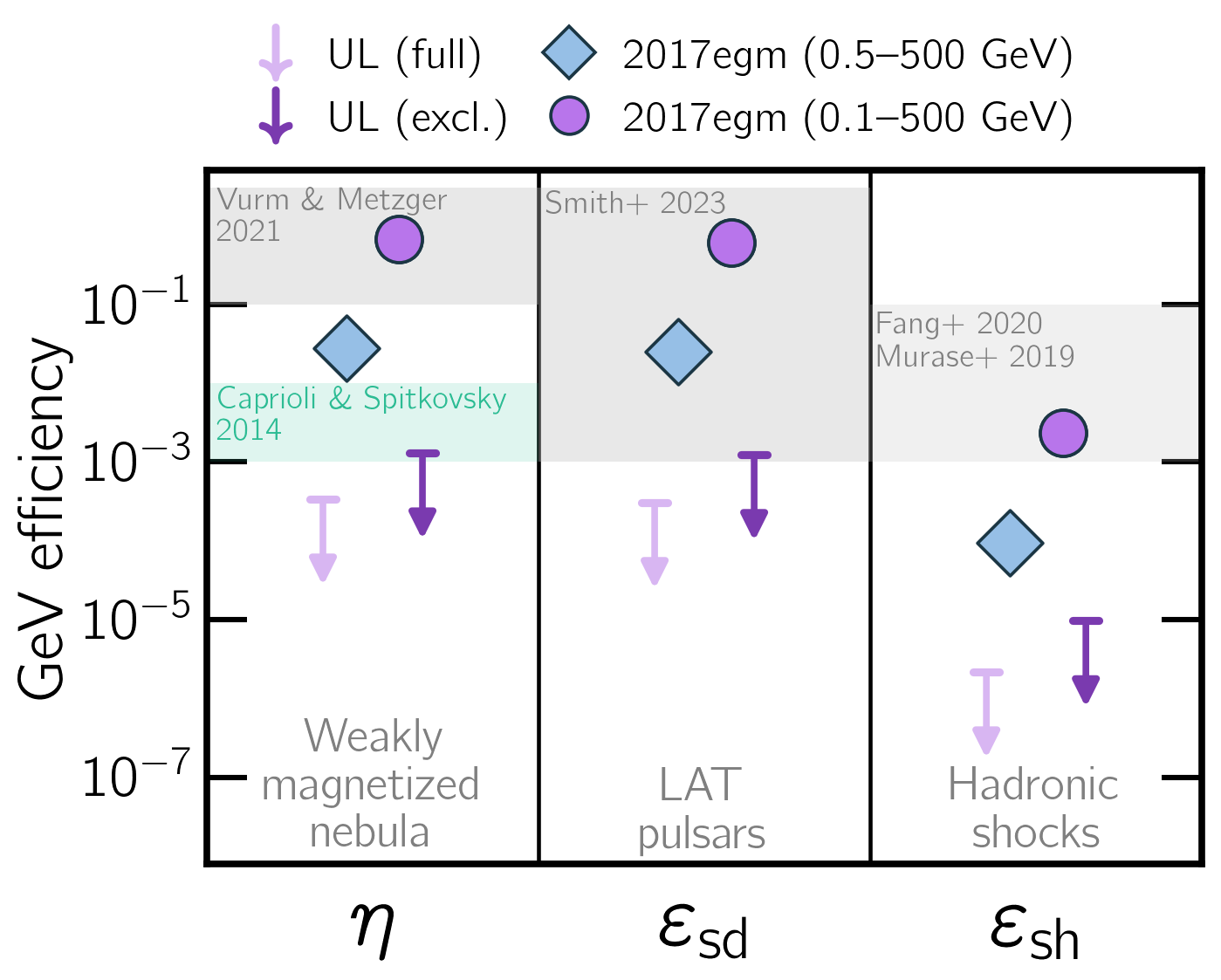}
\caption{GeV efficiency of SLSNe-I across three metrics: the
GeV-to-optical ratio $\eta \equiv L_\gamma/L_{\rm opt}$ (left), the
spin-down efficiency $\varepsilon_{\rm sd} \equiv L_\gamma/L_{\rm sd}$
(center), and the shock efficiency
$\varepsilon_{\rm sh} \equiv L_\gamma\,t_{\rm rise}/E_k$ (right). Gray
bands indicate the ranges expected for weakly magnetized magnetar
nebulae \citep{Vurm&Metzger21}, \textit{Fermi}-LAT rotation-powered
pulsars \citep{Fermi-LAT:2023zzt}, and hadronic CSM shocks
\citep{Fang+20,Murase+19}; the green band in the left panel marks the
$\eta \sim 10^{-3}$--$10^{-2}$ range expected for CSM-powered emission
\citep{Caprioli&Spitkovsky14}. Downward arrows show 95\% C.L.\
joint-likelihood upper limits from the full 223-source SLSN-I sample
(light purple) and after removing the highest-weight sources (dark
purple). Blue diamonds and magenta circles mark the SN~2017egm best-fit
values in the 0.5--500\,GeV and 0.1--500\,GeV bands, respectively. The
0.1--500\,GeV value for SN~2017egm reaches the hadronic-shock band in
$\varepsilon_{\rm sh}$ (right) and sits near the top of the weakly
magnetized nebula band in $\eta$ (left). However, the corresponding
$\eta \sim 0.68$ exceeds the $\lesssim 10^{-2}$ expected for CSM shocks
by more than an order of magnitude, disfavoring a CSM-interaction
origin and favoring a magnetar IC interpretation.}
\label{fig:gev_efficiency_panel}
\end{figure}

Since the physically motivated joint-likelihood analysis weights are fairly concentrated (Fig.~\ref{fig:jla_lorenz} and Table~\ref{tab:top_sources}), the limits below are driven by a small number of nearby SLSNe-I rather than the full 223-source sample. They should therefore be interpreted as constraints on the LAT-bright end of the sample rather than population averages. The corresponding 95\% C.L.\ upper limits at $\Gamma=2$ are summarized in Table~\ref{tab:jla_results} and Fig.~\ref{fig:gev_efficiency_panel}.

Under the \textbf{standard-candle hypothesis}, $w_i \propto d_{L,i}^{-2}$, we constrain the common intrinsic GeV luminosity integrated over the 0.5--500~GeV energy band assuming power-law index $\Gamma=2$  to
\begin{equation}
\label{eq:Lband_ul}
  L_\gamma(0.5\text{--}500\,\mathrm{GeV}) < 1.5\times10^{40}\,\mathrm{erg\,s^{-1}}.
\end{equation}

This is the broadest of the physically motivated weighting models, with $N_{\rm eff}=19.3$; the largest contributors are SN~2018bsz, SN~2017egm, and SN~2019ieh, which carry 15\%, 11\%, and 10\% of the total weight, respectively. The standard-candle result is fairly stable: removing SN~2018bsz weakens the limit by only a factor of 1.6, to $L_\gamma <2.3\times10^{40}\,\mathrm{erg\,s^{-1}}$, and removing the remainder of the dominant sources leaves the bound essentially unchanged, $L_\gamma<2.4\times10^{40}\,\mathrm{erg\,s^{-1}}$. 

In the \textbf{optical-luminosity model}, where the common parameter is the GeV-to-optical ratio at the transparency epoch,
\begin{equation}
\label{eq:eta_ul}
    \eta \equiv \frac{L_\gamma}{L_{\rm opt}(t_{\rm BH})}
  < 3.4\times10^{-4}
\end{equation}
the weight distribution is moderately concentrated ($N_{\rm eff}=12.8$). SN~2018bsz contributes $\sim$22\% of the total weight, followed by iPTF15eov ($\sim$10\%) and several other nearby sources.  Removing SN~2018bsz weakens the constraint to $\eta<5.3\times10^{-4}$, a factor of 1.6 above the full-sample limit. Removing the remainder of dominant sources results in $\eta<1.3\times10^{-3}$.

For the \textbf{magnetar model}, where $w_i \propto L_{{\rm sd},i}(t_{{\rm BH},i})\,d_{L,i}^{-2}$, we constrain the fraction of the spin-down power at transparency that emerges in GeV $\gamma$-rays:
\begin{equation}
\label{eq:eps_ul}
  \varepsilon_{\rm sd} \equiv \frac{L_\gamma}{L_{\rm sd}(t_{\rm BH})} < 3.1\times10^{-4}.
\end{equation}
This model is similarly distributed to the optical-luminosity model ($N_{\rm eff}=14.3$), led by SN~2018bsz ($\sim$19\%). Removing SN~2018bsz alone weakens the constraint to $\varepsilon_{\rm sd}<4.5\times10^{-4}$, a factor of 1.5 above the full-sample result, and removing the remainder of dominant sources results in $\varepsilon_{\rm sd}<1.3\times10^{-3}$, about 4$\times$ weaker than the original value.

The remaining physically motivated models give
\begin{equation}
\frac{L_\gamma}{E_k} < 4.42\times10^{-12}\,\mathrm{s^{-1}}
\end{equation}
for the \textbf{kinetic-energy weighting}. The kinetic-energy model is also moderately concentrated ($N_{\rm eff}=4.6$): removing SN~2018bsz weakens the bound to $L_\gamma/E_k<3.5\times10^{-11}\,\mathrm{s^{-1}}$, a factor of 7.9. Removing the remaining dominant sources leave the limit effectively unchanged. 

Finally, in the shock-power model, we obtain
\begin{equation}
\label{eq:epssh_ul}
\varepsilon_{\rm sh} \equiv \frac{L_\gamma}{E_k/t_{\rm rise}}< 2.2\times10^{-6}.
\end{equation} 
This is the most concentrated weighting tested, with $N_{\rm eff}=2.4$ and 64\% of the total weight carried by SN~2018bsz alone. Removing SN~2018bsz weakens the limit to $\varepsilon_{\rm sh}<9.7\times10^{-6}$. 

These dominant source exclusions confirm that the physically motivated JLA limits are set by a handful of nearby SLSNe-I---especially SN~2018bsz---rather than by the full sample. However, even with this caveat, the constraints remain astrophysically stringent. In weakly magnetized IC nebula models one expects $\eta\sim1$ near $t\sim t_{\rm BH}$, so our most conservative limit, $\eta<1.3\times10^{-3}$, lies three orders of magnitude below that expectation. Likewise, the shock-power limit $\varepsilon_{\rm sh}<9.7\times10^{-6}$ is well below the $\sim10^{-3}$--$10^{-1}$ efficiencies expected if a substantial fraction of the shock power were converted into hadronic $\gamma$-rays.

\subsection{Population Constraints on the GeV Efficiency}
\label{subsec:hierarchical}

The joint-likelihood limits in Sec.~\ref{subsec:jla_results} assume a common efficiency $\eta$ across all SLSNe-I, and therefore cannot distinguish between a uniformly faint population and one in which most sources are weak $\gamma$-ray emitters while a small subset is bright. To probe this possibility, we relax the common-efficiency assumption and instead model each source as having its own efficiency $\eta_i$, drawn from a shared population distribution. In particular, we adopt a hierarchical model in which $\eta_i$ follows a log-normal distribution with mean $\mu_\eta$ and dispersion $\sigma_\eta$, i.e.\ $\ln \eta \sim \mathcal{N}(\mu_\eta, \sigma_\eta^2)$. Using optical-luminosity weights ($w_i \propto L_{{\rm opt},i} d_{L,i}^{-2}$), which directly encode $\eta \equiv L_\gamma/L_{\rm opt}$, we construct the full joint likelihood by convolving the LAT likelihood profile of each source with the population prior and summing over all sources. This yields a two-dimensional likelihood surface in $(\mu_\eta, \sigma_\eta)$.

The resulting likelihood is maximized at $\eta \rightarrow 0$, consistent with no detectable population-level GeV emission. We therefore proceed directly to upper limits on the bright fraction: fewer than $0.7\%$ of SLSNe-I can have $\eta > 10^{-2}$, and fewer than $0.4\%$ can have $\eta > 10^{-1}$ (95\% C.L.). These limits disfavor scenarios in which a small but non-negligible fraction of SLSNe-I are bright GeV emitters while the majority remain undetected.

\section{Additional ZTF sources}
\label{sec:ztf}

The \citet{Gomez:2024xce} catalog includes SLSNe-I discovered through December~2022. Since then, ZTF has continued to identify nearby SLSNe-I that, owing to their proximity, are promising targets for the $\gamma$-ray search presented here. We therefore compile a supplementary sample of 11 ZTF-discovered SLSNe-I with $z < 0.1$ (Table~\ref{tab:ztf_sources}). We treat these sources as a supplementary sample, analyze them individually and keep separate from the main joint-likelihood analysis. 

We attempt \texttt{MOSFiT} magnetar-plus-radioactive-decay fits for these sources using the same configuration as \citet{Gomez:2024xce}. However, the available ZTF alert-stream photometry---which primarily captures the peak and lacks both the rise and late-time decline---does not reliably constrain the ejecta and magnetar parameters. We therefore do not adopt source-specific search windows. Instead, we conduct a uniform search for each source using a window starting at the individual peak optical time, $t_{\rm peak}$, and ending on March 1, 2026 (MJD 61100). This choice of $t_{\rm peak}$ as the window start is conservative: any GeV emission powered by a central engine or CSM interaction should persist or increase after the optical peak, while the ejecta become progressively more transparent. All 11 sources pass the Galactic latitude ($|b| > 10^{\circ}$) and 4FGL proximity cuts applied to the main sample (Sec.~\ref{sec:data-selection}). 

\begin{table*}
\centering
\caption{Properties of 11 additional ZTF SLSN-I at $z < 0.1$ not in the \citet{Gomez:2024xce} catalog.
  $d_L$ is the luminosity distance, $\Delta t$ the observer-frame window duration
  $[\,t_{\rm peak},\,\max(t_{\rm peak}+365(1+z),\,\mathrm{MJD}\,60860)\,]$,
  and $\mathrm{TS}_{\rm pred}$ the predicted test statistic under the standard-candle hypothesis
  (Eq.~\ref{eq:ts_scaling_sc}). Sources are ordered by $d_L$.}
\label{tab:ztf_sources}
\resizebox{0.9\textwidth}{!}{%
\begin{tabular}{lccccccccc}
\hline\hline
Name & $z$ & RA & Dec & $d_L$ & Signal MJD window & Control MJD window & $\mathrm{TS}^{\rm sc}$ & $\mathrm{TS}_{\rm obs}$ & $\mathrm{TS}_{\rm ctrl}$ \\
 & & [deg] & [deg] & (Mpc) & & & & & \\
\hline
SN2024jlc   & 0.0390 & 230.50 & 62.81  & 177.8 & [60504.8, 61100.0] & [59909.6, 60504.8] & 9.04 & 7.07 & 0.00 \\
AT2021ahpl  & 0.0510 & 229.00 & $-$19.29 & 234.5 & [59626.0, 61100.0] & [58151.9, 59626.0] & 3.23 & 0.00  & 0.00  \\
SN2024dde   & 0.0540 & 222.48 & 13.90  & 248.9 & [60422.8, 61100.0] & [59745.6, 60422.8] & 2.55 & 0.00 & 0.00 \\
SN2020jhm   & 0.0600 & 233.26 & 67.91  & 277.7 & [58992.8, 61100.0] & [56885.7, 58992.8] & 1.66 & 0.85 & 0.09 \\
SN2024amf   & 0.0670 & 233.60 & $-$0.14  & 311.6 & [60398.9, 61100.0] & [59697.9, 60398.9] & 1.05 & 3.01 & 0.00 \\
SN2025esr   & 0.0770 & 111.72 & 69.64  & 360.6 & [60774.7, 61100.0] & [60449.5, 60774.7] & 0.13 & 0.00 & 0.00 \\
SN2024dbb   & 0.0790 & 24.15  & 33.40  & 370.5 & [60465.0, 61100.0] & [59829.9, 60465.0] & 0.53 & 1.83 & 0.00 \\
SN2023acco  & 0.0888 & 33.72  & 9.43   & 419.3 & [60326.8, 61100.0] & [59553.5, 60326.8] & 0.33 & 0.04 & 0.74 \\
SN2025imr   & 0.0900 & 340.40 & $-$3.18  & 425.3 & [60807.0, 61100.0] & [60514.0, 60807.0] & 0.04 & 0.00 & 0.00 \\
SN2022vxc   & 0.1000 & 307.29 & $-$6.77  & 475.7 & [59873.7, 61100.0] & [58647.4, 59873.7] & 0.20 & 2.79 & 0.13 \\
SN2024ahr   & 0.1000 & 215.50 & $-$12.51 & 475.7 & [60413.9, 61100.0] & [59727.8, 60413.9] & 0.20 & 0.00 & 2.98 \\
\hline
\end{tabular}%
}
\end{table*}

We report the individual-source results in Table~\ref{tab:ztf_sources}. No source shows statistically significant GeV emission. The most notable case is SN~2024jlc ($z=0.039$), which yields $\mathrm{TS}_{\rm obs}=7.07$ (with $\Gamma_{\rm best}=2.11$ ) and $\mathrm{TS}_{\rm ctrl}=0$. The remaining 10 sources are all consistent with background, with observed TS values ranging from 0 to 3.0 and no systematic excess relative to their control windows. We note that only SN~2024jlc is predicted to have $\mathrm{TS}^{\rm sc}>9$ under the standard candle hypothesis for the time windows outlined in Table~\ref{tab:ztf_sources}. 

\noindent{\textbf{SN~2024jlc (ZTF24aapadbb).}}~The result for this source is sensitive to the treatment of the background model. Fixing the normalizations of nearby catalog sources rather than allowing them to float increases the TS to $\sim 17$. However, this difference is driven almost entirely by the lowest energy bin (500--667~MeV), where PSF leakage from the nearby blazar 4FGL~J1543.0+6130 (2.78$^{\circ}$ away, within the $\sim$3.2$^{\circ}$ 68\% containment radius at 500 MeV) and diffuse model residuals are most severe. We therefore adopt the more conservative result with freed background parameters. 

The current excess is suggestive but not conclusive. It is likely that the current \FermiLAT\ is capturing only a portion of the transparency window for this source. Continued \textit{Fermi}-LAT monitoring will extend the integration time and enable a source-specific analysis with a properly constrained $t_{\rm BH}$ window once late-time photometry becomes available, providing a more definitive test.

\section{Discussion}
\label{sec:disc}
We have conducted the most comprehensive search to date for GeV emission from hydrogen-poor SLSNe-I with the \textit{Fermi}-LAT, using 17 years of data and physically motivated, source-specific transparency windows tied to the BH optical depth. No statistically significant ($\geq 5\sigma$) individual detection or population-level signal is found. Our main result is therefore a set of population-level upper limits on the GeV-to-optical luminosity ratio at $t\sim t_{\rm BH}$ and on related efficiencies defined relative to the magnetar spin-down power or the ejecta kinetic energy.

\subsection{Implications for magnetar-powered models}
\label{subsec:disc_magnetar}

In the magnetar spin-down picture, the rotational energy injected by the central engine is reprocessed in a nebula of relativistic pairs confined by the expanding ejecta, and the partition of this energy between synchrotron and IC channels is set by the nebular magnetization $\epsilon_B$. For weakly magnetized nebulae ($\epsilon_B \lesssim 10^{-3}$), IC scattering of thermal optical photons dominates, and the bulk of the non-thermal power emerges in the GeV band with a predicted efficiency $\eta \sim 1$ at $t \sim t_{\rm BH}$ \citep{Vurm&Metzger21}. Our population-level limit $\eta < 3.4\times10^{-4}$ (Eq.~\ref{eq:eta_ul}; Table~\ref{tab:jla_results}), the conservative bound $\eta < 1.3\times10^{-3}$ obtained after removing the highest-weight sources, and the hierarchical constraint $f(\eta > 10^{-2}) < 0.7\%$ together rule out this scenario for essentially all SLSNe-I in our sample. The data require one of the following:

\begin{enumerate}
  \item \textbf{Strong nebular magnetization} ($\epsilon_B \gtrsim 0.1$).  In this regime the nebular emission shifts to synchrotron radiation peaking in the X-ray band, which is reprocessed by the ejecta into optical and UV photons. The GeV IC component is correspondingly suppressed by a factor $\sim(1+\epsilon_B/\epsilon_{\rm IC})^{-2}$, easily accounting for $\eta \ll 1$. This is consistent with the non-detection of non-thermal X-ray emission from SLSNe at late times \citep{Margutti+18}, which also favors a strongly magnetized nebula.

  \item \textbf{Efficient internal absorption.}  Even for low $\epsilon_B$, GeV photons produced within the nebula can be absorbed by pair production on the dense optical/UV radiation field of the ejecta before escaping. The optical depth to $\gamma\gamma \to e^+e^-$ depends on the ejecta geometry and the photon energy, and for compact ejecta at $t \lesssim t_{\rm BH}$ it can exceed unity at $E_\gamma \gtrsim$ a few GeV, thermalizing the non-thermal emission. Our analysis window ($0.5$--$3\,t_{\rm BH}$) is designed to minimize this effect, but we cannot exclude residual absorption at the window onset.

  \item \textbf{Engine shutdown before transparency.}  If the physical spin-down timescale satisfies $T_{\rm sd} \ll t_{\rm BH}$, the magnetar exhausts its rotational energy before the ejecta become transparent and no GeV emission is expected \citep{Metzger+18b}. For the sources dominating our constraints (SN~2018bsz, SN~2017egm), the \texttt{MOSFiT} posteriors give $T_{\rm sd} > t_{\rm BH}$, so the engine is expected to remain active at transparency. Across the full sample, however, some fraction of sources may have short-lived engines.
\end{enumerate}


The magnetar spin-down efficiency limit $\varepsilon_{\rm sd} < 3.1\times10^{-4}$ (Fig.~\ref{fig:gev_efficiency}) lies at the low end of the range spanned by \textit{Fermi}-LAT rotation-powered pulsars, $L_\gamma/\dot{E} \sim 10^{-4}$--$10^{-1}$ \citep{Fermi-LAT:2023zzt}. While the pulsar magnetosphere and the SLSN nebular environment are not directly comparable, the constraint rules out scenarios in which a substantial fraction of the spin-down power at transparency escapes as GeV $\gamma$-rays. The near-equality $\eta \approx \varepsilon_{\rm sd}$ reflects the fact that at $t \sim t_{\rm BH}$ the ejecta thermalization efficiency remains high ($L_{\rm opt} \approx 0.9\,L_{\rm sd}$), so the optical and spin-down luminosities are closely related.

\subsection{Implications for CSM-interaction models}
\label{subsec:disc_csm}
 
In the CSM-interaction scenario, the supernova ejecta drives a strong shock into dense CSM, accelerating protons and heavier nuclei to relativistic energies. Hadronic collisions ($pp \to \pi^0 \to \gamma\gamma$) then produce GeV--TeV $\gamma$-rays with an efficiency set by the fraction $\varepsilon_{\rm rel}$ of shock power channeled into relativistic particles, typically estimated at $\varepsilon_{\rm rel} \lesssim 0.01$--$0.1$ \citep{Fang+20,Murase+19}. Our shock-power model constrains
\begin{equation*}
  \varepsilon_{\rm sh} = \frac{L_\gamma}{E_k/t_{\rm rise}} < 2.2\times10^{-6}
  \qquad (\Gamma = 2,\; 95\%\;\text{C.L.}),
\end{equation*}
nominally several orders of magnitude below this range. As discussed in Sec.~\ref{subsec:jla}, however, a direct comparison between our $\varepsilon_{\rm sh}$ and the intrinsic acceleration efficiency $\varepsilon_{\rm rel}$ of \citet{Fang+20} is not straightforward. The calorimetric framework predicts $\gamma$-ray production at optical peak, when the shocks are radiative and hadronic interactions are efficient but the ejecta are opaque to GeV photons. Our measurement, by contrast, probes the escaping flux at $t \sim t_{\rm BH}$, when the ejecta are transparent but the shock power has declined. The observed $\varepsilon_{\rm sh}$ therefore reflects the convolution of a declining shock power with increasing ejecta transparency, and may underestimate the intrinsic acceleration efficiency at peak.
 
Nevertheless, the non-detection remains informative. If SLSNe-I are powered by CSM interaction, the low $\varepsilon_{\rm sh}$ implies that either (i) the particle acceleration efficiency at $t \sim t_{\rm BH}$ is far lower than at peak, as expected if the shock transitions from radiative to non-radiative as the CSM density drops, or (ii) the shock is radiation-mediated during the luminous phase, in which case the post-shock temperature is insufficient to accelerate particles to relativistic energies and the bulk of the dissipated energy is emitted as thermal photons. Even if particle acceleration operates at conventional efficiency, $\gamma\gamma$ absorption on the intense optical radiation field can strongly attenuate the escaping $\gamma$-ray flux, particularly at $t \lesssim t_{\rm BH}$. A quantitative assessment of these competing effects requires a time-dependent model coupling shock evolution, particle acceleration, and radiative transfer through the ejecta, which is beyond the scope of this work.

\subsection{Interpreting the SN~2017egm excess and the emerging ZTF sample}
\label{subsec:disc_2017egm}

The $\mathrm{TS} = 19.5$ excess at the position of SN~2017egm is the most significant feature in our dataset. Although it does not reach the conventional $5\sigma$ threshold, the cross-checks of Sec.~\ref{subsec:2017egm} disfavor instrumental artifacts, diffuse-model bias, and blazar contamination. We briefly discuss its implications if interpreted as real emission.

From the best-fit RoI power-law fit ($\Gamma = 2.11\pm0.32$, $N_0 = 1.82\times10^{-14}\,\mathrm{ph\,cm^{-2}\,s^{-1}\,MeV^{-1}}$), the implied band-integrated GeV luminosity is $L_\gamma \approx 3.5\times10^{41}\,\mathrm{erg\,s^{-1}}$ (0.5--500\,GeV), yielding $\eta \equiv L_\gamma/\langle L_{\rm opt}\rangle \approx 2.8\times10^{-2}$. While below the $\eta \sim 1$ prediction of weakly magnetized nebulae, this is roughly two orders of magnitude \emph{above} the population-level upper limit $\eta < 3.4\times10^{-4}$ (Table~\ref{tab:jla_results}), implying that if the excess is real, SN~2017egm is an outlier. The inferred efficiency is consistent with a moderately magnetized nebula \mbox{($\epsilon_B \sim 10^{-2}$--$10^{-1}$)} in which most of the spin-down power emerges as synchrotron radiation, with only a small IC tail escaping in the GeV band.

Extending the analysis to 100\,MeV increases the inferred $L_\gamma$ for SN~2017egm, bringing the best-fit $\varepsilon_{\rm sh}$ closer to the range predicted by the calorimetric framework of \citet{Fang+20}. As discussed in Secs.~\ref{subsec:jla} and~\ref{subsec:disc_csm}, however, this efficiency is complicated by the mismatch between the time of peak shock power and the time when GeV photons can escape. A more robust diagnostic is the GeV-to-optical luminosity ratio $\eta$. In the CSM-interaction scenario, both the $\gamma$-ray and optical emission are ultimately powered by the same shock, so the $\gamma$-ray-to-optical ratio should satisfy $\eta \lesssim \varepsilon_{\rm rel} \sim 10^{-2}$ \citep{Fang+20, Cheung:2022joh, Marti-Devesa:2024hic}. The SN~2017egm excess, if real, implies $\eta \approx 2.8 \times 10^{-2}$ in the 0.5--500\,GeV band---already at the upper edge of this expectation---and $\eta\sim0.68$ when the energy band is extended down to 100~MeV, well above what CSM shocks can accommodate. The event is therefore more naturally explained in the magnetar IC scenario, where the $\gamma$-ray and optical luminosities are powered by distinct channels. The spectral shape above ${\sim}10$\,GeV, where $\gamma\gamma$ absorption would suppress hadronic emission but not the magnetar IC component, could provide a future discriminant with deeper exposure or CTA Observatory observations \citep{Fang+20,Vurm&Metzger21}.

The non-detection of SN~2018bsz---closer than SN~2017egm and with a ${\sim}5\times$ more powerful inferred engine---imposes a strong consistency test. Under a constant-efficiency model, SN~2018bsz would yield $\mathrm{TS} \approx 9$--300 (depending on the $L_0$ posterior), yet we observe $\mathrm{TS}_{\rm obs} = 0$. This disfavors uniform GeV emission and requires either (i) that the SN~2017egm excess is a background fluctuation, or (ii) that the GeV efficiency varies by at least an order of magnitude from source to source. Such large scatter could arise from differences in nebular magnetization, ejecta geometry, or the viewing angle of a jetted outflow, but is difficult to reconcile with the relatively narrow range of optical properties exhibited by the SLSN-I class.

We note that SN~2017egm is also unusual in its host environment: it occurred in the massive, near-solar metallicity spiral NGC~3191 \citep{Nicholl2017,Izzo2018}, in contrast to the low-mass, low-metallicity dwarf galaxies that typically host SLSNe-I. Whether this atypical environment influences the central engine properties or the ejecta conditions that favor GeV escape remains an open question.

The supplementary ZTF sample (Sec.~\ref{sec:ztf}) provides an early look at the next generation of nearby targets. Of 11 additional SLSNe-I at $z<0.1$ discovered since the end of the \citet{Gomez:2024xce} catalog, none reaches the detection threshold. The nearest new source, SN~2024jlc ($z=0.039$), yields $\mathrm{TS}_{\rm obs}=7.07$ with $\mathrm{TS}_{\rm ctrl}=0$ and a best-fit index $\Gamma\approx 2$---the second source in our combined dataset, after SN~2017egm, to show a temporally coincident excess at a well-motivated spectral index. The two cases share several features: both are among the nearest SLSNe-I in their respective samples, both have negligible control-window TS, and both have best-fit indices consistent with IC or hadronic emission. Neither individually reaches $5\sigma$, and the SN~2024jlc result is sensitive to the background treatment at low energies (Sec.~\ref{sec:ztf}). These two marginal excesses from the two nearest available SLSNe-I are suggestive but do not constitute a detection.

Unlike the main-sample sources, SN~2024jlc is likely still within its transparency window: the available data cover only $\sim$600~days post-peak, and the absence of late-time photometry prevents a definitive determination of $t_{\rm BH}$. Continued \textit{Fermi}-LAT monitoring, combined with optical follow-up to constrain the ejecta parameters, will provide a more definitive test. If the excess strengthens with additional exposure, it would suggest that GeV emission may be detectable from the nearest SLSNe-I at efficiencies far below the weakly magnetized nebula prediction but above our current population-level upper limits. Conversely, if it fades, it would further tighten the constraints on even rare GeV-bright events.

An independent confirmation or refutation of the SN~2017egm signal will likely require a comparably nearby ($z \lesssim 0.03$) SLSN-I discovered during the operational lifetime of \textit{Fermi}-LAT. At such low redshifts, SLSNe-I reach apparent magnitudes $m_r \lesssim 18$--$19$, well within the reach of current wide-field surveys (ZTF, ATLAS, and their successors), so discovery completeness is not the limiting factor. The constraining quantity is instead the intrinsic volumetric rate. Adopting $\mathcal{R} \approx 100\,\mathrm{Gpc^{-3}\,yr^{-1}}$ from Refs.~\citep{Quimby:2013jb, Prajs:2016cjj} and assuming effectively full sky coverage from the combination of ongoing surveys, the expected rate of SLSNe-I at $z < 0.03$ is $\sim$0.2--0.5~yr$^{-1}$. A decisive test of the SN~2017egm signal may therefore require up to a decade of continued monitoring, absent a fortuitous nearby event. At higher redshifts ($z < 0.1$), the discovery rate rises to $\sim$10--15~yr$^{-1}$, and LSST will substantially increase the completeness of SLSN-I identification through its deeper photometry and systematic spectroscopic follow-up programs. While individual sources at $z \sim 0.1$ are far less constraining than those at $z \sim 0.03$ (owing to the $d_L^{-4}$ scaling of TS), the accumulation of $\sim$100--150 such events over a decade of LSST operation would meaningfully improve population-level characterization.

The Cherenkov Telescope Array (CTA) offers $\sim$10$\times$ improved sensitivity above 100~GeV relative to \textit{Fermi}-LAT, but its pointed observing mode limits integration times to $\sim$50~hours per target. For the spectra predicted by nebular IC models, the bulk of the GeV flux falls below 10~GeV, where \textit{Fermi}-LAT's all-sky survey mode provides far superior time-integrated exposure. CTA is therefore most valuable as a complement for sources with hard spectra extending above 100~GeV, rather than as a primary discovery instrument for the canonical nebular IC signal. A proposed next-generation all-sky survey telescope such as VLAST, with $\sim$5$\times$ the effective area of \textit{Fermi}-LAT, would substantially advance this field.

\section{Summary}
\label{sec:concl}

We have searched for GeV $\gamma$-ray emission from 223 spectroscopically confirmed SLSNe-I using 17~years of \textit{Fermi}-LAT data, defining physically motivated search windows based on the ejecta-specific BH transparency time for each source. Our main findings are:

\begin{enumerate}
\item \textbf{No population-level GeV signal.} The joint-likelihood analysis yields $\mathrm{TS}_{\rm max} = 0$ across all six weighting models and all spectral indices $\Gamma \in [1.5,\,4.0]$. The GeV-to-optical efficiency is constrained to $\eta < 3.4\times10^{-4}$ ($\Gamma=2$, 95\%~C.L.; Table~\ref{tab:jla_results}), more than three orders of magnitude below weakly magnetized nebula predictions.

\item \textbf{First constraint on the GeV-active fraction.} A hierarchical population model limits the fraction of SLSNe-I with $\eta > 10^{-2}$ to less than $0.7\%$ (95\% C.L.), disfavoring even a small GeV-bright subpopulation.

\item \textbf{Suggestive excesses at SN~2017egm and SN~2024jlc.} The two nearest SLSNe-I in our combined dataset show temporally coincident excesses ($\mathrm{TS}=19.5$ and $7.1$, respectively) with physically motivated spectral indices and negligible control-window TS. Neither individually reaches $5\sigma$, and the SN~2024jlc result is sensitive to low-energy background modeling. If the SN~2017egm excess is real, the implied efficiency is $\eta \approx 2.8\times10^{-2}$ (0.5--500\,GeV) and $\eta \sim 0.68$ (0.1--500\,GeV), consistent with a moderately magnetized nebula. The elevated $L_\gamma/L_{\rm opt}$ ratio---well above the $\lesssim 10^{-2}$ expected for CSM shocks---disfavors a hadronic origin and instead points to a magnetar IC scenario.
  
\item \textbf{The SN~2017egm excess is unlikely universal.} The non-detection of SN~2018bsz---closer and with a ${\sim}5\times$ more powerful inferred engine---requires either (1) a statistical fluctuation, (2) intrinsic scatter in $\eta$ exceeding an order of magnitude, or (3) different powering engines. 

\item \textbf{Future prospects.} At $z \lesssim 0.03$, SLSNe-I are bright enough for discovery by current all-sky surveys, but the volumetric rate limits the expected yield to $\sim$0.2--0.5~yr$^{-1}$. A decisive test of the SN~2017egm signal may therefore require up to a decade of continued \textit{Fermi}-LAT monitoring. A next-generation all-sky $\gamma$-ray telescope such as VLAST would accelerate this program by a factor of $\sim$5.
\end{enumerate}

These results establish that GeV emission from SLSNe-I is either strongly suppressed or confined to a rare subclass. Definitive confirmation or exclusion of the SN~2017egm signal awaits the next comparably nearby SLSN-I.

\acknowledgements

MC and TL acknowledge support from the Swedish Research Council under contract 2022-04283 and the Swedish National Space Agency under contract 117/19. MC, TL, and AG acknowledge support from the EDUCATE Excellence Centre funded by the Swedish Research Council through grant Dnr 2022-06627. TL also acknowledges sabbatical support from the Wenner-Gren foundation under contract SSh2024-0037.  AG has also received support from the Swedish Research Council through grant Dnr 2025-03692.
BDM acknowledges support from the National Science Foundation (grant AST-2406637) and the Simons Foundation (grant 727700).  The Flatiron Institute is supported by the Simons Foundation.

Parts of this work were performed using computing resources provided by the National Academic Infrastructure for Supercomputing in Sweden~(NAISS) under Projects~\mbox{2023/3-21}, \mbox{2023/6-297}, \mbox{2024/5-666} and~\mbox{2024/6-339}, which is partially funded by the Swedish Research Council through Grant~\mbox{2022-06725}.

\noindent \textbf{Software and Data.}~This project made use of \texttt{Astropy} \cite{astropy:2022}, \texttt{Numpy} \citep{numpy:2020}, \texttt{Matplotlib} \cite{matplotlib:2007}, and \texttt{Pandas} \cite{pandas:2020} Python packages. The authors also acknowledge the use of public data from the \textit{Fermi Science Support Center} data archive. 

\noindent \textbf{AI Usage Statement.}~Claude Code was used to assist with coding and data analysis workflows. ChatGPT was used for grammar and language correction. AI tools were not used to generate scientific results or draw conclusions. All analysis, code, and the final text were produced, checked, and verified by the authors, who take the full responsibility for the content of this work.

\bibliography{bibliograpy}

@PREAMBLE{
 "\providecommand{\noopsort}[1]{}" 
 # "\providecommand{\singleletter}[1]{#1}%" 
}

@ARTICLE{Fang+20,
       author = {{Fang}, Ke and {Metzger}, Brian D. and {Vurm}, Indrek and {Aydi}, Elias and {Chomiuk}, Laura},
        title = "{High-energy Neutrinos and Gamma Rays from Nonrelativistic Shock-powered Transients}",
      journal = {\apj},
         year = 2020,
        month = nov,
       volume = {904},
       number = {1},
          eid = {4},
        pages = {4},
          doi = {10.3847/1538-4357/abbc6e},
archivePrefix = {arXiv},
       eprint = {2007.15742},
 primaryClass = {astro-ph.HE},
       adsurl = {https://ui.adsabs.harvard.edu/abs/2020ApJ...904....4F}
}

@ARTICLE{SDSSII-1,
       author = {{Frieman}, Joshua A. and {Bassett}, Bruce and {Becker}, Andrew and {Choi}, Changsu and {Cinabro}, David and {DeJongh}, Fritz and {Depoy}, Darren L. and {Dilday}, Ben and {Doi}, Mamoru and {Garnavich}, Peter M. and {Hogan}, Craig J. and {Holtzman}, Jon and {Im}, Myungshin and {Jha}, Saurabh and {Kessler}, Richard and {Konishi}, Kohki and {Lampeitl}, Hubert and {Marriner}, John and {Marshall}, Jennifer L. and {McGinnis}, David and {Miknaitis}, Gajus and {Nichol}, Robert C. and {Prieto}, Jose Luis and {Riess}, Adam G. and {Richmond}, Michael W. and {Romani}, Roger and {Sako}, Masao and {Schneider}, Donald P. and {Smith}, Mathew and {Takanashi}, Naohiro and {Tokita}, Kouichi and {van der Heyden}, Kurt and {Yasuda}, Naoki and {Zheng}, Chen and {Adelman-McCarthy}, Jennifer and {Annis}, James and {Assef}, Roberto J. and {Barentine}, John and {Bender}, Ralf and {Blandford}, Roger D. and {Boroski}, William N. and {Bremer}, Malcolm and {Brewington}, Howard and {Collins}, Chris A. and {Crotts}, Arlin and {Dembicky}, Jack and {Eastman}, Jason and {Edge}, Alastair and {Edmondson}, Edmond and {Elson}, Edward and {Eyler}, Michael E. and {Filippenko}, Alexei V. and {Foley}, Ryan J. and {Frank}, Stephan and {Goobar}, Ariel and {Gueth}, Tina and {Gunn}, James E. and {Harvanek}, Michael and {Hopp}, Ulrich and {Ihara}, Yutaka and {Ivezi{\'c}}, {\v{Z}}elko and {Kahn}, Steven and {Kaplan}, Jared and {Kent}, Stephen and {Ketzeback}, William and {Kleinman}, Scott J. and {Kollatschny}, Wolfram and {Kron}, Richard G. and {Krzesi{\'n}ski}, Jurek and {Lamenti}, Dennis and {Leloudas}, Giorgos and {Lin}, Huan and {Long}, Daniel C. and {Lucey}, John and {Lupton}, Robert H. and {Malanushenko}, Elena and {Malanushenko}, Viktor and {McMillan}, Russet J. and {Mendez}, Javier and {Morgan}, Christopher W. and {Morokuma}, Tomoki and {Nitta}, Atsuko and {Ostman}, Linda and {Pan}, Kaike and {Rockosi}, Constance M. and {Romer}, A. Kathy and {Ruiz-Lapuente}, Pilar and {Saurage}, Gabrelle and {Schlesinger}, Katie and {Snedden}, Stephanie A. and {Sollerman}, Jesper and {Stoughton}, Chris and {Stritzinger}, Maximilian and {Subba Rao}, Mark and {Tucker}, Douglas and {Vaisanen}, Petri and {Watson}, Linda C. and {Watters}, Shannon and {Wheeler}, J. Craig and {Yanny}, Brian and {York}, Donald},
        title = "{The Sloan Digital Sky Survey-II Supernova Survey: Technical Summary}",
      journal = {\aj},
         year = 2008,
        month = jan,
       volume = {135},
       number = {1},
        pages = {338-347},
          doi = {10.1088/0004-6256/135/1/338},
archivePrefix = {arXiv},
       eprint = {0708.2749},
 primaryClass = {astro-ph},
       adsurl = {https://ui.adsabs.harvard.edu/abs/2008AJ....135..338F}
}

@ARTICLE{SDSSII-2,
       author = {{Sako}, Masao and {Bassett}, Bruce and {Becker}, Andrew and {Cinabro}, David and {DeJongh}, Fritz and {Depoy}, D.~L. and {Dilday}, Ben and {Doi}, Mamoru and {Frieman}, Joshua A. and {Garnavich}, Peter M. and {Hogan}, Craig J. and {Holtzman}, Jon and {Jha}, Saurabh and {Kessler}, Richard and {Konishi}, Kohki and {Lampeitl}, Hubert and {Marriner}, John and {Miknaitis}, Gajus and {Nichol}, Robert C. and {Prieto}, Jose Luis and {Riess}, Adam G. and {Richmond}, Michael W. and {Romani}, Roger and {Schneider}, Donald P. and {Smith}, Mathew and {SubbaRao}, Mark and {Takanashi}, Naohiro and {Tokita}, Kouichi and {van der Heyden}, Kurt and {Yasuda}, Naoki and {Zheng}, Chen and {Barentine}, John and {Brewington}, Howard and {Choi}, Changsu and {Dembicky}, Jack and {Harnavek}, Michael and {Ihara}, Yutaka and {Im}, Myungshin and {Ketzeback}, William and {Kleinman}, Scott J. and {Krzesi{\'n}ski}, Jurek and {Long}, Daniel C. and {Malanushenko}, Elena and {Malanushenko}, Viktor and {McMillan}, Russet J. and {Morokuma}, Tomoki and {Nitta}, Atsuko and {Pan}, Kaike and {Saurage}, Gabrelle and {Snedden}, Stephanie A.},
        title = "{The Sloan Digital Sky Survey-II Supernova Survey: Search Algorithm and Follow-up Observations}",
      journal = {\aj},
         year = 2008,
        month = jan,
       volume = {135},
       number = {1},
        pages = {348-373},
          doi = {10.1088/0004-6256/135/1/348},
archivePrefix = {arXiv},
       eprint = {0708.2750},
 primaryClass = {astro-ph},
       adsurl = {https://ui.adsabs.harvard.edu/abs/2008AJ....135..348S}
}

@ARTICLE{OGLE,
       author = {{Udalski}, A. and {Szyma{\'n}ski}, M.~K. and {Szyma{\'n}ski}, G.},
        title = "{OGLE-IV: Fourth Phase of the Optical Gravitational Lensing Experiment}",
      journal = {\actaa},
         year = 2015,
        month = mar,
       volume = {65},
       number = {1},
        pages = {1-38},
          doi = {10.48550/arXiv.1504.05966},
archivePrefix = {arXiv},
       eprint = {1504.05966},
 primaryClass = {astro-ph.SR},
       adsurl = {https://ui.adsabs.harvard.edu/abs/2015AcA....65....1U}
}

@ARTICLE{Palomar,
       author = {{Rau}, Arne and {Kulkarni}, Shrinivas R. and {Law}, Nicholas M. and {Bloom}, Joshua S. and {Ciardi}, David and {Djorgovski}, George S. and {Fox}, Derek B. and {Gal-Yam}, Avishay and {Grillmair}, Carl C. and {Kasliwal}, Mansi M. and {Nugent}, Peter E. and {Ofek}, Eran O. and {Quimby}, Robert M. and {Reach}, William T. and {Shara}, Michael and {Bildsten}, Lars and {Cenko}, S. Bradley and {Drake}, Andrew J. and {Filippenko}, Alexei V. and {Helfand}, David J. and {Helou}, George and {Howell}, D. Andrew and {Poznanski}, Dovi and {Sullivan}, Mark},
        title = "{Exploring the Optical Transient Sky with the Palomar Transient Factory}",
      journal = {\pasp},
         year = 2009,
        month = dec,
       volume = {121},
       number = {886},
        pages = {1334},
          doi = {10.1086/605911},
archivePrefix = {arXiv},
       eprint = {0906.5355},
 primaryClass = {astro-ph.CO},
       adsurl = {https://ui.adsabs.harvard.edu/abs/2009PASP..121.1334R}
}

@INPROCEEDINGS{Panstarss.7733E..0EK,
       author = {{Kaiser}, Nick and {Burgett}, William and {Chambers}, Ken and {Denneau}, Larry and {Heasley}, Jim and {Jedicke}, Robert and {Magnier}, Eugene and {Morgan}, Jeff and {Onaka}, Peter and {Tonry}, John},
        title = "{The Pan-STARRS wide-field optical/NIR imaging survey}",
    booktitle = {Ground-based and Airborne Telescopes III},
         year = 2010,
       editor = {{Stepp}, Larry M. and {Gilmozzi}, Roberto and {Hall}, Helen J.},
       series = {Society of Photo-Optical Instrumentation Engineers (SPIE) Conference Series},
       volume = {7733},
        month = jul,
          eid = {77330E},
        pages = {77330E},
          doi = {10.1117/12.859188},
       adsurl = {https://ui.adsabs.harvard.edu/abs/2010SPIE.7733E..0EK}
}

@ARTICLE{Catalina,
       author = {{Drake}, A.~J. and {Djorgovski}, S.~G. and {Mahabal}, A. and {Beshore}, E. and {Larson}, S. and {Graham}, M.~J. and {Williams}, R. and {Christensen}, E. and {Catelan}, M. and {Boattini}, A. and {Gibbs}, A. and {Hill}, R. and {Kowalski}, R.},
        title = "{First Results from the Catalina Real-Time Transient Survey}",
      journal = {\apj},
         year = 2009,
        month = may,
       volume = {696},
       number = {1},
        pages = {870-884},
          doi = {10.1088/0004-637X/696/1/870},
archivePrefix = {arXiv},
       eprint = {0809.1394},
 primaryClass = {astro-ph},
       adsurl = {https://ui.adsabs.harvard.edu/abs/2009ApJ...696..870D}
}

@article{SNLS:2013qua,
    author = "Howell, D. A. and others",
    title = "{Two superluminous supernovae from the early universe discovered by the Supernova Legacy Survey}",
    eprint = "1310.0470",
    archivePrefix = "arXiv",
    primaryClass = "astro-ph.CO",
    doi = "10.1088/0004-637X/779/2/98",
    journal = "Astrophys. J.",
    volume = "779",
    pages = "98",
    year = "2013"
}

@ARTICLE{Gaia2021A&A...649A...1G,
       author = {{Gaia Collaboration} and {Brown}, A.~G.~A. and {Vallenari}, A. and {Prusti}, T. and {de Bruijne}, J.~H.~J. and {Babusiaux}, C. and {Biermann}, M. and {Creevey}, O.~L. and {Evans}, D.~W. and {Eyer}, L. and {Hutton}, A. and {Jansen}, F. and {Jordi}, C. and {Klioner}, S.~A. and {Lammers}, U. and {Lindegren}, L. and {Luri}, X. and {Mignard}, F. and {Panem}, C. and {Pourbaix}, D. and {Randich}, S. and {Sartoretti}, P. and {Soubiran}, C. and {Walton}, N.~A. and {Arenou}, F. and {Bailer-Jones}, C.~A.~L. and {Bastian}, U. and {Cropper}, M. and {Drimmel}, R. and {Katz}, D. and {Lattanzi}, M.~G. and {van Leeuwen}, F. and {Bakker}, J. and {Cacciari}, C. and {Casta{\~n}eda}, J. and {De Angeli}, F. and {Ducourant}, C. and {Fabricius}, C. and {Fouesneau}, M. and {Fr{\'e}mat}, Y. and {Guerra}, R. and {Guerrier}, A. and {Guiraud}, J. and {Jean-Antoine Piccolo}, A. and {Masana}, E. and {Messineo}, R. and {Mowlavi}, N. and {Nicolas}, C. and {Nienartowicz}, K. and {Pailler}, F. and {Panuzzo}, P. and {Riclet}, F. and {Roux}, W. and {Seabroke}, G.~M. and {Sordo}, R. and {Tanga}, P. and {Th{\'e}venin}, F. and {Gracia-Abril}, G. and {Portell}, J. and {Teyssier}, D. and {Altmann}, M. and {Andrae}, R. and {Bellas-Velidis}, I. and {Benson}, K. and {Berthier}, J. and {Blomme}, R. and {Brugaletta}, E. and {Burgess}, P.~W. and {Busso}, G. and {Carry}, B. and {Cellino}, A. and {Cheek}, N. and {Clementini}, G. and {Damerdji}, Y. and {Davidson}, M. and {Delchambre}, L. and {Dell'Oro}, A. and {Fern{\'a}ndez-Hern{\'a}ndez}, J. and {Galluccio}, L. and {Garc{\'\i}a-Lario}, P. and {Garcia-Reinaldos}, M. and {Gonz{\'a}lez-N{\'u}{\~n}ez}, J. and {Gosset}, E. and {Haigron}, R. and {Halbwachs}, J. -L. and {Hambly}, N.~C. and {Harrison}, D.~L. and {Hatzidimitriou}, D. and {Heiter}, U. and {Hern{\'a}ndez}, J. and {Hestroffer}, D. and {Hodgkin}, S.~T. and {Holl}, B. and {Jan{\ss}en}, K. and {Jevardat de Fombelle}, G. and {Jordan}, S. and {Krone-Martins}, A. and {Lanzafame}, A.~C. and {L{\"o}ffler}, W. and {Lorca}, A. and {Manteiga}, M. and {Marchal}, O. and {Marrese}, P.~M. and {Moitinho}, A. and {Mora}, A. and {Muinonen}, K. and {Osborne}, P. and {Pancino}, E. and {Pauwels}, T. and {Petit}, J. -M. and {Recio-Blanco}, A. and {Richards}, P.~J. and {Riello}, M. and {Rimoldini}, L. and {Robin}, A.~C. and {Roegiers}, T. and {Rybizki}, J. and {Sarro}, L.~M. and {Siopis}, C. and {Smith}, M. and {Sozzetti}, A. and {Ulla}, A. and {Utrilla}, E. and {van Leeuwen}, M. and {van Reeven}, W. and {Abbas}, U. and {Abreu Aramburu}, A. and {Accart}, S. and {Aerts}, C. and {Aguado}, J.~J. and {Ajaj}, M. and {Altavilla}, G. and {{\'A}lvarez}, M.~A. and {{\'A}lvarez Cid-Fuentes}, J. and {Alves}, J. and {Anderson}, R.~I. and {Anglada Varela}, E. and {Antoja}, T. and {Audard}, M. and {Baines}, D. and {Baker}, S.~G. and {Balaguer-N{\'u}{\~n}ez}, L. and {Balbinot}, E. and {Balog}, Z. and {Barache}, C. and {Barbato}, D. and {Barros}, M. and {Barstow}, M.~A. and {Bartolom{\'e}}, S. and {Bassilana}, J. -L. and {Bauchet}, N. and {Baudesson-Stella}, A. and {Becciani}, U. and {Bellazzini}, M. and {Bernet}, M. and {Bertone}, S. and {Bianchi}, L. and {Blanco-Cuaresma}, S. and {Boch}, T. and {Bombrun}, A. and {Bossini}, D. and {Bouquillon}, S. and {Bragaglia}, A. and {Bramante}, L. and {Breedt}, E. and {Bressan}, A. and {Brouillet}, N. and {Bucciarelli}, B. and {Burlacu}, A. and {Busonero}, D. and {Butkevich}, A.~G. and {Buzzi}, R. and {Caffau}, E. and {Cancelliere}, R. and {C{\'a}novas}, H. and {Cantat-Gaudin}, T. and {Carballo}, R. and {Carlucci}, T. and {Carnerero}, M.~I. and {Carrasco}, J.~M. and {Casamiquela}, L. and {Castellani}, M. and {Castro-Ginard}, A. and {Castro Sampol}, P. and {Chaoul}, L. and {Charlot}, P. and {Chemin}, L. and {Chiavassa}, A. and {Cioni}, M. -R.~L. and {Comoretto}, G. and {Cooper}, W.~J. and {Cornez}, T. and {Cowell}, S. and {Crifo}, F. and {Crosta}, M. and {Crowley}, C. and {Dafonte}, C. and {Dapergolas}, A. and {David}, M. and {David}, P. and {de Laverny}, P. and {De Luise}, F. and {De March}, R. and {De Ridder}, J. and {de Souza}, R. and {de Teodoro}, P. and {de Torres}, A. and {del Peloso}, E.~F. and {del Pozo}, E. and {Delbo}, M. and {Delgado}, A. and {Delgado}, H.~E. and {Delisle}, J. -B. and {Di Matteo}, P. and {Diakite}, S. and {Diener}, C. and {Distefano}, E. and {Dolding}, C. and {Eappachen}, D. and {Edvardsson}, B. and {Enke}, H. and {Esquej}, P. and {Fabre}, C. and {Fabrizio}, M. and {Faigler}, S. and {Fedorets}, G. and {Fernique}, P. and {Fienga}, A. and {Figueras}, F. and {Fouron}, C. and {Fragkoudi}, F. and {Fraile}, E. and {Franke}, F. and {Gai}, M. and {Garabato}, D. and {Garcia-Gutierrez}, A. and {Garc{\'\i}a-Torres}, M. and {Garofalo}, A. and {Gavras}, P. and {Gerlach}, E. and {Geyer}, R. and {Giacobbe}, P. and {Gilmore}, G. and {Girona}, S. and {Giuffrida}, G. and {Gomel}, R. and {Gomez}, A. and {Gonzalez-Santamaria}, I. and {Gonz{\'a}lez-Vidal}, J.~J. and {Granvik}, M. and {Guti{\'e}rrez-S{\'a}nchez}, R. and {Guy}, L.~P. and {Hauser}, M. and {Haywood}, M. and {Helmi}, A. and {Hidalgo}, S.~L. and {Hilger}, T. and {H{\l}adczuk}, N. and {Hobbs}, D. and {Holland}, G. and {Huckle}, H.~E. and {Jasniewicz}, G. and {Jonker}, P.~G. and {Juaristi Campillo}, J. and {Julbe}, F. and {Karbevska}, L. and {Kervella}, P. and {Khanna}, S. and {Kochoska}, A. and {Kontizas}, M. and {Kordopatis}, G. and {Korn}, A.~J. and {Kostrzewa-Rutkowska}, Z. and {Kruszy{\'n}ska}, K. and {Lambert}, S. and {Lanza}, A.~F. and {Lasne}, Y. and {Le Campion}, J. -F. and {Le Fustec}, Y. and {Lebreton}, Y. and {Lebzelter}, T. and {Leccia}, S. and {Leclerc}, N. and {Lecoeur-Taibi}, I. and {Liao}, S. and {Licata}, E. and {Lindstr{\o}m}, E.~P. and {Lister}, T.~A. and {Livanou}, E. and {Lobel}, A. and {Madrero Pardo}, P. and {Managau}, S. and {Mann}, R.~G. and {Marchant}, J.~M. and {Marconi}, M. and {Marcos Santos}, M.~M.~S. and {Marinoni}, S. and {Marocco}, F. and {Marshall}, D.~J. and {Martin Polo}, L. and {Mart{\'\i}n-Fleitas}, J.~M. and {Masip}, A. and {Massari}, D. and {Mastrobuono-Battisti}, A. and {Mazeh}, T. and {McMillan}, P.~J. and {Messina}, S. and {Michalik}, D. and {Millar}, N.~R. and {Mints}, A. and {Molina}, D. and {Molinaro}, R. and {Moln{\'a}r}, L. and {Montegriffo}, P. and {Mor}, R. and {Morbidelli}, R. and {Morel}, T. and {Morris}, D. and {Mulone}, A.~F. and {Munoz}, D. and {Muraveva}, T. and {Murphy}, C.~P. and {Musella}, I. and {Noval}, L. and {Ord{\'e}novic}, C. and {Orr{\`u}}, G. and {Osinde}, J. and {Pagani}, C. and {Pagano}, I. and {Palaversa}, L. and {Palicio}, P.~A. and {Panahi}, A. and {Pawlak}, M. and {Pe{\~n}alosa Esteller}, X. and {Penttil{\"a}}, A. and {Piersimoni}, A.~M. and {Pineau}, F. -X. and {Plachy}, E. and {Plum}, G. and {Poggio}, E. and {Poretti}, E. and {Poujoulet}, E. and {Pr{\v{s}}a}, A. and {Pulone}, L. and {Racero}, E. and {Ragaini}, S. and {Rainer}, M. and {Raiteri}, C.~M. and {Rambaux}, N. and {Ramos}, P. and {Ramos-Lerate}, M. and {Re Fiorentin}, P. and {Regibo}, S. and {Reyl{\'e}}, C. and {Ripepi}, V. and {Riva}, A. and {Rixon}, G. and {Robichon}, N. and {Robin}, C. and {Roelens}, M. and {Rohrbasser}, L. and {Romero-G{\'o}mez}, M. and {Rowell}, N. and {Royer}, F. and {Rybicki}, K.~A. and {Sadowski}, G. and {Sagrist{\`a} Sell{\'e}s}, A. and {Sahlmann}, J. and {Salgado}, J. and {Salguero}, E. and {Samaras}, N. and {Sanchez Gimenez}, V. and {Sanna}, N. and {Santove{\~n}a}, R. and {Sarasso}, M. and {Schultheis}, M. and {Sciacca}, E. and {Segol}, M. and {Segovia}, J.~C. and {S{\'e}gransan}, D. and {Semeux}, D. and {Shahaf}, S. and {Siddiqui}, H.~I. and {Siebert}, A. and {Siltala}, L. and {Slezak}, E. and {Smart}, R.~L. and {Solano}, E. and {Solitro}, F. and {Souami}, D. and {Souchay}, J. and {Spagna}, A. and {Spoto}, F. and {Steele}, I.~A. and {Steidelm{\"u}ller}, H. and {Stephenson}, C.~A. and {S{\"u}veges}, M. and {Szabados}, L. and {Szegedi-Elek}, E. and {Taris}, F. and {Tauran}, G. and {Taylor}, M.~B. and {Teixeira}, R. and {Thuillot}, W. and {Tonello}, N. and {Torra}, F. and {Torra}, J. and {Turon}, C. and {Unger}, N. and {Vaillant}, M. and {van Dillen}, E. and {Vanel}, O. and {Vecchiato}, A. and {Viala}, Y. and {Vicente}, D. and {Voutsinas}, S. and {Weiler}, M. and {Wevers}, T. and {Wyrzykowski}, {\L}. and {Yoldas}, A. and {Yvard}, P. and {Zhao}, H. and {Zorec}, J. and {Zucker}, S. and {Zurbach}, C. and {Zwitter}, T.},
        title = "{Gaia Early Data Release 3. Summary of the contents and survey properties}",
      journal = {\aap},
         year = 2021,
        month = may,
       volume = {649},
          eid = {A1},
        pages = {A1},
          doi = {10.1051/0004-6361/202039657},
archivePrefix = {arXiv},
       eprint = {2012.01533},
 primaryClass = {astro-ph.GA},
       adsurl = {https://ui.adsabs.harvard.edu/abs/2021A&A...649A...1G}
}

@ARTICLE{DES_2016MNRAS.460.1270D,
       author = {{Dark Energy Survey Collaboration} and {Abbott}, T. and {Abdalla}, F.~B. and {Aleksi{\'c}}, J. and {Allam}, S. and {Amara}, A. and {Bacon}, D. and {Balbinot}, E. and {Banerji}, M. and {Bechtol}, K. and {Benoit-L{\'e}vy}, A. and {Bernstein}, G.~M. and {Bertin}, E. and {Blazek}, J. and {Bonnett}, C. and {Bridle}, S. and {Brooks}, D. and {Brunner}, R.~J. and {Buckley-Geer}, E. and {Burke}, D.~L. and {Caminha}, G.~B. and {Capozzi}, D. and {Carlsen}, J. and {Carnero-Rosell}, A. and {Carollo}, M. and {Carrasco-Kind}, M. and {Carretero}, J. and {Castander}, F.~J. and {Clerkin}, L. and {Collett}, T. and {Conselice}, C. and {Crocce}, M. and {Cunha}, C.~E. and {D'Andrea}, C.~B. and {da Costa}, L.~N. and {Davis}, T.~M. and {Desai}, S. and {Diehl}, H.~T. and {Dietrich}, J.~P. and {Dodelson}, S. and {Doel}, P. and {Drlica-Wagner}, A. and {Estrada}, J. and {Etherington}, J. and {Evrard}, A.~E. and {Fabbri}, J. and {Finley}, D.~A. and {Flaugher}, B. and {Foley}, R.~J. and {Fosalba}, P. and {Frieman}, J. and {Garc{\'\i}a-Bellido}, J. and {Gaztanaga}, E. and {Gerdes}, D.~W. and {Giannantonio}, T. and {Goldstein}, D.~A. and {Gruen}, D. and {Gruendl}, R.~A. and {Guarnieri}, P. and {Gutierrez}, G. and {Hartley}, W. and {Honscheid}, K. and {Jain}, B. and {James}, D.~J. and {Jeltema}, T. and {Jouvel}, S. and {Kessler}, R. and {King}, A. and {Kirk}, D. and {Kron}, R. and {Kuehn}, K. and {Kuropatkin}, N. and {Lahav}, O. and {Li}, T.~S. and {Lima}, M. and {Lin}, H. and {Maia}, M.~A.~G. and {Makler}, M. and {Manera}, M. and {Maraston}, C. and {Marshall}, J.~L. and {Martini}, P. and {McMahon}, R.~G. and {Melchior}, P. and {Merson}, A. and {Miller}, C.~J. and {Miquel}, R. and {Mohr}, J.~J. and {Morice-Atkinson}, X. and {Naidoo}, K. and {Neilsen}, E. and {Nichol}, R.~C. and {Nord}, B. and {Ogando}, R. and {Ostrovski}, F. and {Palmese}, A. and {Papadopoulos}, A. and {Peiris}, H.~V. and {Peoples}, J. and {Percival}, W.~J. and {Plazas}, A.~A. and {Reed}, S.~L. and {Refregier}, A. and {Romer}, A.~K. and {Roodman}, A. and {Ross}, A. and {Rozo}, E. and {Rykoff}, E.~S. and {Sadeh}, I. and {Sako}, M. and {S{\'a}nchez}, C. and {Sanchez}, E. and {Santiago}, B. and {Scarpine}, V. and {Schubnell}, M. and {Sevilla-Noarbe}, I. and {Sheldon}, E. and {Smith}, M. and {Smith}, R.~C. and {Soares-Santos}, M. and {Sobreira}, F. and {Soumagnac}, M. and {Suchyta}, E. and {Sullivan}, M. and {Swanson}, M. and {Tarle}, G. and {Thaler}, J. and {Thomas}, D. and {Thomas}, R.~C. and {Tucker}, D. and {Vieira}, J.~D. and {Vikram}, V. and {Walker}, A.~R. and {Wechsler}, R.~H. and {Weller}, J. and {Wester}, W. and {Whiteway}, L. and {Wilcox}, H. and {Yanny}, B. and {Zhang}, Y. and {Zuntz}, J.},
        title = "{The Dark Energy Survey: more than dark energy - an overview}",
      journal = {\mnras},
         year = 2016,
        month = aug,
       volume = {460},
       number = {2},
        pages = {1270-1299},
          doi = {10.1093/mnras/stw641},
archivePrefix = {arXiv},
       eprint = {1601.00329},
 primaryClass = {astro-ph.CO},
       adsurl = {https://ui.adsabs.harvard.edu/abs/2016MNRAS.460.1270D}
}

@ARTICLE{ZTF_2019PASP..131a8002B,
       author = {{Bellm}, Eric C. and {Kulkarni}, Shrinivas R. and {Graham}, Matthew J. and {Dekany}, Richard and {Smith}, Roger M. and {Riddle}, Reed and {Masci}, Frank J. and {Helou}, George and {Prince}, Thomas A. and {Adams}, Scott M. and {Barbarino}, C. and {Barlow}, Tom and {Bauer}, James and {Beck}, Ron and {Belicki}, Justin and {Biswas}, Rahul and {Blagorodnova}, Nadejda and {Bodewits}, Dennis and {Bolin}, Bryce and {Brinnel}, Valery and {Brooke}, Tim and {Bue}, Brian and {Bulla}, Mattia and {Burruss}, Rick and {Cenko}, S. Bradley and {Chang}, Chan-Kao and {Connolly}, Andrew and {Coughlin}, Michael and {Cromer}, John and {Cunningham}, Virginia and {De}, Kishalay and {Delacroix}, Alex and {Desai}, Vandana and {Duev}, Dmitry A. and {Eadie}, Gwendolyn and {Farnham}, Tony L. and {Feeney}, Michael and {Feindt}, Ulrich and {Flynn}, David and {Franckowiak}, Anna and {Frederick}, S. and {Fremling}, C. and {Gal-Yam}, Avishay and {Gezari}, Suvi and {Giomi}, Matteo and {Goldstein}, Daniel A. and {Golkhou}, V. Zach and {Goobar}, Ariel and {Groom}, Steven and {Hacopians}, Eugean and {Hale}, David and {Henning}, John and {Ho}, Anna Y.~Q. and {Hover}, David and {Howell}, Justin and {Hung}, Tiara and {Huppenkothen}, Daniela and {Imel}, David and {Ip}, Wing-Huen and {Ivezi{\'c}}, {\v{Z}}eljko and {Jackson}, Edward and {Jones}, Lynne and {Juric}, Mario and {Kasliwal}, Mansi M. and {Kaspi}, S. and {Kaye}, Stephen and {Kelley}, Michael S.~P. and {Kowalski}, Marek and {Kramer}, Emily and {Kupfer}, Thomas and {Landry}, Walter and {Laher}, Russ R. and {Lee}, Chien-De and {Lin}, Hsing Wen and {Lin}, Zhong-Yi and {Lunnan}, Ragnhild and {Giomi}, Matteo and {Mahabal}, Ashish and {Mao}, Peter and {Miller}, Adam A. and {Monkewitz}, Serge and {Murphy}, Patrick and {Ngeow}, Chow-Choong and {Nordin}, Jakob and {Nugent}, Peter and {Ofek}, Eran and {Patterson}, Maria T. and {Penprase}, Bryan and {Porter}, Michael and {Rauch}, Ludwig and {Rebbapragada}, Umaa and {Reiley}, Dan and {Rigault}, Mickael and {Rodriguez}, Hector and {van Roestel}, Jan and {Rusholme}, Ben and {van Santen}, Jakob and {Schulze}, S. and {Shupe}, David L. and {Singer}, Leo P. and {Soumagnac}, Maayane T. and {Stein}, Robert and {Surace}, Jason and {Sollerman}, Jesper and {Szkody}, Paula and {Taddia}, F. and {Terek}, Scott and {Van Sistine}, Angela and {van Velzen}, Sjoert and {Vestrand}, W. Thomas and {Walters}, Richard and {Ward}, Charlotte and {Ye}, Quan-Zhi and {Yu}, Po-Chieh and {Yan}, Lin and {Zolkower}, Jeffry},
        title = "{The Zwicky Transient Facility: System Overview, Performance, and First Results}",
      journal = {\pasp},
         year = 2019,
        month = jan,
       volume = {131},
       number = {995},
        pages = {018002},
          doi = {10.1088/1538-3873/aaecbe},
archivePrefix = {arXiv},
       eprint = {1902.01932},
 primaryClass = {astro-ph.IM},
       adsurl = {https://ui.adsabs.harvard.edu/abs/2019PASP..131a8002B}
}

@ARTICLE{RT_2018A&A...611A..45R,
       author = {{Renault-Tinacci}, N. and {Kotera}, K. and {Neronov}, A. and {Ando}, S.},
        title = "{Search for {\ensuremath{\gamma}}-ray emission from superluminous supernovae with the Fermi-LAT}",
      journal = {\aap},
         year = 2018,
        month = mar,
       volume = {611},
          eid = {A45},
        pages = {A45},
          doi = {10.1051/0004-6361/201730741},
archivePrefix = {arXiv},
       eprint = {1708.08971},
 primaryClass = {astro-ph.HE},
       adsurl = {https://ui.adsabs.harvard.edu/abs/2018A&A...611A..45R}
}

@misc{Li:2024ics,
    author = "Li, Shang and Liang, Yun-Feng and Liao, Neng-Hui and Lei, Lei and Fan, Yi-Zhong",
    title = "{Fermi-LAT discovery of the GeV emission of the superluminous supernovae SN 2017egm}",
    eprint = "2407.05968",
    archivePrefix = "arXiv",
    primaryClass = "astro-ph.HE",
    month = jul,
    year = "2024"
}

@article{VERITAS:2023msf,
    author = "Acharyya, A. and others",
    collaboration = "VERITAS",
    title = "{VERITAS and Fermi-LAT Constraints on the Gamma-Ray Emission from Superluminous Supernovae SN2015bn and SN2017egm}",
    eprint = "2302.06686",
    archivePrefix = "arXiv",
    primaryClass = "astro-ph.HE",
    doi = "10.3847/1538-4357/acb7e6",
    journal = "Astrophys. J.",
    volume = "945",
    number = "1",
    pages = "30",
    year = "2023"
}

@article{Gomez:2024xce,
    author = "Gomez, Sebastian and others",
    title = "{The Type I superluminous supernova catalogue I: light-curve properties, models, and catalogue description}",
    eprint = "2407.07946",
    archivePrefix = "arXiv",
    primaryClass = "astro-ph.HE",
    doi = "10.1093/mnras/stae2270",
    journal = "Mon. Not. Roy. Astron. Soc.",
    volume = "535",
    number = "1",
    pages = "471--515",
    year = "2024"
}

@article{Guillochon:2016rhj,
    author = "Guillochon, James and Parrent, Jerod and Kelley, Luke Zoltan and Margutti, Raffaella",
    title = "{An Open Catalog for Supernova Data}",
    eprint = "1605.01054",
    archivePrefix = "arXiv",
    primaryClass = "astro-ph.SR",
    doi = "10.3847/1538-4357/835/1/64",
    journal = "Astrophys. J.",
    volume = "835",
    number = "1",
    pages = "64",
    year = "2017"
}

@article{Gomez:2020jyq,
    author = "Gomez, Sebastian and Berger, Edo and Blanchard, Peter K. and Hosseinzadeh, Griffin and Nicholl, Matt and Villar, V. Ashley and Yin, Yao",
    title = "{FLEET: A Redshift-Agnostic Machine Learning Pipeline to Rapidly Identify Hydrogen-Poor Superluminous Supernovae}",
    eprint = "2009.01853",
    archivePrefix = "arXiv",
    primaryClass = "astro-ph.HE",
    doi = "10.3847/1538-4357/abbf49",
    journal = "Astrophys. J.",
    volume = "904",
    number = "1",
    pages = "74",
    year = "2020"
}

@article{Gomez:2022xsq,
    author = "Gomez, Sebastian and Berger, Edo and Blanchard, Peter K. and Hosseinzadeh, Griffin and Nicholl, Matt and Hiramatsu, Daichi and Villar, V. Ashley and Yin, Yao",
    title = "{The First Two Years of FLEET: An Active Search for Superluminous Supernovae}",
    eprint = "2210.10811",
    archivePrefix = "arXiv",
    primaryClass = "astro-ph.HE",
    doi = "10.3847/1538-4357/acc536",
    journal = "Astrophys. J.",
    volume = "949",
    number = "2",
    pages = "114",
    year = "2023"
}

@ARTICLE{FermiMission2009,
       author = {Fermi-LAT Collaboration},
        title = "{The Large Area Telescope on the Fermi Gamma-Ray Space Telescope Mission}",
      journal = {\apj},
         year = 2009,
        month = jun,
       volume = {697},
       number = {2},
        pages = {1071-1102},
          doi = {10.1088/0004-637X/697/2/1071},
archivePrefix = {arXiv},
       eprint = {0902.1089},
 primaryClass = {astro-ph.IM},
       adsurl = {https://ui.adsabs.harvard.edu/abs/2009ApJ...697.1071A}
}

@article{Fermi-LAT:2022byn,
    author = "Abdollahi, Soheila and others",
    collaboration = "Fermi-LAT",
    title = "{Incremental Fermi Large Area Telescope Fourth Source Catalog}",
    eprint = "2201.11184",
    archivePrefix = "arXiv",
    primaryClass = "astro-ph.HE",
    doi = "10.3847/1538-4365/ac6751",
    journal = "Astrophys. J. Supp.",
    volume = "260",
    number = "2",
    pages = "53",
    year = "2022"
}

@ARTICLE{Nicholl2017,
  author = {{Nicholl}, M. and {Berger}, E. and {Margutti}, R. and et al.},
  title = {The Superluminous Supernova SN 2017egm in the Nearby Spiral Galaxy NGC 3191},
  journal = {ApJL},
  year = {2017},
  volume = {845},
  pages = {L8},
  doi = {10.3847/2041-8213/aa82b1},
  adsurl = {https://ui.adsabs.harvard.edu/abs/2017ApJ...845L...8N}
}

@ARTICLE{Bose2018,
       author = {{Bose}, Subhash and {Dong}, Subo and {Pastorello}, A. and {Filippenko}, Alexei V. and {Kochanek}, C.~S. and {Mauerhan}, Jon and {Romero-Ca{\~n}izales}, C. and {Brink}, Thomas G. and {Chen}, Ping and {Prieto}, J.~L. and {Post}, R. and {Ashall}, Christopher and {Grupe}, Dirk and {Tomasella}, L. and {Benetti}, Stefano and {Shappee}, B.~J. and {Stanek}, K.~Z. and {Cai}, Zheng and {Falco}, E. and {Lundqvist}, Peter and {Mattila}, Seppo and {Mutel}, Robert and {Ochner}, Paolo and {Pooley}, David and {Stritzinger}, M.~D. and {Villanueva}, Jr., S. and {Zheng}, WeiKang and {Beswick}, R.~J. and {Brown}, Peter J. and {Cappellaro}, E. and {Davis}, Scott and {Fraser}, Morgan and {de Jaeger}, Thomas and {Elias-Rosa}, N. and {Gall}, C. and {Gaudi}, B. Scott and {Herczeg}, Gregory J. and {Hestenes}, Julia and {Holoien}, T.~W.-S. and {Hosseinzadeh}, Griffin and {Hsiao}, E.~Y. and {Hu}, Shaoming and {Jaejin}, Shin and {Jeffers}, Ben and {Koff}, R.~A. and {Kumar}, Sahana and {Kurtenkov}, Alexander and {Lau}, Marie Wingyee and {Prentice}, Simon and {Reynolds}, T. and {Rudy}, Richard J. and {Shahbandeh}, Melissa and {Somero}, Auni and {Stassun}, Keivan G. and {Thompson}, Todd A. and {Valenti}, Stefano and {Woo}, Jong-Hak and {Yunus}, Sameen},
        title = "{Gaia17biu/SN 2017egm in NGC 3191: The Closest Hydrogen-poor Superluminous Supernova to Date Is in a {\textquotedblleft}Normal,{\textquotedblright} Massive, Metal-rich Spiral Galaxy}",
      journal = {\apj},
         year = 2018,
        month = jan,
       volume = {853},
       number = {1},
          eid = {57},
        pages = {57},
          doi = {10.3847/1538-4357/aaa298},
archivePrefix = {arXiv},
       eprint = {1708.00864},
 primaryClass = {astro-ph.HE},
       adsurl = {https://ui.adsabs.harvard.edu/abs/2018ApJ...853...57B}
}

@ARTICLE{Izzo2018,
       author = {{Izzo}, L. and {Th{\"o}ne}, C.~C. and {Garc{\'\i}a-Benito}, R. and {de Ugarte Postigo}, A. and {Cano}, Z. and {Kann}, D.~A. and {Bensch}, K. and {Della Valle}, M. and {Galad{\'\i}-Enr{\'\i}quez}, D. and {Hedrosa}, R.~P.},
        title = "{The host of the Type I SLSN 2017egm. A young, sub-solar metallicity environment in a massive spiral galaxy}",
      journal = {\aap},
         year = 2018,
        month = feb,
       volume = {610},
          eid = {A11},
        pages = {A11},
          doi = {10.1051/0004-6361/201731766},
archivePrefix = {arXiv},
       eprint = {1708.03856},
 primaryClass = {astro-ph.HE},
       adsurl = {https://ui.adsabs.harvard.edu/abs/2018A&A...610A..11I}
}

@ARTICLE{Pursiainen2022,
       author = {{Pursiainen}, M. and {Leloudas}, G. and {Paraskeva}, E. and {Cikota}, A. and {Anderson}, J.~P. and {Angus}, C.~R. and {Brennan}, S. and {Bulla}, M. and {Camacho-I{\~n}iguez}, E. and {Charalampopoulos}, P. and {Chen}, T.-W. and {Delgado Manche{\~n}o}, M. and {Fraser}, M. and {Frohmaier}, C. and {Galbany}, L. and {Guti{\'e}rrez}, C.~P. and {Gromadzki}, M. and {Inserra}, C. and {Maund}, J. and {M{\"u}ller-Bravo}, T.~E. and {Mu{\~n}oz Torres}, S. and {Nicholl}, M. and {Onori}, F. and {Patat}, F. and {Pessi}, P.~J. and {Roy}, R. and {Spyromilio}, J. and {Wiseman}, P. and {Young}, D.~R.},
        title = "{SN 2018bsz: A Type I superluminous supernova with aspherical circumstellar material}",
      journal = {\aap},
         year = 2022,
        month = oct,
       volume = {666},
          eid = {A30},
        pages = {A30},
          doi = {10.1051/0004-6361/202243256},
archivePrefix = {arXiv},
       eprint = {2202.01635},
 primaryClass = {astro-ph.HE},
       adsurl = {https://ui.adsabs.harvard.edu/abs/2022A&A...666A..30P}
}

@ARTICLE{Schulze2018,
       author = {{Schulze}, S. and {Kr{\"u}hler}, T. and {Leloudas}, G. and {Gorosabel}, J. and {Mehner}, A. and {Buchner}, J. and {Kim}, S. and {Ibar}, E. and {Amor{\'\i}n}, R. and {Herrero-Illana}, R. and {Anderson}, J.~P. and {Bauer}, F.~E. and {Christensen}, L. and {de Pasquale}, M. and {de Ugarte Postigo}, A. and {Gallazzi}, A. and {Hjorth}, J. and {Morrell}, N. and {Malesani}, D. and {Sparre}, M. and {Stalder}, B. and {Stark}, A.~A. and {Th{\"o}ne}, C.~C. and {Wheeler}, J.~C.},
        title = "{Cosmic evolution and metal aversion in superluminous supernova host galaxies}",
      journal = {\mnras},
         year = 2018,
        month = jan,
       volume = {473},
       number = {1},
        pages = {1258-1285},
          doi = {10.1093/mnras/stx2352},
archivePrefix = {arXiv},
       eprint = {1612.05978},
 primaryClass = {astro-ph.GA},
       adsurl = {https://ui.adsabs.harvard.edu/abs/2018MNRAS.473.1258S}
}

@article{Fermi-LAT:2016uux,
    author = "Albert, A. and others",
    collaboration = "Fermi-LAT, DES",
    title = "{Searching for Dark Matter Annihilation in Recently Discovered Milky Way Satellites with Fermi-LAT}",
    eprint = "1611.03184",
    archivePrefix = "arXiv",
    primaryClass = "astro-ph.HE",
    reportNumber = "FERMILAB-PUB-16-073-AE",
    doi = "10.3847/1538-4357/834/2/110",
    journal = "Astrophys. J.",
    volume = "834",
    number = "2",
    pages = "110",
    year = "2017"
}

@article{McDaniel:2023bju,
    author = "McDaniel, Alex and Ajello, Marco and Karwin, Christopher M. and Di Mauro, Mattia and Drlica-Wagner, Alex and S\'anchez-Conde, Miguel A.",
    title = "{Legacy analysis of dark matter annihilation from the Milky~Way dwarf spheroidal galaxies with 14~years of Fermi-LAT data}",
    eprint = "2311.04982",
    archivePrefix = "arXiv",
    primaryClass = "astro-ph.HE",
    reportNumber = "FERMILAB-PUB-23-686-PPD",
    doi = "10.1103/PhysRevD.109.063024",
    journal = "Phys. Rev. D",
    volume = "109",
    number = "6",
    pages = "063024",
    year = "2024"
}

@misc{Crnogorcevic:2025gxd,
    author = "Crnogor{\v{c}}evi{\'c}, Milena and Linden, Tim and Peter, Annika H. G.",
    title = "{Are X-Ray Detected Active Galactic Nuclei in Dwarf Galaxies Gamma-Ray Bright?}",
    eprint = "2509.18239",
    archivePrefix = "arXiv",
    primaryClass = "astro-ph.HE",
    month = sep,
    year = "2025"
}

@article{wilks1938,
author = "Wilks, S. S.",
doi = "10.1214/aoms/1177732360",
fjournal = "The Annals of Mathematical Statistics",
journal = "Ann. Math. Statist.",
month = mar,
number = "1",
pages = "60--62",
publisher = "The Institute of Mathematical Statistics",
title = "The Large-Sample Distribution of the Likelihood Ratio for Testing Composite Hypotheses",
url = "https://doi.org/10.1214/aoms/1177732360",
volume = "9",
year = "1938"
}

@ARTICLE{astropy:2022,
       author = {{Astropy Collaboration}},
        title = "{The Astropy Project: Sustaining and Growing a Community-oriented Open-source Project and the Latest Major Release (v5.0) of the Core Package}",
      journal = {\apj},
         year = 2022,
        month = aug,
       volume = {935},
       number = {2},
          eid = {167},
        pages = {167},
          doi = {10.3847/1538-4357/ac7c74},
archivePrefix = {arXiv},
       eprint = {2206.14220},
 primaryClass = {astro-ph.IM},
       adsurl = {https://ui.adsabs.harvard.edu/abs/2022ApJ...935..167A}
}

@Article{         numpy:2020,
 title         = {Array programming with {NumPy}},
 author        = {Charles R. Harris and K. Jarrod Millman and St{\'{e}}fan J.
                 van der Walt and Ralf Gommers and Pauli Virtanen and David
                 Cournapeau and Eric Wieser and Julian Taylor and Sebastian
                 Berg and Nathaniel J. Smith and Robert Kern and Matti Picus
                 and Stephan Hoyer and Marten H. van Kerkwijk and Matthew
                 Brett and Allan Haldane and Jaime Fern{\'{a}}ndez del
                 R{\'{i}}o and Mark Wiebe and Pearu Peterson and Pierre
                 G{\'{e}}rard-Marchant and Kevin Sheppard and Tyler Reddy and
                 Warren Weckesser and Hameer Abbasi and Christoph Gohlke and
                 Travis E. Oliphant},
 year          = {2020},
 month         = sep,
 journal       = {Nature},
 volume        = {585},
 number        = {7825},
 pages         = {357--362},
 doi           = {10.1038/s41586-020-2649-2},
 publisher     = {Springer Science and Business Media {LLC}},
 url           = {https://doi.org/10.1038/s41586-020-2649-2}
}

@Article{matplotlib:2007,
  Author    = {Hunter, J. D.},
  Title     = {Matplotlib: A 2D graphics environment},
  Journal   = {Computing in Science \& Engineering},
  Volume    = {9},
  Number    = {3},
  Pages     = {90--95},
  publisher = {IEEE COMPUTER SOC},
  doi       = {10.1109/MCSE.2007.55},
  year      = 2007
}

@software{pandas:2020,
    author       = {The Pandas Development Team},
    title        = {pandas-dev/pandas: Pandas},
    month        = feb,
    year         = 2020,
    publisher    = {Zenodo},
    version      = {latest},
    doi          = {10.5281/zenodo.3509134},
    url          = {https://doi.org/10.5281/zenodo.3509134}
}

@article{Anderson:2018yst,
    author = "Anderson, J. P. and others",
    title = "{A nearby super-luminous supernova with a long pre-maximum {\textbackslash}{\&} {\textquotedblleft}plateau{\textquotedblright} and strong CII features}",
    eprint = "1806.10609",
    archivePrefix = "arXiv",
    primaryClass = "astro-ph.HE",
    doi = "10.1051/0004-6361/201833725",
    journal = "Astron. Astrophys.",
    volume = "620",
    pages = "A67",
    year = "2018"
}

@article{Hosseinzadeh:2021uep,
    author = "Hosseinzadeh, Griffin and Berger, Edo and Metzger, Brian D. and Gomez, Sebastian and Nicholl, Matt and Blanchard, Peter",
    title = "{Bumpy Declining Light Curves Are Common in Hydrogen-poor Superluminous Supernovae}",
    eprint = "2109.09743",
    archivePrefix = "arXiv",
    primaryClass = "astro-ph.HE",
    doi = "10.3847/1538-4357/ac67dd",
    journal = "Astrophys. J.",
    volume = "933",
    number = "1",
    pages = "14",
    year = "2022"
}

@article{Cheung:2022joh,
    author = "Cheung, C. C. and others",
    title = "{Fermi LAT Gamma-ray Detection of the Recurrent Nova RS Ophiuchi during its 2021 Outburst}",
    eprint = "2207.02921",
    archivePrefix = "arXiv",
    primaryClass = "astro-ph.HE",
    doi = "10.3847/1538-4357/ac7eb7",
    journal = "Astrophys. J.",
    volume = "935",
    number = "1",
    pages = "44",
    year = "2022"
}

@article{Marti-Devesa:2024hic,
    author = "Mart{\'\i}-Devesa, G. and Cheung, C. C. and Di Lalla, N. and Renaud, M. and Principe, G. and Omodei, N. and Acero, F.",
    title = "{Early-time gamma-ray constraints on cosmic-ray acceleration in the core-collapse SN 2023ixf with the Fermi Large Area Telescope}",
    eprint = "2404.10487",
    archivePrefix = "arXiv",
    primaryClass = "astro-ph.HE",
    doi = "10.1051/0004-6361/202349061",
    journal = "Astron. Astrophys.",
    volume = "686",
    pages = "A254",
    year = "2024"
}

@article{Acero:2026slsn,
      author         = {Fabio Acero et al.},
      title          = {Gamma-ray signature of superluminous supernovae: {Fermi}-{LAT} {GeV} detection of {SN}~2017egm and evidence for a central engine},
      journal        = {Astronomy \& Astrophysics},
      year           = {2026},
      note           = {accepted},
      collaboration  = {Fermi-LAT},
}

@article{Zhu:2023ntt,
    author = "Zhu, Jiazheng and others",
    title = "{SN 2017egm: A Helium-rich Superluminous Supernova with Multiple Bumps in the Light Curves}",
    eprint = "2303.03424",
    archivePrefix = "arXiv",
    primaryClass = "astro-ph.HE",
    doi = "10.3847/1538-4357/acc2c3",
    journal = "Astrophys. J.",
    volume = "949",
    number = "1",
    pages = "23",
    year = "2023"
}

@article{Mazzali:2016pgx,
    author = "Mazzali, P. A. and Sullivan, M. and Pian, E. and Greiner, J. and Kann, D. A.",
    title = "{Spectrum formation in Superluminous Supernovae (Type I)}",
    eprint = "1603.00388",
    archivePrefix = "arXiv",
    primaryClass = "astro-ph.HE",
    doi = "10.1093/mnras/stw512",
    journal = "Mon. Not. Roy. Astron. Soc.",
    volume = "458",
    number = "4",
    pages = "3455--3465",
    year = "2016"
}

@article{Quimby:2013jb,
    author = "Quimby, Robert M. and Yuan, Fang and Akerlof, Carl and Wheeler, J. Craig",
    title = "{Rates of Superluminous Supernovae at z{\textasciitilde}0.2}",
    eprint = "1302.0911",
    archivePrefix = "arXiv",
    primaryClass = "astro-ph.CO",
    doi = "10.1093/mnras/stt213",
    journal = "Mon. Not. Roy. Astron. Soc.",
    volume = "431",
    pages = "912",
    year = "2013"
}

@article{Prajs:2016cjj,
    author = "Prajs, S. and others",
    title = "{The volumetric rate of superluminous supernovae at z {\ensuremath{\sim}} 1}",
    eprint = "1605.05250",
    archivePrefix = "arXiv",
    primaryClass = "astro-ph.HE",
    doi = "10.1093/mnras/stw1942",
    journal = "Mon. Not. Roy. Astron. Soc.",
    volume = "464",
    number = "3",
    pages = "3568--3579",
    year = "2017"
}

@misc{Margutti_priv,
  author = {Margutti, Raffaella},
  note   = {private communication}
}

@article{Fermi-LAT:2023zzt,
    author = "Smith, D. A. and others",
    collaboration = "Fermi-LAT",
    title = "{The Third Fermi Large Area Telescope Catalog of Gamma-Ray Pulsars}",
    eprint = "2307.11132",
    archivePrefix = "arXiv",
    primaryClass = "astro-ph.HE",
    doi = "10.3847/1538-4357/acee67",
    journal = "Astrophys. J.",
    volume = "958",
    number = "2",
    pages = "191",
    year = "2023"
}

@ARTICLE{Quimby+07,
   author = {{Quimby}, R.~M. and {Aldering}, G. and {Wheeler}, J.~C. and 
	{H{\"o}flich}, P. and {Akerlof}, C.~W. and {Rykoff}, E.~S.},
    title = "{SN 2005ap: A Most Brilliant Explosion}",
  journal = {\apjl},
archivePrefix = "arXiv",
   eprint = {0709.0302},
     year = 2007,
    month = oct,
   volume = 668,
    pages = {L99-L102},
      doi = {10.1086/522862},
   adsurl = {http://adsabs.harvard.edu/abs/2007ApJ...668L..99Q}
}

@ARTICLE{GalYam18,
       author = {{Gal-Yam}, Avishay},
        title = "{The Most Luminous Supernovae}",
      journal = {\araa},
         year = 2019,
        month = aug,
       volume = {57},
        pages = {305-333},
          doi = {10.1146/annurev-astro-081817-051819},
archivePrefix = {arXiv},
       eprint = {1812.01428},
 primaryClass = {astro-ph.HE},
       adsurl = {https://ui.adsabs.harvard.edu/abs/2019ARA&A..57..305G}
}

@ARTICLE{Smith+07,
   author = {{Smith}, N. and {Li}, W. and {Foley}, R.~J. and {Wheeler}, J.~C. and 
	{Pooley}, D. and {Chornock}, R. and {Filippenko}, A.~V. and 
	{Silverman}, J.~M. and {Quimby}, R. and {Bloom}, J.~S. and {Hansen}, C.
	},
    title = "{SN 2006gy: Discovery of the Most Luminous Supernova Ever Recorded, Powered by the Death of an Extremely Massive Star like {$\eta$} Carinae}",
  journal = {\apj},
   eprint = {arXiv:astro-ph/0612617},
     year = 2007,
    month = sep,
   volume = 666,
    pages = {1116-1128},
      doi = {10.1086/519949},
   adsurl = {http://adsabs.harvard.edu/abs/2007ApJ...666.1116S}
}

@ARTICLE{Chevalier&Irwin11,
       author = {{Chevalier}, Roger A. and {Irwin}, Christopher M.},
        title = "{Shock Breakout in Dense Mass Loss: Luminous Supernovae}",
      journal = {\apjl},
         year = 2011,
        month = mar,
       volume = {729},
       number = {1},
          eid = {L6},
        pages = {L6},
          doi = {10.1088/2041-8205/729/1/L6},
archivePrefix = {arXiv},
       eprint = {1101.1111},
 primaryClass = {astro-ph.HE},
       adsurl = {https://ui.adsabs.harvard.edu/abs/2011ApJ...729L...6C}
}

@ARTICLE{Kasen&Bildsten10,
   author = {{Kasen}, D. and {Bildsten}, L.},
    title = "{Supernova Light Curves Powered by Young Magnetars}",
  journal = {\apj},
archivePrefix = "arXiv",
   eprint = {0911.0680},
 primaryClass = "astro-ph.HE",
     year = 2010,
    month = jul,
   volume = 717,
    pages = {245-249},
      doi = {10.1088/0004-637X/717/1/245},
   adsurl = {http://adsabs.harvard.edu/abs/2010ApJ...717..245K}
}

@ARTICLE{Woosley10,
   author = {{Woosley}, S.~E.},
    title = "{Bright Supernovae from Magnetar Birth}",
  journal = {\apjl},
archivePrefix = "arXiv",
   eprint = {0911.0698},
 primaryClass = "astro-ph.HE",
     year = 2010,
    month = aug,
   volume = 719,
    pages = {L204-L207},
      doi = {10.1088/2041-8205/719/2/L204},
   adsurl = {http://adsabs.harvard.edu/abs/2010ApJ...719L.204W}
}

@ARTICLE{Metzger+15,
   author = {{Metzger}, B.~D. and {Margalit}, B. and {Kasen}, D. and {Quataert}, E.
	},
    title = "{The diversity of transients from magnetar birth in core collapse supernovae}",
  journal = {\mnras},
archivePrefix = "arXiv",
   eprint = {1508.02712},
 primaryClass = "astro-ph.HE",
     year = 2015,
    month = dec,
   volume = 454,
    pages = {3311-3316},
      doi = {10.1093/mnras/stv2224},
   adsurl = {http://adsabs.harvard.edu/abs/2015MNRAS.454.3311M}
}

@ARTICLE{Vurm&Metzger21,
       author = {{Vurm}, Indrek and {Metzger}, Brian D.},
        title = "{Gamma-Ray Thermalization and Leakage from Millisecond Magnetar Nebulae: Toward a Self-consistent Model for Superluminous Supernovae}",
      journal = {\apj},
         year = 2021,
        month = aug,
       volume = {917},
       number = {2},
          eid = {77},
        pages = {77},
          doi = {10.3847/1538-4357/ac0826},
archivePrefix = {arXiv},
       eprint = {2101.05299},
 primaryClass = {astro-ph.HE},
       adsurl = {https://ui.adsabs.harvard.edu/abs/2021ApJ...917...77V}
}

@ARTICLE{Caprioli&Spitkovsky14,
       author = {{Caprioli}, D. and {Spitkovsky}, A.},
        title = "{Simulations of Ion Acceleration at Non-relativistic Shocks. I. Acceleration Efficiency}",
      journal = {\apj},
         year = "2014",
        month = "Mar",
       volume = {783},
       number = {2},
          eid = {91},
        pages = {91},
          doi = {10.1088/0004-637X/783/2/91},
archivePrefix = {arXiv},
       eprint = {1310.2943},
 primaryClass = {astro-ph.HE},
       adsurl = {https://ui.adsabs.harvard.edu/abs/2014ApJ...783...91C}
}

@ARTICLE{Nicholl+17,
   author = {{Nicholl}, M. and {Berger}, E. and {Margutti}, R. and {Blanchard}, P.~K. and 
	{Milisavljevic}, D. and {Challis}, P. and {Metzger}, B.~D. and 
	{Chornock}, R.},
    title = "{An Ultraviolet Excess in the Superluminous Supernova Gaia16apd Reveals a Powerful Central Engine}",
  journal = {\apjl},
archivePrefix = "arXiv",
   eprint = {1611.06993},
 primaryClass = "astro-ph.SR",
     year = 2017,
    month = jan,
   volume = 835,
      eid = {L8},
    pages = {L8},
      doi = {10.3847/2041-8213/aa56c5},
   adsurl = {http://adsabs.harvard.edu/abs/2017ApJ...835L...8N}
}

@ARTICLE{Nicholl+17d,
   author = {{Nicholl}, M. and {Guillochon}, J. and {Berger}, E.},
    title = "{The Magnetar Model for Type I Superluminous Supernovae. I. Bayesian Analysis of the Full Multicolor Light-curve Sample with MOSFiT}",
  journal = {\apj},
archivePrefix = "arXiv",
   eprint = {1706.00825},
 primaryClass = "astro-ph.HE",
     year = 2017,
    month = nov,
   volume = 850,
      eid = {55},
    pages = {55},
      doi = {10.3847/1538-4357/aa9334},
   adsurl = {http://adsabs.harvard.edu/abs/2017ApJ...850...55N}
}

@ARTICLE{Murase+19,
       author = {{Murase}, Kohta and {Franckowiak}, Anna and {Maeda}, Keiichi and
         {Margutti}, Raffaella and {Beacom}, John F.},
        title = "{High-energy Emission from Interacting Supernovae: New Constraints on Cosmic-Ray Acceleration in Dense Circumstellar Environments}",
      journal = {\apj},
         year = "2019",
        month = "Mar",
       volume = {874},
       number = {1},
          eid = {80},
        pages = {80},
          doi = {10.3847/1538-4357/ab0422},
archivePrefix = {arXiv},
       eprint = {1807.01460},
 primaryClass = {astro-ph.HE},
       adsurl = {https://ui.adsabs.harvard.edu/abs/2019ApJ...874...80M}
}

@ARTICLE{Margutti+18,
   author = {{Margutti}, R. and {Alexander}, K.~D. and {Xie}, X. and others},
    title = "{The Binary Neutron Star Event LIGO/Virgo GW170817 160 Days after Merger: Synchrotron Emission across the Electromagnetic Spectrum}",
  journal = {\apjl},
archivePrefix = "arXiv",
   eprint = {1801.03531},
 primaryClass = "astro-ph.HE",
     year = 2018,
    month = mar,
   volume = 856,
      eid = {L18},
    pages = {L18},
      doi = {10.3847/2041-8213/aab2ad},
   adsurl = {http://adsabs.harvard.edu/abs/2018ApJ...856L..18M}
}

@ARTICLE{Metzger+18b,
   author = {{Metzger}, B.~D. and {Beniamini}, P. and {Giannios}, D.},
    title = "{Effects of Fall-Back Accretion on Proto-Magnetar Outflows in Gamma-Ray Bursts and Superluminous Supernovae}",
  journal = {ArXiv e-prints},
archivePrefix = "arXiv",
   eprint = {1802.07750},
 primaryClass = "astro-ph.HE",
     year = 2018,
    month = feb,
   adsurl = {http://adsabs.harvard.edu/abs/2018arXiv180207750M}
}

@STRING{apj	= {ApJ} }

\appendix
\section{List of SLSNe}
\label{app:A}
{\footnotesize
\setlength{\tabcolsep}{8pt}
\begin{longtable*}{lllllllll}
\caption{Properties of the 223 SLSNe-I in the final \textit{Fermi}-LAT analysis sample.}
\label{tab:sample} \\
\toprule
Name & R.A. & Dec. & $z$ & $t_{\rm BH}$ & $\Delta t_{\rm LAT}$ & $|b|$ & $d_{\rm 4FGL}$ & Survey \\
 & (J2000) & (J2000) & & (d) & (d) & ($^\circ$) & ($^\circ$) & \\
\midrule
\endfirsthead
\toprule
Name & R.A. & Dec. & $z$ & $t_{\rm BH}$ & $\Delta t_{\rm LAT}$ & $|b|$ & $d_{\rm 4FGL}$ & Survey \\
 & (J2000) & (J2000) & & (d) & (d) & ($^\circ$) & ($^\circ$) & \\
\midrule
\endhead
\midrule
\multicolumn{9}{r}{\textit{Continued on next page}} \\
\endfoot
\bottomrule
\endlastfoot
SN\,2009cb & 12:59:15.85 & +27:16:41.3 & 0.1867 & 62.5 & 156 & 88.3 & 2.83 & PTF \\
SN\,2009jh & 14:49:10.09 & +29:25:10.4 & 0.3499 & 176.4 & 441 & 64.0 & 1.88 & PTF \\
SN\,2010gx & 11:25:46.71 & -08:49:41.4 & 0.2297 & 34.6 & 86 & 48.5 & 1.11 & PS1 \\
SN\,2010kd & 12:08:01.11 & +49:13:31.1 & 0.1010 & 146.8 & 367 & 66.4 & 1.41 & Other \\
SN\,2010md & 16:37:47.04 & +06:12:32.3 & 0.0987 & 137.7 & 344 & 32.4 & 0.67 & PTF \\
SN\,2011ke & 13:50:57.79 & +26:16:42.2 & 0.1428 & 168.3 & 421 & 76.7 & 4.10 & PS1 \\
SN\,2011kf & 14:36:57.53 & +16:30:56.6 & 0.2450 & 54.5 & 136 & 63.4 & 2.10 & CSS \\
SN\,2011kg & 01:39:45.51 & +29:55:27.0 & 0.1924 & 91.0 & 228 & 31.8 & 3.03 & PTF \\
SN\,2011kl & 00:57:22.64 & -46:48:03.6 & 0.6770 & 100.6 & 252 & 70.3 & 4.58 & CSS \\
SN\,2012aa & 14:52:33.48 & -03:31:54.0 & 0.0830 & 440.6 & 1102 & 47.6 & 1.87 & CSS \\
SN\,2012il & 09:46:12.85 & +19:50:28.2 & 0.1750 & 155.9 & 390 & 47.1 & 2.51 & PS1 \\
SN\,2013dg & 13:18:41.35 & -07:04:43.0 & 0.2650 & 68.0 & 170 & 55.2 & 0.83 & CSS \\
SN\,2013hy & 02:42:32.82 & -01:21:30.1 & 0.6630 & 80.9 & 202 & 52.9 & 1.35 & DES \\
SN\,2015bn & 11:33:41.55 & +00:43:33.5 & 0.1136 & 139.7 & 349 & 57.7 & 0.27 & PSST \\
SN\,2016aj & 12:59:00.84 & -26:07:40.6 & 0.4850 & 82.0 & 205 & 36.7 & 2.25 & PSST \\
SN\,2016ard & 14:10:44.56 & -10:09:35.4 & 0.2025 & 72.7 & 182 & 48.0 & 2.34 & PSST \\
SN\,2016eay & 12:02:51.70 & +44:15:27.4 & 0.1013 & 87.5 & 219 & 70.3 & 3.18 & Gaia \\
SN\,2016els & 20:30:13.92 & -10:57:01.8 & 0.2170 & 52.6 & 131 & 26.9 & 1.77 & PSST \\
SN\,2016inl & 02:44:25.83 & +19:10:42.7 & 0.3057 & 223.5 & 559 & 36.2 & 1.64 & PSST \\
SN\,2016wi & 07:58:50.67 & +66:07:39.2 & 0.2240 & 189.0 & 472 & 31.4 & 0.48 & iPTF \\
SN\,2017dwh & 14:34:42.93 & +31:29:16.7 & 0.1300 & 45.0 & 113 & 67.1 & 2.50 & ATLAS \\
SN\,2017egm & 10:19:05.62 & +46:27:14.1 & 0.0307 & 114.9 & 287 & 54.4 & 3.05 & iPTF \\
SN\,2017ens & 12:04:09.37 & -01:55:52.2 & 0.1086 & 195.7 & 489 & 58.8 & 1.21 & ATLAS \\
SN\,2017gci & 06:46:45.03 & -27:14:55.9 & 0.0870 & 145.7 & 364 & 13.0 & 1.70 & Gaia \\
SN\,2017hbx & 17:49:12.64 & +25:22:20.9 & 0.1652 & 79.2 & 198 & 24.3 & 1.26 & Gaia \\
SN\,2017jan & 03:07:22.57 & -64:23:01.0 & 0.3960 & 151.0 & 378 & 46.9 & 0.98 & OGLE \\
SN\,2018avk & 13:11:27.75 & +65:38:17.2 & 0.1320 & 122.9 & 307 & 51.4 & 2.37 & ZTF \\
SN\,2018beh & 09:31:23.03 & +17:48:28.0 & 0.0600 & 135.5 & 339 & 43.2 & 1.57 & ZTF \\
SN\,2018bgv & 11:02:30.29 & +55:35:55.8 & 0.0795 & 72.6 & 182 & 55.4 & 0.85 & ZTF \\
SN\,2018bsz & 16:09:39.11 & -32:03:45.6 & 0.0267 & 49.6 & 124 & 14.3 & 1.21 & ATLAS \\
SN\,2018bym & 18:43:13.43 & +45:12:28.2 & 0.2740 & 50.3 & 126 & 20.2 & 2.63 & ZTF \\
SN\,2018cxa & 22:28:34.59 & +11:37:05.6 & 0.1900 & 90.4 & 226 & 38.0 & 0.99 & ZTF \\
SN\,2018don & 13:55:08.64 & +58:29:42.0 & 0.0734 & 112.7 & 282 & 56.7 & 2.80 & ZTF \\
SN\,2018fd & 09:10:36.36 & +35:43:18.4 & 0.2630 & 461.3 & 1153 & 42.7 & 2.23 & ZTF \\
SN\,2018ffj & 02:30:59.81 & -17:20:27.2 & 0.2340 & 92.0 & 230 & 65.0 & 2.24 & ZTF \\
SN\,2018ffs & 20:54:37.15 & +22:04:51.8 & 0.1410 & 69.1 & 173 & 14.5 & 2.43 & ZTF \\
SN\,2018gbw & 15:55:38.02 & +28:21:38.0 & 0.3454 & 117.6 & 294 & 49.5 & 2.88 & ZTF \\
SN\,2018gft & 23:57:17.94 & -15:37:53.3 & 0.2320 & 106.4 & 266 & 73.0 & 1.68 & ZTF \\
SN\,2018gkz & 07:58:11.55 & +19:31:07.9 & 0.2405 & 358.3 & 896 & 23.2 & 2.15 & ZTF \\
SN\,2018hpq & 18:28:41.26 & +75:48:47.4 & 0.1240 & 137.1 & 343 & 27.6 & 2.63 & ZTF \\
SN\,2018hti & 03:40:53.76 & +11:46:37.4 & 0.0612 & 159.7 & 399 & 33.4 & 1.39 & ATLAS \\
SN\,2018kyt & 12:27:56.24 & +56:23:35.6 & 0.1080 & 101.4 & 253 & 60.4 & 2.48 & ZTF \\
SN\,2018lfd & 23:14:59.32 & +48:45:27.8 & 0.2686 & 129.3 & 323 & 11.1 & 2.65 & ZTF \\
SN\,2018lfe & 09:33:29.56 & +00:03:08.4 & 0.3500 & 86.3 & 216 & 35.4 & 0.80 & ZTF \\
SN\,2018lzv & 12:44:02.32 & +56:01:44.5 & 0.4340 & 159.7 & 399 & 61.1 & 1.14 & ZTF \\
SN\,2018lzw & 07:39:32.76 & +27:44:02.7 & 0.3198 & 247.7 & 619 & 22.1 & 2.49 & ZTF \\
SN\,2018lzx & 22:29:27.24 & +13:10:39.8 & 0.4373 & 347.3 & 868 & 37.0 & 1.64 & ZTF \\
SN\,2019J & 10:03:46.77 & +06:46:24.4 & 0.1346 & 195.9 & 490 & 45.4 & 1.14 & ZTF \\
SN\,2019aamp & 14:37:49.27 & +20:18:16.6 & 0.4040 & 71.2 & 178 & 64.8 & 0.54 & ZTF \\
SN\,2019aamq & 20:55:36.14 & -08:40:31.4 & 0.3860 & 179.5 & 449 & 31.5 & 1.15 & ZTF \\
SN\,2019aamr & 15:29:23.55 & +38:06:12.6 & 0.4200 & 87.5 & 219 & 55.3 & 1.34 & ZTF \\
SN\,2019aams & 23:43:36.16 & +12:29:01.0 & 0.6360 & 41.1 & 103 & 47.1 & 2.95 & ZTF \\
SN\,2019aamt & 21:15:08.00 & +32:43:01.3 & 0.2138 & 116.1 & 290 & 11.1 & 0.97 & ZTF \\
SN\,2019aamu & 02:55:08.89 & +11:27:22.4 & 0.2590 & 110.5 & 276 & 41.1 & 3.16 & ZTF \\
SN\,2019aamv & 12:45:01.65 & +33:33:14.1 & 0.3996 & 295.2 & 738 & 83.4 & 2.58 & ZTF \\
SN\,2019aamw & 23:48:54.54 & +24:59:59.8 & 0.2200 & 155.2 & 388 & 35.7 & 2.99 & ZTF \\
SN\,2019aamx & 15:57:48.27 & +27:28:03.5 & 0.4100 & 101.1 & 253 & 48.8 & 3.53 & ZTF \\
SN\,2019bgu & 09:57:15.34 & +32:00:05.6 & 0.1480 & 80.7 & 202 & 52.1 & 0.41 & ZTF \\
SN\,2019cca & 12:02:51.40 & -16:39:47.7 & 0.4103 & 100.8 & 252 & 44.7 & 1.15 & ZTF \\
SN\,2019cdt & 08:17:53.93 & +65:28:46.5 & 0.1530 & 36.4 & 91 & 33.4 & 0.63 & ZTF \\
SN\,2019cwu & 14:51:37.30 & +48:59:13.7 & 0.3200 & 146.4 & 366 & 58.2 & 1.88 & ZTF \\
SN\,2019dgr & 09:45:32.68 & +04:56:02.2 & 0.3815 & 64.4 & 161 & 40.5 & 1.77 & ZTF \\
SN\,2019dlr & 11:17:34.18 & +00:30:02.8 & 0.2600 & 160.0 & 400 & 55.1 & 3.11 & ZTF \\
SN\,2019dwa & 15:38:57.48 & +56:36:18.2 & 0.0820 & 47.0 & 118 & 48.2 & 1.48 & ZTF \\
SN\,2019enz & 13:57:06.08 & +27:59:38.1 & 0.2550 & 34.8 & 87 & 75.4 & 2.90 & ATLAS \\
SN\,2019eot & 18:00:29.94 & +50:17:43.3 & 0.3057 & 67.8 & 170 & 28.4 & 3.38 & ZTF \\
SN\,2019gam & 10:19:18.33 & +17:12:43.4 & 0.1235 & 205.1 & 513 & 53.6 & 1.90 & ZTF \\
SN\,2019gfm & 15:35:46.59 & +24:03:44.9 & 0.1817 & 135.9 & 340 & 53.0 & 3.77 & ZTF \\
SN\,2019gqi & 14:21:12.01 & +28:54:05.9 & 0.3642 & 77.0 & 192 & 70.1 & 2.42 & ZTF \\
SN\,2019hge & 22:24:21.21 & +24:47:17.1 & 0.0866 & 120.3 & 301 & 27.1 & 1.74 & ZTF \\
SN\,2019hno & 19:39:12.96 & +62:43:40.9 & 0.2600 & 74.5 & 186 & 18.7 & 1.65 & ZTF \\
SN\,2019ieh & 16:42:10.84 & +06:59:02.5 & 0.0320 & 32.2 & 80 & 31.8 & 0.69 & ZTF \\
SN\,2019itq & 21:54:28.76 & -08:20:50.2 & 0.4810 & 142.8 & 357 & 44.2 & 0.82 & ZTF \\
SN\,2019kcy & 14:08:19.78 & +08:58:01.0 & 0.3990 & 96.5 & 241 & 64.3 & 4.36 & ZTF \\
SN\,2019kwq & 17:07:58.84 & +58:42:03.9 & 0.4900 & 198.3 & 496 & 36.3 & 2.02 & ZTF \\
SN\,2019kws & 14:15:04.46 & +50:39:06.8 & 0.1977 & 76.3 & 191 & 61.6 & 2.14 & ZTF \\
SN\,2019kwt & 19:39:22.59 & +78:45:43.6 & 0.3562 & 111.9 & 280 & 24.3 & 1.59 & ZTF \\
SN\,2019kwu & 13:57:39.77 & +64:21:18.6 & 0.6000 & 78.0 & 195 & 51.3 & 1.72 & ZTF \\
SN\,2019lsq & 00:04:40.60 & +42:52:11.3 & 0.1295 & 139.3 & 348 & 19.2 & 2.75 & ZTF \\
SN\,2019neq & 17:54:26.74 & +47:15:40.6 & 0.1059 & 28.0 & 70 & 28.9 & 2.45 & ZTF \\
SN\,2019nhs & 00:52:01.45 & +07:36:59.8 & 0.1890 & 50.0 & 125 & 55.3 & 2.07 & ZTF \\
SN\,2019obk & 22:33:54.09 & -02:09:42.3 & 0.1656 & 56.6 & 141 & 48.7 & 3.15 & ZTF \\
SN\,2019otl & 02:52:21.63 & -17:48:12.5 & 0.5140 & 151.6 & 379 & 60.6 & 0.76 & ZTF \\
SN\,2019pud & 21:12:55.01 & -16:38:07.1 & 0.1136 & 39.1 & 98 & 38.6 & 4.16 & ZTF \\
SN\,2019pvs & 21:18:26.79 & -20:44:00.2 & 0.1670 & 203.2 & 508 & 41.3 & 3.84 & ZTF \\
SN\,2019qgk & 22:29:57.55 & -04:06:02.2 & 0.3468 & 94.8 & 237 & 49.2 & 1.36 & ZTF \\
SN\,2019sgg & 01:01:11.77 & +14:01:35.4 & 0.5726 & 138.5 & 346 & 48.8 & 3.37 & ZTF \\
SN\,2019sgh & 01:12:39.42 & +36:28:24.8 & 0.3440 & 145.0 & 363 & 26.2 & 3.01 & ZTF \\
SN\,2019szu & 00:10:13.14 & -19:41:32.5 & 0.2120 & 561.4 & 1404 & 78.0 & 1.17 & ZTF \\
SN\,2019ujb & 09:03:15.17 & +40:14:32.5 & 0.2008 & 55.1 & 138 & 41.6 & 1.92 & ZTF \\
SN\,2019unb & 09:47:57.01 & +00:49:35.9 & 0.0635 & 221.5 & 554 & 38.8 & 0.49 & ZTF \\
SN\,2019vvc & 09:13:30.13 & +44:46:26.2 & 0.3314 & 196.0 & 490 & 43.5 & 1.32 & ZTF \\
SN\,2019xaq & 13:54:46.82 & +84:04:01.6 & 0.2000 & 114.2 & 286 & 32.8 & 1.19 & ZTF \\
SN\,2019xdy & 08:24:51.33 & +22:10:46.0 & 0.2206 & 64.8 & 162 & 29.9 & 0.41 & ZTF \\
SN\,2019zbv & 10:15:01.12 & +43:24:53.7 & 0.3785 & 230.9 & 577 & 54.6 & 1.02 & ZTF \\
SN\,2019zeu & 10:13:57.43 & +08:16:48.6 & 0.3900 & 72.6 & 181 & 48.3 & 1.83 & ZTF \\
SN\,2020abjx & 02:15:02.30 & -08:37:43.7 & 0.3900 & 213.3 & 533 & 62.9 & 0.56 & ZTF \\
SN\,2020afag & 00:15:46.25 & +47:00:08.5 & 0.3815 & 77.5 & 194 & 15.4 & 1.33 & ZTF \\
SN\,2020afah & 10:20:18.32 & +53:19:21.4 & 0.3754 & 211.0 & 528 & 51.8 & 2.65 & ZTF \\
SN\,2020ank & 08:16:14.65 & +04:19:26.9 & 0.2485 & 35.4 & 88 & 20.8 & 1.99 & ZTF \\
SN\,2020aup & 13:09:44.44 & +12:29:13.4 & 0.3100 & 100.0 & 250 & 74.7 & 0.59 & ZTF \\
SN\,2020auv & 16:34:12.53 & +37:05:51.8 & 0.2800 & 130.0 & 325 & 42.5 & 0.69 & ZTF \\
SN\,2020dlb & 08:08:34.14 & +34:44:12.9 & 0.3980 & 76.3 & 191 & 30.1 & 0.30 & ZTF \\
SN\,2020exj & 14:42:40.01 & +30:14:39.2 & 0.1216 & 121.9 & 305 & 65.5 & 1.24 & ZTF \\
SN\,2020fvm & 14:12:45.93 & +34:44:16.2 & 0.2428 & 144.2 & 360 & 71.0 & 1.53 & ZTF \\
SN\,2020fyq & 14:46:10.44 & +23:48:02.0 & 0.1765 & 297.2 & 743 & 63.9 & 1.33 & ZTF \\
SN\,2020htd & 17:44:17.28 & +38:55:30.4 & 0.3515 & 113.3 & 283 & 29.1 & 0.95 & ZTF \\
SN\,2020iyj & 09:15:36.64 & +53:27:32.0 & 0.3690 & 114.3 & 286 & 42.7 & 0.83 & ZTF \\
SN\,2020jii & 15:34:55.29 & +02:51:11.7 & 0.3960 & 68.0 & 170 & 44.0 & 1.33 & ZTF \\
SN\,2020kox & 11:06:04.97 & +26:17:28.7 & 0.4560 & 158.4 & 396 & 66.5 & 1.93 & ZTF \\
SN\,2020myh & 23:37:30.55 & +21:42:42.8 & 0.2830 & 68.1 & 170 & 38.0 & 0.46 & ZTF \\
SN\,2020onb & 14:23:00.60 & +49:10:40.7 & 0.1530 & 60.9 & 152 & 61.8 & 1.39 & ZTF \\
SN\,2020qef & 22:56:10.53 & +28:45:53.2 & 0.1831 & 83.3 & 208 & 27.7 & 2.99 & ZTF \\
SN\,2020qlb & 19:07:49.58 & +62:57:49.6 & 0.1585 & 206.9 & 517 & 22.2 & 2.44 & ZTF \\
SN\,2020rmv & 00:40:00.19 & -14:35:25.1 & 0.2621 & 370.3 & 926 & 77.2 & 3.22 & ZTF \\
SN\,2020tcw & 15:28:17.08 & +39:56:50.5 & 0.0640 & 90.0 & 225 & 55.2 & 1.52 & ZTF \\
SN\,2020uew & 03:19:23.87 & -29:59:05.7 & 0.2237 & 431.3 & 1078 & 57.5 & 3.37 & ATLAS \\
SN\,2020vpg & 01:38:04.19 & +34:43:27.9 & 0.2570 & 107.2 & 268 & 27.2 & 1.70 & ZTF \\
SN\,2020xga & 03:46:39.37 & -11:14:33.9 & 0.4400 & 63.1 & 158 & 46.0 & 1.01 & ZTF \\
SN\,2020xgd & 00:19:45.83 & +05:08:18.6 & 0.4540 & 55.8 & 139 & 56.8 & 1.21 & ZTF \\
SN\,2020xkv & 22:37:46.00 & +23:31:37.4 & 0.2410 & 211.5 & 529 & 29.9 & 2.18 & ZTF \\
SN\,2020zbf & 01:58:01.67 & -41:20:51.8 & 0.3500 & 189.2 & 473 & 70.3 & 0.47 & ATLAS \\
SN\,2020znr & 07:19:06.42 & +23:53:07.4 & 0.1000 & 108.0 & 270 & 16.4 & 2.59 & ZTF \\
SN\,2020zzb & 00:10:09.05 & +09:29:35.6 & 0.1659 & 185.4 & 463 & 52.0 & 3.02 & ZTF \\
SN\,2021bnw & 10:53:52.17 & +12:33:29.0 & 0.0980 & 71.1 & 178 & 58.9 & 2.16 & ZTF \\
SN\,2021een & 20:26:44.42 & +16:33:06.2 & 0.1600 & 58.1 & 145 & 12.4 & 2.03 & ZTF \\
SN\,2021ejo & 10:31:38.99 & +31:00:04.5 & 0.4400 & 242.8 & 607 & 59.3 & 5.29 & ZTF \\
SN\,2021ek & 03:23:49.91 & -10:02:41.2 & 0.1930 & 105.6 & 264 & 50.3 & 2.00 & ZTF \\
SN\,2021fpl & 20:14:18.62 & -18:10:56.6 & 0.1210 & 176.6 & 441 & 26.2 & 1.49 & ZTF \\
SN\,2021gtr & 17:38:28.03 & +48:29:47.6 & 0.3030 & 269.0 & 672 & 31.7 & 0.92 & ZTF \\
SN\,2021hpc & 13:33:01.07 & +43:10:33.9 & 0.2400 & 122.0 & 305 & 71.9 & 1.59 & ZTF \\
SN\,2021hpx & 09:31:06.26 & -19:31:04.8 & 0.2130 & 78.2 & 195 & 22.7 & 0.40 & ZTF \\
SN\,2021kty & 14:18:48.77 & +15:00:50.8 & 0.1590 & 152.0 & 380 & 66.4 & 1.44 & ZTF \\
SN\,2021lwz & 09:44:47.39 & +34:42:44.2 & 0.0650 & 75.9 & 190 & 49.6 & 2.71 & ZTF \\
SN\,2021mkr & 14:56:52.57 & +60:50:05.1 & 0.2800 & 67.2 & 168 & 50.2 & 3.15 & ZTF \\
SN\,2021nxq & 14:25:20.68 & +37:45:44.7 & 0.1500 & 250.7 & 627 & 67.6 & 0.45 & ZTF \\
SN\,2021rwz & 23:09:50.16 & +10:01:42.4 & 0.1900 & 182.7 & 457 & 45.4 & 3.76 & ZTF \\
SN\,2021txk & 21:27:49.27 & -08:04:58.3 & 0.4600 & 85.2 & 213 & 38.3 & 1.50 & ZTF \\
SN\,2021vuw & 06:40:47.46 & +27:47:01.8 & 0.2000 & 143.0 & 358 & 10.1 & 2.89 & ZTF \\
SN\,2021ynn & 02:50:35.44 & -19:51:47.2 & 0.2200 & 59.6 & 149 & 61.8 & 1.36 & ZTF \\
SN\,2021yrp & 00:23:17.57 & +50:08:43.5 & 0.3000 & 231.6 & 579 & 12.5 & 2.25 & ZTF \\
SN\,2021zcl & 05:09:14.45 & -06:03:13.9 & 0.1170 & 101.2 & 253 & 25.5 & 1.93 & ZTF \\
SN\,2022aawb & 01:44:52.38 & +23:00:43.9 & 0.1300 & 72.9 & 182 & 38.2 & 1.59 & ZTF \\
SN\,2022abdu & 03:06:04.91 & -46:43:17.0 & 0.1300 & 90.9 & 227 & 57.1 & 3.44 & ATLAS \\
SN\,2022acsx & 06:12:59.13 & +68:48:45.4 & 0.2800 & 177.6 & 444 & 21.8 & 1.42 & ZTF \\
SN\,2022ful & 19:20:10.67 & +50:23:42.4 & 0.1500 & 165.2 & 413 & 16.3 & 2.39 & ZTF \\
SN\,2022gyv & 12:57:59.61 & +41:04:42.1 & 0.3900 & 166.2 & 415 & 76.0 & 0.69 & ZTF \\
SN\,2022le & 10:50:27.86 & -06:17:48.1 & 0.2491 & 109.2 & 273 & 45.6 & 2.15 & ZTF \\
SN\,2022ljr & 15:43:02.08 & +18:04:36.0 & 0.2000 & 87.6 & 219 & 49.6 & 2.47 & ZTF \\
SN\,2022lxd & 17:36:38.68 & +61:33:18.7 & 0.5400 & 156.0 & 390 & 32.5 & 1.95 & ZTF \\
SN\,2022npq & 16:21:03.73 & +14:50:54.7 & 0.2600 & 75.5 & 189 & 39.9 & 2.28 & ZTF \\
SN\,2022pjq & 21:48:58.16 & -26:22:36.8 & 0.1700 & 40.3 & 101 & 49.5 & 1.52 & ZTF \\
SN\,2022ued & 08:54:45.99 & +78:49:48.6 & 0.1087 & 244.3 & 611 & 32.4 & 3.05 & ZTF \\
SN\,2023cmx & 07:06:54.65 & +37:58:35.5 & 0.2330 & 91.8 & 229 & 19.1 & 0.25 & ZTF \\
SN\,CSS160710 & 16:04:19.54 & +39:28:12.7 & 0.1800 & 38.5 & 96 & 48.4 & 2.89 & CSS \\
SN\,DES14C1fi & 03:33:49.80 & -27:03:31.6 & 1.3020 & 69.1 & 173 & 54.0 & 2.08 & DES \\
SN\,DES14C1rhg & 03:38:07.27 & -27:42:45.7 & 0.4810 & 80.5 & 201 & 53.2 & 1.10 & DES \\
SN\,DES14E2slp & 00:33:04.08 & -44:11:42.8 & 0.5700 & 188.2 & 471 & 72.5 & 1.85 & DES \\
SN\,DES14S2qri & 02:43:32.14 & -01:07:34.2 & 1.5000 & 130.8 & 327 & 52.6 & 1.14 & DES \\
SN\,DES14X2byo & 02:23:46.93 & -06:08:12.3 & 0.8680 & 29.3 & 73 & 59.6 & 0.76 & DES \\
SN\,DES14X3taz & 02:28:04.46 & -04:05:12.7 & 0.6080 & 121.6 & 304 & 57.4 & 1.83 & DES \\
SN\,DES15C3hav & 03:31:52.17 & -28:15:09.5 & 0.3920 & 102.2 & 256 & 54.6 & 1.65 & DES \\
SN\,DES15E2mlf & 00:41:33.40 & -43:27:17.2 & 1.8610 & 59.8 & 149 & 73.6 & 2.13 & DES \\
SN\,DES15S1nog & 02:52:14.98 & -00:44:36.3 & 0.5650 & 82.7 & 207 & 50.8 & 0.71 & DES \\
SN\,DES15X1noe & 02:14:41.93 & -04:52:54.5 & 1.1880 & 128.3 & 321 & 60.2 & 2.09 & DES \\
SN\,DES15X3hm & 02:26:54.96 & -05:03:38.0 & 0.8600 & 38.7 & 97 & 58.3 & 0.83 & DES \\
SN\,DES16C2aix & 03:40:41.17 & -29:22:48.4 & 1.0680 & 89.5 & 224 & 52.9 & 0.69 & DES \\
SN\,DES16C2nm & 03:40:14.83 & -29:05:53.5 & 1.9980 & 232.4 & 581 & 52.9 & 0.41 & DES \\
SN\,DES16C3dmp & 03:31:28.35 & -28:32:28.3 & 0.5620 & 89.2 & 223 & 54.8 & 1.67 & DES \\
SN\,DES16C3ggu & 03:31:12.00 & -28:34:38.7 & 0.9490 & 248.6 & 621 & 54.8 & 1.72 & DES \\
SN\,DES17C3gyp & 03:27:51.87 & -28:23:44.3 & 0.4700 & 170.2 & 426 & 55.5 & 2.47 & DES \\
SN\,DES17X1amf & 02:17:46.70 & -05:36:01.0 & 0.9200 & 158.8 & 397 & 60.2 & 2.03 & DES \\
SN\,DES17X1blv & 02:20:59.64 & -04:29:00.8 & 0.6900 & 98.9 & 247 & 58.9 & 1.99 & DES \\
SN\,LSQ12dlf & 01:50:29.80 & -21:48:45.4 & 0.2550 & 124.0 & 310 & 75.6 & 2.61 & LSQ \\
SN\,LSQ14an & 12:53:47.83 & -29:31:27.2 & 0.1637 & 234.7 & 587 & 33.3 & 1.69 & LSQ \\
SN\,LSQ14bdq & 10:01:41.60 & -12:22:13.4 & 0.3450 & 134.9 & 337 & 32.9 & 1.79 & LSQ \\
SN\,LSQ14mo & 10:22:41.53 & -16:55:14.4 & 0.2560 & 41.2 & 103 & 33.1 & 1.36 & LSQ \\
SN\,OGLE15qz & 03:08:35.88 & -70:30:41.6 & 0.6800 & 175.2 & 438 & 42.5 & 1.54 & OGLE \\
SN\,OGLE15sd & 01:42:21.46 & -71:47:15.6 & 0.5650 & 98.7 & 247 & 44.8 & 3.84 & OGLE \\
SN\,OGLE15xl & 04:20:53.45 & -73:36:20.2 & 0.1980 & 29.4 & 73 & 36.2 & 1.22 & OGLE \\
SN\,OGLE15xx & 05:35:40.35 & -66:20:48.5 & 0.2100 & 34.0 & 85 & 32.2 & 1.26 & OGLE \\
SN\,OGLE16dmu & 04:48:26.34 & -62:20:10.6 & 0.4260 & 254.3 & 636 & 38.0 & 1.35 & OGLE \\
SN\,PS110awh & 22:14:29.83 & -00:04:03.6 & 0.9084 & 57.8 & 144 & 43.6 & 1.97 & PS1 \\
SN\,PS110bzj & 03:31:39.83 & -27:47:42.2 & 0.6500 & 66.2 & 166 & 54.6 & 1.90 & PS1 \\
SN\,PS110ky & 22:13:37.85 & +01:14:23.6 & 0.9558 & 60.0 & 150 & 42.6 & 2.46 & PS1 \\
SN\,PS110pm & 12:12:42.20 & +46:59:29.5 & 1.2060 & 54.2 & 135 & 68.7 & 2.77 & PS1 \\
SN\,PS111afv & 12:15:37.77 & +48:10:48.6 & 1.4070 & 42.5 & 106 & 67.8 & 1.87 & PS1 \\
SN\,PS111aib & 22:18:12.22 & +01:33:32.0 & 0.9970 & 255.5 & 639 & 43.3 & 2.61 & PS1 \\
SN\,PS111ap & 10:48:27.73 & +57:09:09.2 & 0.5240 & 84.4 & 211 & 52.9 & 1.46 & PS1 \\
SN\,PS111bam & 08:41:14.19 & +44:01:56.9 & 1.5650 & 96.4 & 241 & 37.7 & 1.18 & PS1 \\
SN\,PS111bdn & 02:25:46.29 & -05:03:56.6 & 0.7380 & 61.9 & 155 & 58.5 & 0.85 & PS1 \\
SN\,PS111tt & 16:12:45.78 & +54:04:17.0 & 1.2830 & 78.7 & 197 & 44.6 & 1.10 & PS1 \\
SN\,PS112bmy & 03:34:13.12 & -26:31:17.2 & 1.5720 & 135.1 & 338 & 53.8 & 2.28 & PS1 \\
SN\,PS112bqf & 02:24:54.62 & -04:50:22.7 & 0.5220 & 117.2 & 293 & 58.5 & 1.13 & PS1 \\
SN\,PS112cil & 08:40:56.17 & +45:24:41.9 & 0.3200 & 44.8 & 112 & 37.7 & 1.76 & PS1 \\
SN\,PS113gt & 12:18:02.04 & +47:34:46.0 & 0.8840 & 99.3 & 248 & 68.5 & 2.26 & PS1 \\
SN\,PS113or & 09:54:40.30 & +02:11:42.2 & 1.5200 & 79.4 & 199 & 40.9 & 2.31 & PS1 \\
SN\,PS114bj & 10:02:08.43 & +03:39:19.0 & 0.5215 & 204.1 & 510 & 43.3 & 3.07 & PS1 \\
SN\,PS15cjz & 02:40:44.62 & -00:53:26.4 & 0.2200 & 103.2 & 258 & 52.9 & 1.01 & DES \\
SN\,PTF09atu & 16:30:24.55 & +23:38:25.0 & 0.5015 & 163.4 & 409 & 40.8 & 2.56 & PTF \\
SN\,PTF09cnd & 16:12:08.94 & +51:29:16.1 & 0.2584 & 103.2 & 258 & 45.4 & 1.87 & PTF \\
SN\,PTF10aagc & 09:39:56.92 & +21:43:17.1 & 0.2060 & 71.1 & 178 & 46.3 & 3.03 & PTF \\
SN\,PTF10bfz & 12:54:41.27 & +15:24:17.0 & 0.1701 & 101.6 & 254 & 78.3 & 1.80 & PTF \\
SN\,PTF10bjp & 10:06:34.30 & +67:59:19.0 & 0.3584 & 146.5 & 366 & 42.4 & 1.94 & PTF \\
SN\,PTF10iam & 15:45:30.85 & +54:02:33.0 & 0.1090 & 83.3 & 208 & 48.4 & 1.26 & PTF \\
SN\,PTF10nmn & 15:50:02.81 & -07:24:42.4 & 0.1237 & 320.3 & 801 & 34.7 & 0.41 & PTF \\
SN\,PTF10uhf & 16:52:46.70 & +47:36:21.8 & 0.2882 & 61.0 & 152 & 39.3 & 0.99 & PTF \\
SN\,PTF10vqv & 03:03:06.80 & -01:32:34.9 & 0.4518 & 127.7 & 319 & 49.4 & 2.47 & PTF \\
SN\,PTF12dam & 14:24:46.20 & +46:13:48.3 & 0.1073 & 102.0 & 255 & 63.5 & 2.76 & PTF \\
SN\,PTF12gty & 16:01:15.23 & +21:23:17.4 & 0.1768 & 123.9 & 310 & 46.6 & 1.72 & PTF \\
SN\,PTF12hni & 22:31:55.86 & -06:47:49.0 & 0.1056 & 109.0 & 272 & 51.2 & 1.83 & PTF \\
SN\,PTF12mxx & 22:30:16.73 & +27:58:22.0 & 0.3296 & 70.2 & 176 & 25.3 & 1.44 & PTF \\
SN\,SSS120810 & 23:18:01.80 & -56:09:25.6 & 0.1560 & 109.2 & 273 & 56.5 & 2.23 & Other \\
SN\,iPTF13ajg & 16:39:03.95 & +37:01:38.4 & 0.7403 & 80.0 & 200 & 41.5 & 0.87 & iPTF \\
SN\,iPTF13bdl & 12:36:56.14 & +13:07:45.5 & 0.4030 & 457.6 & 1144 & 75.6 & 1.04 & iPTF \\
SN\,iPTF13bjz & 10:38:19.83 & +24:24:51.3 & 0.2712 & 121.8 & 305 & 60.0 & 1.15 & iPTF \\
SN\,iPTF13cjq & 00:14:27.18 & +24:17:08.8 & 0.3962 & 87.7 & 219 & 37.8 & 0.52 & iPTF \\
SN\,iPTF13dcc & 02:57:02.49 & -00:18:44.1 & 0.4305 & 233.4 & 583 & 49.7 & 1.44 & iPTF \\
SN\,iPTF13ehe & 06:53:21.50 & +67:07:56.0 & 0.3434 & 173.6 & 434 & 25.0 & 1.72 & iPTF \\
SN\,iPTF15eov & 04:05:41.57 & +28:56:40.7 & 0.0535 & 46.5 & 116 & 17.1 & 3.19 & iPTF \\
SN\,iPTF16asu & 12:59:09.36 & +13:48:10.8 & 0.1870 & 47.7 & 119 & 76.6 & 0.57 & iPTF \\
SN\,iPTF16bad & 17:16:39.73 & +28:22:12.6 & 0.2467 & 121.2 & 303 & 32.1 & 1.45 & iPTF \\
SN\,iPTF16eh & 12:41:06.21 & +32:48:30.9 & 0.4270 & 160.7 & 402 & 83.9 & 2.57 & iPTF \\
\end{longtable*}}

\end{document}